\documentclass[a4paper,titlepage]{article}
\usepackage[pdftex,hidelinks]{hyperref} 
\usepackage{times}
\usepackage{mathptmx}
\usepackage{comment}
\usepackage{url}
\usepackage[utf8]{inputenc}
\usepackage[authoryear,round]{natbib}
\usepackage{doi}

\usepackage{graphicx}
\usepackage{epstopdf}

\usepackage{array}
\usepackage{amsfonts}
\usepackage{amsmath}
\usepackage{amsthm}
\usepackage{amssymb}
\usepackage{bm}
\usepackage{enumerate}

\newcommand{\figref}[1]{Fig.~\ref{#1}}
\newcommand{\sectref}[1]{Sect.~\ref{#1}}

\newcommand{\myvec}[1]{{\bm {#1}}}  
\newcommand{\myqed}[0]{\qed}             
\newcommand{\polar}[0]{polar}

\newcounter{ex}
\newcommand{\qreff}[2]{%
	\refstepcounter{ex}%
	\textbf{Example~\theex}\label{#2}%
	\space(#1)%
}

\DeclareMathOperator{\argmin}{arg~min}
\DeclareMathOperator{\tr}{tr}
\DeclareMathOperator{\Cov}{Cov}

\DeclareMathOperator{\Col}{Col}

\DeclareMathOperator{\Conic}{Conic}

\begin{document}

\title{Information Processing by Neuron Populations in the Central Nervous System: \\
    A Theory of the Mathematical Structure \\ of Data and Operations \thispagestyle{empty}	
}

\author{Martin N. P. Nilsson \\ 
	RISE Research Institutes of Sweden, \\
	P.O.B. 1263, SE-164 29 Kista, Sweden \\
	{\tt Email:~martin.nilsson@ri.se} \\
	ORCID: 0000-0002-7504-0328 \\
}

\hypersetup{
	pdftitle={Information Processing by Neuron Populations in the Central Nervous System: A Theory of the Mathematical Structure of Data and Operations},
	pdfauthor={Martin N. P. Nilsson},
	pdfkeywords={
		population code, 
		algebra of convex cones,
		sparsity,
		mechanistic model,
		invariant, 
		matrix embedding,
		Moreau decomposition,
		non-negative weights,
		adaptive filter,
		correlation,
		plasticity}
}

\maketitle

\begin{abstract}
\setcounter{page}{2}

	In the mammalian central nervous system, neurons are organized into populations communicating
	by spike trains propagating along axonal bundles. How such populations
	encode and transform information is only partially understood.
	
In this study we introduce a mathematical framework derived from a mechanistic
model of a single plastic neuron. Within this framework, an algebra of convex cones
can rigorously characterize population-level activity. This algebra provides a natural
language describing information representation and processing.
	
	Neuron populations are thereby interpreted not as passive transmitters but as operators acting within this algebraic structure. When interconnected, such populations realize compact algebraic expressions whose functional repertoire includes specialization, generalization, novelty detection, dimensionality reduction, inverse modeling, prediction, and associative memory.
	
	Finally, the approach highlights the role of matrix embeddings in extending representational capacity beyond that afforded by vector-based models. In particular, such embeddings support hierarchical concept formation and structured information processing, with potential implications for both cognitive neuroscience and artificial intelligence.
	
	This paper assumes familiarity with elementary functional analysis and algebras of operators.
	\vskip 1cm
	{\bf Significance statement:}
	This research studies the intriguing interplay between mathematics,
	neurobiology, and cognitive processes. The paper investigates how
	the brain might represent and organize knowledge using mathematical structures
	known as convex cones. These structures are proposed as a bridge between
	the raw, biological workings of the brain and our high-level cognitive
	processes. The findings highlight the potential of neuron populations in the
	brain to carry out complex operations, much like how a computer processes
	information. Moreover, the study touches on the possible existence of a
	``brain language'' spoken by neuron circuits, offering insights into both
	artificial intelligence and our understanding of the human mind.
\end{abstract}


{\bf MSC 2020 terms:}
Mathematical modeling or simulation for problems pertaining to biology [92-10],
Neural biology [92C20].  	
Convex sets and cones of operators [47L07],
Applications of operator algebras to the sciences [47L90],  	
Systems biology, networks [92C42],
Neural networks in biological studies [92B20],
Learning and adaptive systems in artificial intelligence [68T05].
	
{\bf MeSH 2023 terms:}
Neurons [A08.675],
Neurological Models [E05.599.395.642],
Concept Formation [F02.463.785.233],
Learning [F02.463.425],
Computational Biology [H01.158.273.180],
Cognitive Neuroscience [F04.096.628.255.500, H01.158.610.030],
Artificial Intelligence [G17.035.250,L01.224.050.375].

{\bf PhySH 2023 terms:}
Biological information processing,
Neural basis of learning \& memory,
Neural encoding,
Neuroplasticity,
Biological neural networks.

\setcounter{page}{3}

\section{Introduction}
\subsection{The quest for the elusive invariant}
Neurons in the mammalian central nervous system (CNS) form populations whose collective activity
encodes and transforms information. These ensembles communicate via bundles of axons, conveying
information by spike trains. The pioneering scientist John von Neumann initially emphasized
information in individual spike trains. However, at the end of his final work \emph{The Computer and the Brain}
\citep[p. 81]{VonNeumann.1958tca} he recognized the plausibility of encodings based on
statistical relationships \emph{between} spike trains: \emph{``It is [...] perfectly plausible that
	certain (statistical) relationships between [...] trains of pulses should also transmit information.
	In this connection it is natural to think of various correlation-coefficients, and the like.''}
It has indeed become increasingly clear that resolving neuronal communication requires understanding
how \emph{populations} encode information
\citep{Averbeck.et.al.2006ncp,Sakurai.1996pcb,Sanger.2003npc}, and accumulating evidence suggests that
population codes are shaped not only by marginal firing rates but also by correlations
\citep{Panzeri.et.al.2022tsa,Kohn.et.al.2016can}.

A central question is what aspects of population activity remain \emph{stable} as signals propagate
through noisy and heterogeneous biological circuitry. Such stability is often formulated in terms of
\emph{invariants}: properties of a family of activity patterns that are preserved, or preserved up to
a principled equivalence, by the transformations implemented by neuronal populations. The challenge of
maintaining stable invariants in an unreliable system like the nervous system was already pondered by
\citet{VonNeumann.1956pla}. \citet{Perkel.Bullock.1968nc} coined the term
\emph{transformation invariance}. \citet{Pellionisz.Llinas.1985tnt} proposed tensorial invariants under
coordinate transformations as one possible mathematical formalization of information transmission
through populations. Other researchers emphasizing invariance include \citet{MacLennan.1995cca} and
\citet{Kumar.et.al.2010sap}.

This paper focuses on invariants that arise naturally when population activity is represented in a
high-dimensional \emph{message space} and transformed by neuron populations. Our goal is not to claim a
single universal coding mechanism across all CNS circuits. Rather, we ask a more specific question:
\emph{given a widely used and analytically tractable representation of population activity (rates or
	low-pass filtered rates) and a conservative class of neuron and synapse constraints, what
	representational invariants and algebraic operations follow at the population level?} Under these
assumptions we derive a mathematical structure with a computational interpretation.

\subsection{A simple population model}
\label{sect:simplistic-neuron-model}
We begin with the classical linear model of a neuron that outputs a weighted sum of its inputs. In
this article, we represent inter-neuronal communication---including inputs $x_k(t)$ and output $z(t)$---
as \emph{spike rates} or, more generally, as low-pass filtered proxies of spiking activity. Using
standard signal processing terminology, $x_k(t)$ and $z(t)$ are \emph{signals}, {\it i.e.}, time-varying
functions.

\begin{figure}[!hbt]
	\includegraphics[width=\textwidth]{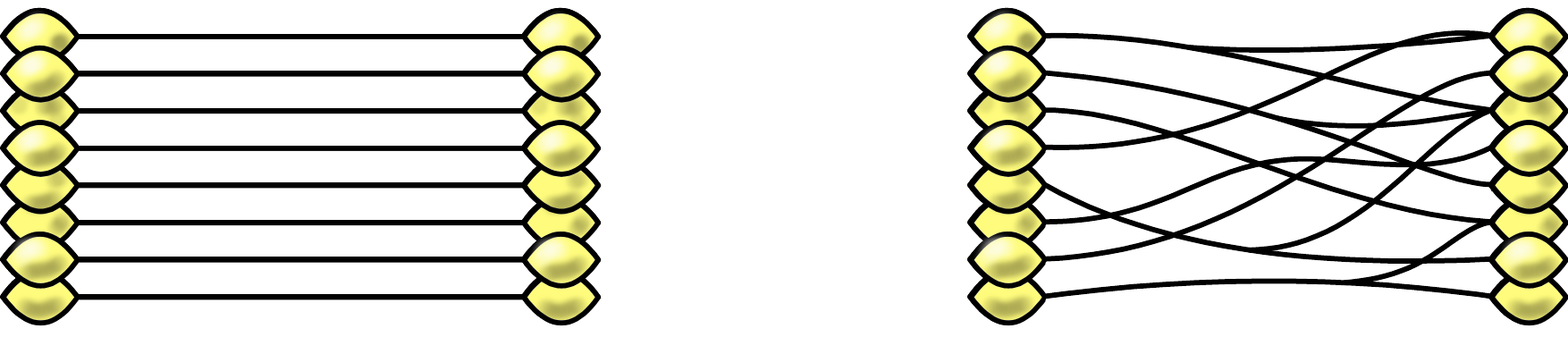}
	\caption{{\bf Interconnected neuron populations.}
		Population-to-population communication can be described abstractly as mappings between
		high-dimensional activity patterns. Left: an idealized topographic mapping, in which
		neighboring source neurons tend to connect to neighboring targets. Right: a diffuse and
		divergent mapping, in which spatial adjacency does not determine the target neurons. Such
		mappings are compatible with weak and heterogeneous correlations between nearby neurons in
		many central populations.
		\label{fig:axons-and-populations}}
\end{figure}

For the introductory development, we assume signals are \emph{pseudo-static}, meaning they vary
slowly enough to be treated as piecewise constant over the timescale of interest. We will often omit
the time parameter $t$ unless needed. The neuron model is
\begin{equation}
	\label{eq:classical-neuron-model}
	z = \sum_{k=1}^n w_k x_k ,
\end{equation}
where $w_k$ are synaptic weights. Although this model greatly simplifies biological neurons, it
provides a transparent entry point for understanding population-level invariants.
A population of $m$ neurons (\figref{fig:axons-and-populations}) can be described as
\begin{equation}
	\label{eq:simple-pop}
	\myvec{z} = W^T \myvec{x} ,
\end{equation}
where $\myvec{x} = (x_1, x_2,\ldots, x_n)^T$ is the vector of inputs, $\myvec{z}= (z_1, z_2,\ldots,
z_m)^T$ is the vector of outputs, and $W$ is the $n \times m$ matrix of synaptic weights;
the superscript $T$ denotes matrix or vector transpose.

\subsection{Messages}
We use the term \emph{message} to denote an instantaneous population-activity vector $\myvec{x}(t)$.
Depending on context, $\myvec{x}(t)$ may represent sensory-driven (afferent) activity arriving via ascending
pathways, internal cortical/subcortical activity, or a mixture of both. We seek invariants of the
mapping \eqref{eq:simple-pop} that capture what is preserved (or preserved up to equivalence) when
messages pass through a population.

Simple aggregate statistics ({\it e.g.}, local averages) can be stable in systems where nearby channels are
strongly correlated and organization is smooth. Such intuitions are often applicable in early sensory
representations that retain topographic structure across successive processing stages ({\it e.g.},
somatosensory, visual, or auditory pathways). However, many central population representations are
high-dimensional and exhibit only weak correlations between nearby recorded units
\citep{Ecker.et.al.2010dnf,Eggermont.1990tcb,Mogensen.et.al.2019aor,Zohary.et.al.1994cnd}. In such
settings, identifying robust invariants becomes nontrivial, particularly because an input message
$\myvec{x}$ can be mapped by $W$ to a wide range of output messages $\myvec{z}$ depending on synaptic
structure and network state.

\subsection{The invariant}
\begin{figure}[!hbt]
	\centering
	\includegraphics[scale=0.9]{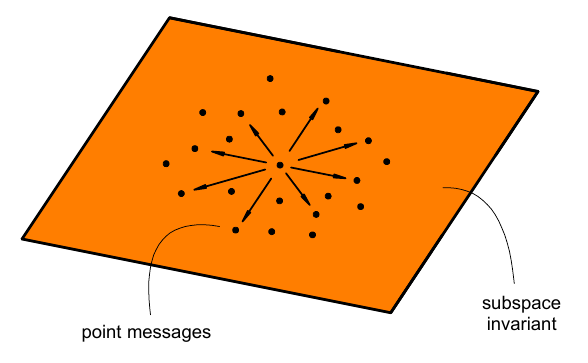}
	\caption{{\bf Messages and invariants.}
		Geometrically, instantaneous spike-rate patterns are points in a high-dimensional space.
		Invariants characterize stable structure associated with \emph{families} of messages, not with
		single messages.
		\label{fig:message-and-concepts}}
\end{figure}

Rather than singular messages, invariants are associated with \emph{sets} or \emph{ensembles} of
messages, represented geometrically as clouds of points. While it may be tempting to interpret such an
information-carrying invariant as a ``concept,'' we use the more neutral term ``invariant,'' leaving
interpretations for the discussion in \sectref{sect:discussion-interpretations}. A simple candidate invariant is the
\emph{linear subspace} spanned by the message cloud (\figref{fig:message-and-concepts}). Under linear
population mappings, subspaces map to subspaces and retain principled relations with other subspaces.
Operationally, the eigenvectors of the message covariance matrix provide a natural basis for such a
subspace.

This subspace viewpoint is a useful first step, but it is not the endpoint of the present paper. When
biophysical constraints impose non-negativity on effective synaptic mixing,
the natural invariant objects generalize from subspaces to \emph{convex
cones}. The main purpose of the paper is to make this transition explicit and to develop the
corresponding algebra of operations.

\subsection{Sparsity}
\label{sect:sparsity}
The importance of sparsity, characterized by many components in the message vector being zero, has
been discussed by several authors
\citep{Ganguli.Sompolinsky.2012css,Olshausen.Field.1997scw,Foldiak.1990fsr,Kanerva.1988sdm}.
It is also observed experimentally in the nervous system
\citep{Ganmor.et.al.2011slo,Olshausen.Field.2004sco,Machens.et.al.2010fbn,QuianQuiroga.et.al.2008sbn,Yu.et.al.2009gpf}
and can support robust representation in high dimensions.

First, sparsity permits a combinatorially large repertoire of messages.
For a message vector $\myvec{v}=(v_i)$, its {\em support} is the set of indices
of its non-zero components:
\begin{equation}
	\operatorname{supp}(\myvec{v})
	\triangleq
	\{i\mid v_i\neq0\}.
\end{equation}
A message with
$|\operatorname{supp}(\myvec{v})|=k$ active components among $n$
therefore has $\binom{n}{k}$ possible {\em support patterns}. If $k$ is a fixed small
fraction of $n$, this number grows exponentially with $n$.
These messages cannot all be linearly independent or exactly orthogonal, but
neither property is required for them to remain distinguishable. The
Johnson--Lindenstrauss lemma
\citep{Johnson.Lindenstrauss.1984eol,Dasgupta.Gupta.2002aep}
provides the complementary geometric result that $N$ points can be
represented in \\ $O(\varepsilon^{-2}\log N)$ dimensions while preserving
their pairwise distances to within a relative error $\varepsilon$.
Equivalently, a space of dimension $n$ can support an exponentially
large set of approximately distance-preserved messages. Sparsity
therefore supplies a combinatorially large message repertoire, while
high-dimensional geometry allows these messages, and the invariants
formed from their sequences, to remain reliably distinguishable.

Second, sparsity helps preserve the geometry---and therefore the meaning---of cone-valued messages
as they pass between populations. We return to this point in
\sectref{sect:activation-function} and \sectref{sect:approximate-cone-invariance}, where we show how
sparse messages can retain approximate equivariance under rectangular, heterogeneous biological
transmission maps.

Additional benefits of sparsity include simplified noise reduction and, for biological neurons,
reduced metabolic load.

\subsection{An algebra of invariants}
The set of linear subspaces of $\mathbb{R}^n$ forms an \emph{algebra of subspaces} equipped with
operations such as \emph{intersection}, \emph{sum}, \emph{projection}, and \emph{rejection}. As we show
in \sectref{sect:mathematical-model}, neuron populations can implement generalizations of these
operations. Under the positivity constraints introduced later, the invariant objects become convex
cones, leading to an \emph{algebra of cones} with a computational interpretation.

In this paper we use the term ``operator'' for the mapping implemented by a population at the
timescale of interest. These operators can be connected feedforward or recurrently. Recurrence
changes the network architecture, not the definition of the population operator.

\subsection{Outline of the paper}
Overall, we explore how neuron populations can implement an algebra of invariants for communication,
processing, and storage of information \emph{within the stated modeling assumptions}.

We begin with a classical linear neuron model with pseudo-static inputs to build intuition using
elementary linear algebra and geometric visualization. In \sectref{sect:subspace-operations} we develop
the algebra of subspaces and its basic operations. This is helpful in its own right and provides a
baseline for the cone generalization.

Introducing a mechanistically motivated neuron model in \sectref{sect:mathematical-model} requires
tools for time-variable signals. In \sectref{sect:algebra-of-cones}, which is the mathematical core of the paper,
we generalize the algebra of
subspaces to convex cones, leveraging positivity constraints on effective synaptic mixing.
\sectref{sect:neuronal-implementation} illustrates how neuron populations can implement basic cone
operations, composite operations, conditionals, and memory operations in this formalism.
The section also illustrates recurrent processing with a
population-level circuit between the red nucleus and spinal circuitry,
consistent with experimental evidence for pathways in both directions,
and shows how this circuit can be described using the cone algebra.

We summarize the results in \sectref{sect:results}, discuss related research, limitations
and extensions in \sectref{sect:discussion}, and conclude the paper in \sectref{sect:conclusions}.

\section{Representation and manipulation of subspaces}
\label{sect:subspace-operations}
This section introduces matrix representations of subspaces and the operations 
later generalized for cones. The matrix formulation also supports numerical experiments and visualization.

In the following, lowercase letters denote subspaces and uppercase letters
matrices while boldface denote vectors. The notation $A \sim B$ says that $A$ and
$B$ represent the same subspace. If $a$ is a subspace and $A$ a matrix
representing it, $a \sim A$ is equivalent to $a = \Col(A) = \{\myvec{x} \in
\mathbb{R}^n | \myvec{x} = A\myvec{\theta}, \myvec{\theta} \in \mathbb{R}^m\}$,
also known as the {\em column space} or {\em range} of the matrix $A$;
here $\myvec{\theta}$ denotes the vector of
linear-combination coefficients.
The {\em linear hull} of a set of vectors $s$ is the set $\operatorname{span}(s)$ of linear
combinations of vectors in $s$.
\subsection{Representations of subspaces}
\begin{figure}[!hbt] 
	\includegraphics[width=\textwidth]{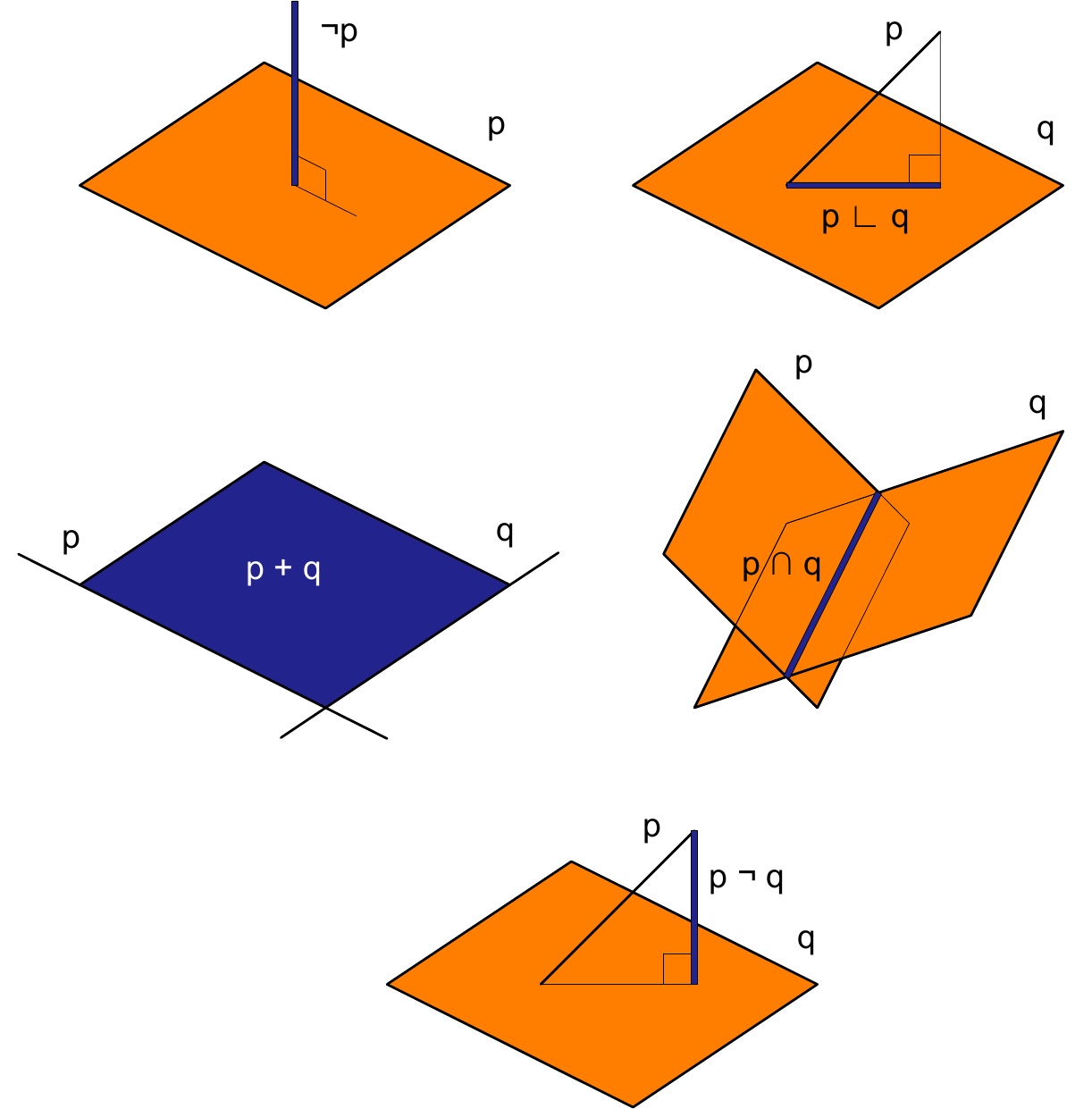}
	\caption{{\bf Basic operations on subspaces.}
		The five basic operations on subspaces are illustrated in
		left-to-right and top-down order:
		{\em orthogonal complement},
		{\em orthogonal projection},
		{\em sum},
		{\em intersection},
		and {\em orthogonal rejection}. 
		\label{fig:basic-operations}}
\end{figure}
One way to represent a subspace $s \subseteq \mathbb{R}^n$
is by an $n \times m$ matrix
\begin{equation}
	\label{eq:datamatrix}
	A = \left(\myvec{a}_1\quad\myvec{a}_2\quad\ldots\quad\myvec{a}_m \right) \in \mathbb{R}^{n \times m}
\end{equation}
of column vectors
$\{\myvec{a}_1,\myvec{a}_2,\ldots,\myvec{a}_m\}\subseteq\mathbb{R}^n$
whose linear hull is the subspace $s$,
\begin{equation}
	s = 
	\operatorname{span}{\{\myvec{a}_1, \myvec{a}_2, \ldots, \myvec{a}_m \}}
	= \Col(A) .
\end{equation}
This representation of $s$ is not unique, but
if we restrict the matrices to
{\em orthogonal projection matrices} $P$, which
satisfy $P^2 = P = P^T$, we obtain a canonical representation.
The matrix $P$ can always be generated
from $A$ by $P = AA^+$, where $A^+$ denotes the Moore-Penrose pseudoinverse
of $A$. 
\sectref{app:representation-of-subspaces} shows that $A \sim AA^T \sim AA^+$.

We can interpret the matrix $A$ 
statistically as a collection of data vectors:
Suppose that the $m$ observation vectors $\myvec{a}_k$ in
\eqref{eq:datamatrix} are independent and identically distributed samples of an 
$n$-dimensional stochastic variable
$\myvec{a}$
with a zero-mean probability distribution.
Then $AA^T$ is approximately proportional to the
{\em covariance matrix},
$AA^T \approx m\, \Cov(\myvec{a})$.
We can think of the subspace as a ``playground''
for messages.
The eigenvectors of the covariance matrix
having the largest eigenvalues
correspond to the most fluctuating directions inside the subspace. 
These eigenvectors are sometimes referred to as
the principal components.

\subsection{Derivation of subspace representation}
\label{app:representation-of-subspaces}
It is well known that any matrix $A$ can be decomposed by 
singular-value decomposition (SVD) as the matrix product $U\Sigma V^T$, where
$U$ and $V$ are orthonormal matrices, and $\Sigma$ is
a diagonal matrix.
\begin{align}
	\Col(A) & = \{\myvec{x}=A\myvec{\theta}|\myvec{\theta} \in \mathbb{R}^m\}  \\
	& = \{\myvec{x}=U \Sigma V^T \myvec{\theta}|\myvec{\theta} \in \mathbb{R}^m\}  \\
	& = \{\myvec{x}=U \Sigma \myvec{\theta}|\myvec{\theta} \in \mathbb{R}^m\}  \\
	& = \{\myvec{x}=U \Sigma \Sigma^T \myvec{\theta}|\myvec{\theta} \in \mathbb{R}^n\}  \\
	& = \{\myvec{x}=U \Sigma \Sigma^T U^T\myvec{\theta}|\myvec{\theta} \in \mathbb{R}^n\}  \\
	& =\Col(AA^T) ,
\end{align}
implying that $A \sim AA^T$. Replacing $\Sigma^T$ with $\Sigma^+$ above similarly leads to $\Col(A) = \Col(AA^+)$
and $A \sim AA^+$.

\subsection{Basic operations on subspaces}
Five basic operations can be performed on subspaces, as shown in \figref{fig:basic-operations}. Each of these operations has an intuitive geometric interpretation. We defer speculative conceptual interpretations of the operations to the discussion in \sectref{sect:discussion-interpretations}.
Below, we describe the operations in detail, introduce symbols for them, and define them in terms of matrix operations.
Given that $P$ and $Q$ are matrices corresponding to subspaces $p$ and $q$, respectively,
the five basic operations are
{\em orthogonal complement} (unary symbol `$\neg$'),
{\em orthogonal projection} (binary `$\llcorner$'),
{\em intersection} (binary `$\cap$'),
{\em sum} (binary `$+$'),
and {\em orthogonal rejection} (binary `$\neg$').
The operator symbols precedence order is assumed to be, from high to low,
`$\neg$', `$\llcorner$', `$\cap$', `$+$', `$=$', `$\subseteq$', and `$\sim$'.
All complements, projections, and rejections in this paper are assumed orthogonal, so this will rarely be explicitly stated.

The basic operations are defined as
\begin{itemize}
	\item{\bf Complement:}
	The {\em complement} of subspace $p$ is the set of all vectors
	perpendicular to $p$. It can be computed by the matrix subtraction
	\begin{equation}
		\neg p \sim I - PP^+ ,
	\end{equation}
	where $I$ is the $n \times n$ identity matrix.

	\item{\bf Sum:}
	The {\em sum} operation, also known as the Minkowski sum, computes all possible sums of one vector from $p$ and one from $q$. 
	{\em Sum} can be computed by matrix addition,
	\begin{equation}
		\begin{split}
			p + q &
			= \Col\left(\left[P\enspace Q \right] \right)
			\sim \left[P\enspace Q \right] \left[
			\begin{array}{c} P^T \\ Q^T \end{array}\right] 
			= PP^T + QQ^T ,
		\end{split}
	\end{equation}
	where $\left[P\enspace Q \right]$ is the matrix created by juxtaposing $P$ and $Q$
	horizontally.	
	{\em Sum} does not distribute over {\em intersection}, so for a third
	subspace $r$, in general,
	\begin{equation}
		(p + q) \cap r \neq (p \cap r) + (q \cap r) .
	\end{equation}
    {\em Sum} and {\em complement} are the only basic operations that can potentially increase the
	subspace dimensionality beyond the maximum dimensionality of their inputs.

	\item{\bf Projection:}
	The {\em projection} operation can be expressed as 
    \begin{equation}
		p\, \llcorner\, q \sim QQ^+PP^+ \sim QQ^+P .
	\end{equation}

	\item{\bf Rejection:}
	The {\em rejection} operation can be defined as the projection of $p$ on the complement of $q$,
	\begin{equation}
		    p \, \neg \, q  =  p\, \llcorner \, (\neg \, q)
		              \sim (I-QQ^+)P .
	\end{equation}
{\em Projection} can be expressed in terms of {\em rejection}. Let
\begin{equation}
	R=(I-QQ^+)P,
\end{equation}
so that $p\neg q\sim R$. Since $RR^+$ is the orthogonal projection
matrix onto $\Col(R)$,
\begin{equation}
	p\neg(p\neg q) \sim (I-RR^+)P = P-RR^+P.
\end{equation}
Here, $RR^+P=R$ because
\begin{equation}
	P=QQ^+P+R,
\end{equation}
where $\Col(QQ^+P)\subseteq q$ is orthogonal to
$\Col(R)\subseteq\neg q$, so	
\begin{align}
	p\neg(p\neg q)
	&\sim P-R \notag\\
	&=P-(I-QQ^+)P \notag\\
	&=QQ^+P \notag\\
	&\sim p\llcorner q.                 
\end{align}

	\item{\bf Intersection:}
	The {\em intersection} operation outputs the vectors 
	belonging to both $p$ and $q$. 
	{\em Intersection} can be defined in terms of {\em complement} and {\em sum}, because the
	complement of the sum of $p$ and $q$ must be perpendicular to both $p$ and $q$:
	\begin{equation}
		\neg (p \,+\, q) = (\neg p) \cap (\neg q),
    \end{equation}	
	so
	\begin{equation}
		\begin{split}
	    p\cap q & = \neg (\neg p \,+\, \neg q) \\
            & \sim I - (2I-PP^+-QQ^+) (2I-PP^+-QQ^+)^+ .
		\end{split}
    \end{equation}	
	Interestingly, {\em intersection} can also be expressed in terms of {\em sum} and {\em rejection} (Proof in \sectref{sect:intersection-by-rejection-and-span}):
		\begin{equation}
		\begin{split}
			p \cap q
			& = (p + q) \neg (q \neg p \,+\, p \neg q) .
		\end{split}
	\end{equation}
\end{itemize}

\subsection{Comparing subspaces}
\label{sect:emptiness-and-normality}
For simplicity, we say that a subspace $a$ is {\em empty} if it contains only the origin, $a = \{\myvec{0}\}$.

\label{sect:partial-order-of-subspaces}
A {\em partial order} of subspaces is crucial for forming hierarchies of invariants.
It can be seen as an {\em is-a} relation, a conditional, or an entailment.
The obvious ordering is to define $p \preceq q$ if and only if
$p$ is a subset (subspace) of $q$, $p \subseteq q$. We can determine this
by checking if the rejection of $q$ from $p$ is empty, {\it i.e.},
$p$ is a subspace of $q$ if and only if $p\, \neg\, q = \{\myvec{0}\}$.

However, such a partial ordering is unsatisfactory in the algebra of 
subspaces because if $p$ is a subspace of $q$,
{\em either $p$ and $q$ are identical, or $q$ must be of higher dimensionality than $p$}.
This limitation renders the partial ordering rather coarse, especially because of the
sparsity requirement, and is an inevitable weakness of subspace invariants.
In \sectref{sect:cone-relations}, we will see that the generalization of invariants from subspaces to convex cones eliminates
this dilemma by allowing a more fine-grained ordering, including same-dimensional invariants. 

Subspaces can be compared for equality using their
orthogonal projection matrices, since these provide canonical
representations: two subspaces are equal if and only if their
projection matrices $P$ and $Q$ are equal.

For a graded measure of similarity, a useful measure can be based on the
Hilbert--Schmidt inner product of their projection matrices,
\begin{equation}
	\langle P,Q\rangle_F
	\triangleq
	\sum_{i,j}p_{ij}q_{ij}
	=
	\tr(P^TQ).
\end{equation}
Although the first expression compares the matrices component by component,
the resulting quantity is independent of the choice of orthonormal basis.
Gleason's theorem \citep{Cooke.et.al.198aep} provides a general justification for
trace-based measures on subspaces.
Under the usual hypotheses of Gleason's theorem, a measure $\mu$ on
subspaces can be represented as
\begin{equation}
	\mu(P)=\operatorname{tr}(\rho P),
\end{equation}
where $P$ is the orthogonal projector onto the measured subspace and
$\rho$ is a positive semidefinite operator of unit trace. The precise
conditions are given in the cited reference.

Gleason's theorem determines the general trace form but does not prescribe a
particular reference operator $\rho$. To compare two non-empty
finite-dimensional subspaces $p$ and $q$, represented by the projectors $P$
and $Q$, we make the additional choice of a reference operator that is uniform
on $q$ and zero on its orthogonal complement:
\begin{equation}
	\rho_Q=\frac{Q}{\operatorname{tr}(Q)}.
\end{equation}
Since $\operatorname{tr}(Q)=d_q$, where $d_q$ is the dimension of $q$,
$\rho_Q$ is positive semidefinite and has unit trace.
This gives
\begin{equation}
	\mu_Q(P)
	=
	\frac{\tr(QP)}{\tr(Q)}
	=
	\frac{\tr(PQ)}{\tr(Q)}
	=
	\frac{\tr(P^TQ)}{\tr(Q)}.
\end{equation}
Here, the second equality follows from the cyclic invariance of the
trace, while the third follows because the orthogonal projection matrix
$P$ is symmetric, so that $P^T=P$.
If $\{\myvec{q}_i\}_{i=1}^{d_q}$ is an orthonormal basis for $q$,
then
\begin{equation}
	\mu_Q(P)
	=
	\frac{1}{d_q}
	\sum_{i=1}^{d_q}\|P\myvec{q}_i\|_2^2.
\end{equation}
Thus, $\mu_Q(P)$ measures the average extent to which directions in
$q$ lie in $p$: it equals one when $q\subseteq p$ and zero when the
two subspaces are orthogonal. Consequently, $\tr(PQ)$ provides a
basis-independent measure of their unnormalized overlap and motivates
the use of the Hilbert--Schmidt inner product for comparing subspaces
\citep{VanRijsbergen.2004tgo}.

The measure $\mu_Q(P)$ is directional because it measures the extent
to which $q$ lies in $p$, normalizing only by the dimension of $q$.
For a symmetric comparison, we instead normalize the same overlap with
respect to both subspaces. The associated Frobenius norm is
$\|A\|_F\triangleq\sqrt{\langle A,A\rangle_F}$.
Analogously to the cosine similarity between vectors, subspaces $p$
and $q$, represented by the projection matrices $P$ and $Q$, can
therefore be compared by
\begin{equation}
	\cos\theta
	=
	\frac{\langle P,Q\rangle_F}
	{\|P\|_F\|Q\|_F},
\end{equation}
where a value near one indicates strong similarity.

\subsection{The activation function and retaining sparsity}
\label{sect:activation-function}
\begin{figure}[!hbt] 
	\includegraphics[width=\textwidth]{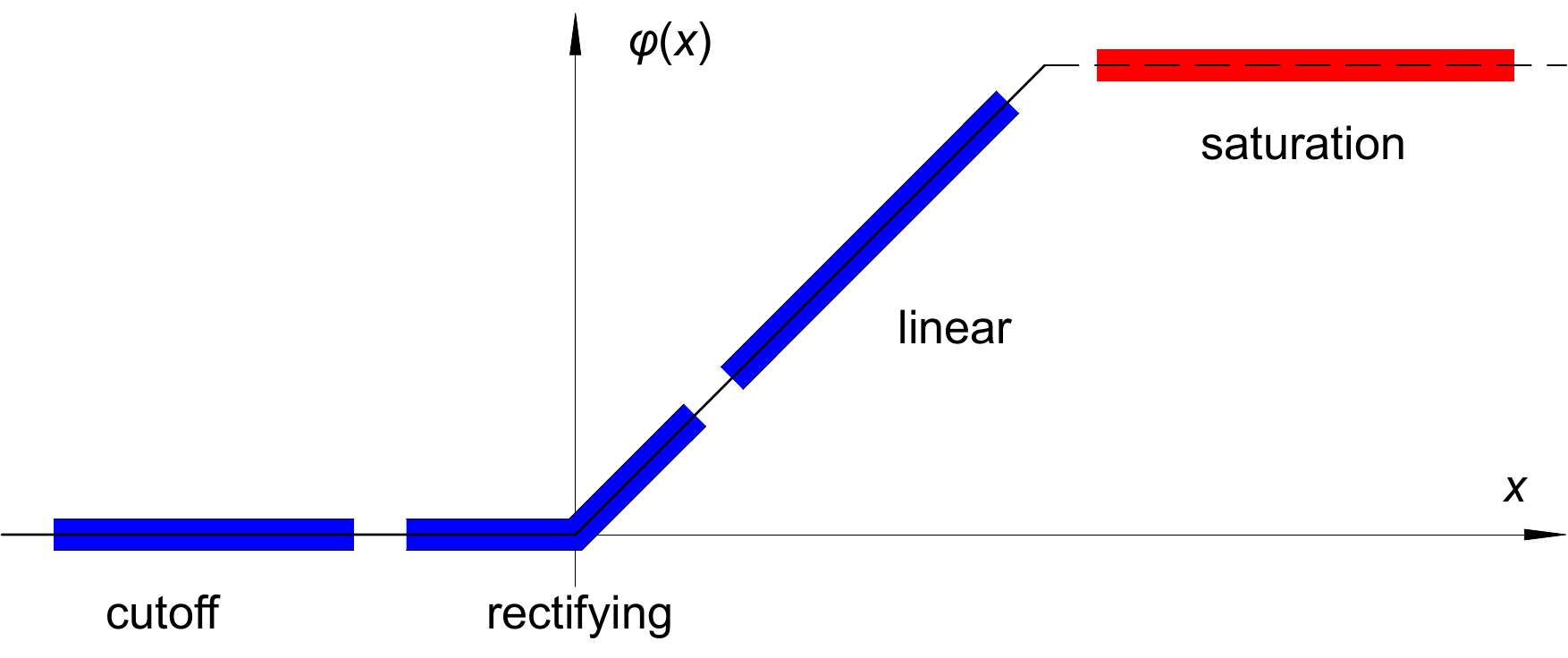}
	\caption{{\bf The activation function.} 
		The activation function $\varphi(\cdot)$ operates in three different ranges: {\em cutoff},
		where it is zero; {\em rectifying}, where it operates as a soft-threshold function; and
		{\em linear}, where it is a purely linear function of the argument. 
	    Although there is a fourth range, {\em saturation}, it is irrelevant in practice 
		due to the potential damage it causes to biological neurons
		through overexcitation (excitotoxicity).
		\label{fig:activation-function}}
\end{figure}

In the classical neuron model \eqref{eq:classical-neuron-model}, an {\em
	activation function} $\varphi(\cdot)$ is often applied to the output. The
activation function is a soft-threshold function and has the three principal
ranges of operation {\em linear}, {\em rectifying}, and {\em cutoff}
(\figref{fig:activation-function}). In a biological neuron, the activation
function is a biophysical consequence of the spike generation mechanism
\citep{Nilsson.Jorntell.2021ccf}.

So far, we considered the neuron to operate predominantly within the
activation function's linear range. The rectifying range, however, has the
essential function to ensure sparse representations. Populations create
representations of invariants by adapting weights based on sequences of
messages. These messages are typically noisy, meaning the invariant
representation may become excessively high-dimensional and needs trimming. A
similar situation exists after a {\em sum} operation. Within the rectifying
range, the activation function applies a soft threshold that only permits the
largest output components to pass through, thereby eliminating noise and preserving the
sparsity of the population's output.

Sparsity is crucial for maintaining the orthogonality of messages when traversing
neuron populations. Consider two interconnected populations, A and B 
(\figref{fig:axons-and-populations}, right) to see this. Intuitively, we can see
the connection as a permutation with some ``softening'' due to the axon
splitting into collaterals terminating at multiple neurons.

More formally, suppose that population B contains $n$ neurons and that A
sends it an $m$-sparse message, where $m\ll n$. Assume that
each active axon makes $p$ collateral connections to distinct neurons in B,
chosen uniformly at random and independently between axons. The expected
number of neurons in B receiving at least one connection is then
\begin{equation}
	q
	=
	n\left[
	1-\left(1-\frac{p}{n}\right)^m
	\right]
	\approx
	n\left[
	1-\exp\left(-\frac{mp}{n}\right)
	\right].
\end{equation}
Under an additional independence approximation, the target sets reached by
two such messages overlap in approximately $q^2/n$ neurons. This occupancy
estimate describes the spreading of message support, but does not by itself
guarantee that angles or orthogonality are preserved.

A quantitative sufficient condition is provided by the
restricted-isometry analysis in
\sectref{sect:approximate-cone-invariance}. If the effective linear
transmission map satisfies an appropriate restricted isometry property (RIP) on the sparse message
model, then pairwise inner products, and hence near orthogonality, are
preserved approximately, given sufficiently uniform total axonal gains and sufficiently
little overlap between collateral target sets.

\section{Mathematical abstraction of neuron populations}
\label{sect:mathematical-model}
\subsection{A generic mechanistic neuron model}

\begin{figure}[!hbt] 
	\includegraphics[width=\textwidth]{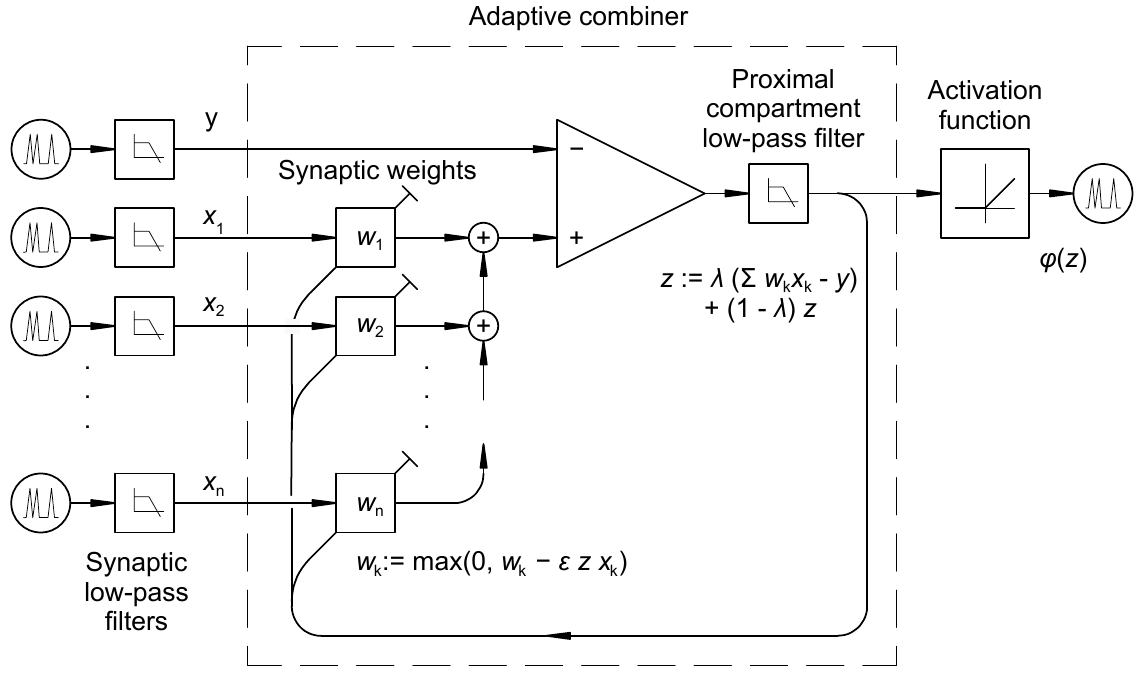}
	\caption{{\bf A generic neuron with plasticity.}
		The neuron operates as an adaptive filter or combiner with internal feedback,
		where weights are non-negative. The componentwise maximum in the
		weight-update path is the projection onto the non-negative orthant;
		combined with the LMS gradient, it gives the projected stochastic-gradient
		interpretation developed in \sectref{sect:nnls-interpretation}.
		The mechanistic model derives from experiments on the widespread class of CNS
		neurons that receive glutamatergic excitation through NMDA and AMPA receptors
		and GABAergic inhibition through GABA receptors, together with known ion-channel
		properties.
		The update factor $\lambda$ characterizes low-pass filtering of
		the output $z$. The proximal compartment comprises the soma and
		proximal sections of the dendrites.
		(Figure adapted from \protect{\citet{Nilsson.2026meo}})
		\label{fig:adaptive-filter}}
\end{figure}

In \figref{fig:adaptive-filter}, $\myvec{x}=(x_1,\ldots,x_n)^T$ denotes the
excitatory inputs,
$y$ the aggregate inhibitory drive formed from multiple inhibitory
inputs whose effective weights are held fixed in the present model,
$\myvec{w}=(w_1,\ldots,w_n)^T$
the excitatory synaptic weights, $z$ the adaptive-combiner output,
$\varphi$ the activation function, $\lambda$ the output-filter update (decay) factor,
and $\varepsilon>0$ the learning rate.  LMS denotes the least-mean-squares
learning rule; the quantities are developed in detail below.

This section reviews a conservative mechanistic model of a generic neuron with
plasticity \citep{Nilsson.2026meo} and uses it to formulate a population model.
Here, ``conservative'' means that the model is derived from established knowledge
of biological neurons and ion-channel properties and contains no speculative or
phenomenological additions. The model represents the widespread class of CNS
neurons that receive glutamatergic excitation through AMPA and NMDA receptors and
GABAergic inhibition through GABA receptors. The extensively studied hippocampal
CA1 pyramidal neuron provides one representative example. The model is not intended
to describe every neuronal phenotype. It provides the foundation for the population
model developed below.

There are at least five significant reasons for relying on a mechanistic and non-speculative neuron model:

{\bf Foundation in Established Principles:} The mechanistic model is rooted in
well-established scientific knowledge of biological neurons. This solid
foundation increases confidence in the model's validity and its explanations.

{\bf Predictive Power:} Since the mechanistic model is grounded in established
principles, it has robust predictive power. It allows predictions about yet
unobserved phenomena, allowing further testing and validating the model.

{\bf Explanatory Insights:} The mechanistic model provides a detailed
understanding of how and why the neuron behaves the way it does. For example,
breaking down the neuron into a formulation as an electrical equivalent circuit
directly reveals its underlying plasticity rule.

{\bf Causal Explanation:} The mechanistic model enables causal explanations. Not
only does it explain the correlations observed in neuron signaling, but it also
elucidates the causal mechanisms behind the observed phenomenon.

{\bf Addressing Complex Systems:} The mechanistic approach is especially
suitable for studying networks of many composite populations whose relations are
intricate. The mechanistic model provides a framework for the mathematical
abstraction of these complex interactions and for understanding the system as a
whole.

\figref{fig:adaptive-filter} shows a schematic of the model. Inhibitory
and excitatory inputs undergo low-pass filtering before reaching an
{\em adaptive combiner}
\citep{Widrow.Stearns.1985asp,Widrow.Hoff.1960asc}, which forms the
neuron's core.

The model allows the neuron to receive multiple inhibitory inputs. Let
$y_j(t)$ denote the filtered signal at inhibitory input $j$, and let
$v_j\geq0$ denote its effective synaptic weight. Their total subtractive
contribution is
\begin{equation}
	y(t)
	=
	\sum_{j=1}^{n_{\mathrm I}} v_j y_j(t).
	\label{eq:aggregate-inhibitory-input}
\end{equation}
The weights $v_j$ are held fixed during the learning process considered
here. Because the model combines their contributions linearly, the
individual inhibitory inputs can therefore be represented exactly by the
single aggregate term $y(t)$ shown in \figref{fig:adaptive-filter}.
Thus, $y(t)$ generally varies with time and represents the combined
inhibitory drive from multiple synapses; it does not represent a single
inhibitory synapse. In contrast, the excitatory inputs $x_k(t)$ have
adjustable weights $w_k$, which are subject to the learning rule. The
adaptive-combiner output $z(t)$ represents the membrane potential and is
subsequently converted into an output spike train with spike rate
$\varphi[z(t)]$, where $\varphi$ is a soft-threshold activation function.

Holding the inhibitory weights fixed is a modeling abstraction inherited
from the mechanistic equivalent-circuit formulation
\citep{Nilsson.2026meo}. It isolates the excitatory learning mechanism
studied here and is not intended to imply that inhibitory synapses are
universally non-plastic or that GABAergic plasticity is absent in the CNS.
The convex-cone structure itself does not depend on this simplification.
It follows from the non-negativity of the effective excitatory weights,
which restricts excitatory mixing to conic combinations.

Contrasted with the basic neuron model introduced in \sectref{sect:simplistic-neuron-model}, this model embodies three distinct new features:  {\em time-variable inputs and outputs},  {\em non-negativity of weights},  and {\em plasticity}. Each enhancement notably amplifies the neuron's computational capabilities, with deeper explorations provided in the subsections below. 

\subsection{Time-variable inputs and outputs}
Whereas we previously assumed slowly changing inputs enabling a pseudo-static approach, we now allow inputs to be vectors of {\em wavelets}---short, time-variable functions \citep{Mallat.2009awt}. Recent experimental findings support this perspective \citep{Markowitz.et.al.2018tso}.
This assumption mandates a fully dynamic approach, which we detail in this section.

When the inputs enter the synapses, they pass through a battery of low-pass filters with diverse time constants.
While summed in the proximal compartment (proximal dendrites and soma), they undergo additional low-pass filtering with the membrane time constant $\tau_m$.
In the frequency (Laplace) domain, the transfer function becomes
\begin{equation}
	\tilde{z}(s) = \frac{1/\tau_m}{s + 1/\tau_m}\left[\myvec{w}^T \, \myvec{\tilde{x}}(s) - \tilde{y}(s)\right],
\end{equation}
where $s$ is the Laplace variable, a tilde denotes a Laplace transform, $z$ is
the adaptive-combiner output, $\myvec{w}$ is the vector of excitatory synapse
weights, and $\myvec{x}$ and $y$ are the filtered excitatory and inhibitory
inputs, respectively.

The formula above neither includes a factor in front of $y$ nor an explicit
representation of the input filters because fixed filter gains can be absorbed
into the effective population operator. Although a population contains only
finitely many neurons, each component of a message is a time-varying signal. We
therefore represent a message from an $n$-neuron population as an element of
$L_2([0,T];\mathbb{R}^n)$. This space is finite-dimensional in its neuronal
coordinate but infinite-dimensional in time. Spatial mixing is described by an
ordinary finite matrix, while synaptic and membrane filtering acts on the
temporal coordinate by convolution.

On a finite observation interval, convolution with the square-integrable
impulse responses considered here is a Hilbert--Schmidt and hence compact
operator \citep{Conway.1990acf}. Compact operators admit arbitrarily accurate
finite-rank approximations. We can therefore use familiar finite-dimensional
matrix geometry to understand the population mapping, with increasing accuracy
as more temporal modes are retained. In the examples below, messages are
restricted to a finite wavelet dictionary, so their geometry is exactly
finite-dimensional. The Laplace transform provides a complementary
frequency-domain description, in which convolution becomes multiplication by a
transfer function.

For the linked examples below, we use one concrete finite dictionary in
$L_2([0,T])$. Let $T=0.20\,\mathrm{s}$ and
\begin{equation*}
 h(t)=
 \begin{cases}
  \sin^2(\pi t/T),&0\leq t\leq T,\\
  0,&\text{otherwise},
 \end{cases}
 \qquad \gamma=\frac{4}{\sqrt{3T}},
\end{equation*}
and define
\begin{align*}
 \phi_1(t)&=\gamma h(t)\cos(2\pi 30t),&
 \phi_2(t)&=\gamma h(t)\sin(2\pi 30t),\\
 \phi_3(t)&=\gamma h(t)\cos(2\pi 60t).&&
\end{align*}
These three wavelets are mutually orthonormal and have zero mean. Their common
envelope $h(t)=\sin^2(\pi t/T)=\tfrac{1}{2}[1-\cos(2\pi t/T)]$ is a Hann window.
It turns each sinusoid on smoothly from zero, reaches full amplitude at the center
of the interval, and returns it to zero at the end. Both the envelope and its
first derivative vanish at the endpoints, avoiding the discontinuities of
abruptly truncated sinusoids. The dictionary is deliberately
idealized: its role is to make every projection and residual in the examples
exact and transparent, not to propose a universal neuronal code. 

The $\phi_i$ represent baseline-subtracted firing-rate signals. Literal
rates may be written as $r_i(t)=r_{0i}+A\phi_i(t)$, but the baseline must
be subtracted, or represented as a separate constant component, before
the inner products and projections below are computed. Biologically,
this baseline correction may be approximated by ionic bias currents
that regulate the membrane potential relative to its resting or adapted
operating point.

The following example illustrates the treatment of signals as Hilbert-space
elements:
\begin{proof}[{\qreff{Neuron inputs as points in Hilbert space}
		{ex:hilbert-space-inputs}}]
	Let
	\begin{equation*}
	 \myvec{x}=(\phi_1,\phi_2)^T\in L_2^2,
	 \qquad y=\phi_1+2\phi_2\in L_2.
	\end{equation*}
	Thus $\myvec{x}$ and $y$ are points in Hilbert spaces, even though their
	components are time-varying functions. The neuron adjusts $\myvec{w}$ so that
	$\myvec{w}^T\myvec{x}\approx y$. Orthonormality of $\phi_1$ and $\phi_2$
	gives the exact optimum $\myvec{w}=(1,2)^T$, for which
	$\myvec{w}^T\myvec{x}=y$.
\end{proof}

Unless stated otherwise, we assume that weight updates are slow relative to feedforward dynamics and that the activation function remains in its linear range. Under these assumptions, the general time-variable model is, up to a constant factor,
\begin{equation}
	\label{eq:defining-equation}
	z(t+\Delta t) = \lambda \left[\myvec{w}^T \myvec{x}(t) - y(t)\right] + (1 - \lambda) z(t) + O(\Delta t^2),
\end{equation}
where $\Delta t$ is the update interval,
$\lambda = 1 - \exp(-\Delta t/\tau_m)$ is the corresponding update factor, and
$O(\Delta t^2)$ collects terms of second and higher order in $\Delta t$.

\subsection{Non-negative weights}
\begin{figure}[!hbt] 
	\centering
	\includegraphics[scale=0.9]{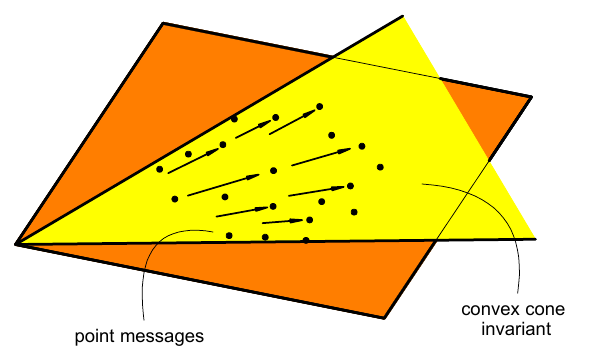}
	\caption{{\bf Convex cone invariants.}
		When the weight matrix is non-negative, subspaces generalize to convex cones. This representation substantially improves the expressive power of the invariants.	
		\label{fig:cones}}
\end{figure}
The biological representation of adaptive excitatory weights as counts
of synaptic AMPA receptors \citep{Nilsson.2026meo} inherently implies
non-negativity. A neuron therefore forms non-negative, rather than
arbitrary signed, linear combinations of its input signals. The set
generated by a family of input signals is consequently a convex cone
rather than a linear subspace (\figref{fig:cones}). This motivates an
algebra of convex cones in Hilbert space.

At first glance, this may seem like a restriction compared with the
earlier algebra of subspaces. However, the resulting cone formulation
not only provides a powerful generalization of the subspace
representation, but also gives rise to an algebra that is surprisingly
similar to that of subspaces, as will be elaborated upon in
\sectref{sect:algebra-of-cones}.

At the population level, the corresponding transmission matrices are
typically sparse and, for excitatory transmission, non-negative.
However, linearity alone is sufficient for the image of a convex cone
to remain a convex cone. Entrywise non-negativity of the transmission
matrix is not required for this mathematical property, but approximate
preservation of the cone operations requires additional geometric
conditions, studied in
\sectref{sect:approximate-cone-invariance}.

\begin{proof}[{\qreff{Non-negative weights}
		{ex:non-negative-weights}}]
	Keep $\myvec{x}=(\phi_1,\phi_2)^T$, but take
	$y_-=\phi_1-2\phi_2$. The unconstrained least-squares coefficient vector
	would be $(1,-2)^T$. With non-negative weights, however, the optimum is
	$\myvec{w}=(1,0)^T$. To see the geometry, let
	\begin{equation*}
	 C=\{\alpha\phi_1+\beta\phi_2\mid \alpha,\beta\geq0\}.
	\end{equation*}
	For a closed convex cone $D$, let $P_Du$ denote
	the closest point in $D$ of a point $u$, and let $\neg D$ denote the cone consisting of
	all vectors having non-positive inner product with every vector in $D$.
	Then
	\begin{equation*}
	 P_Cy_-=\phi_1,
	 \qquad y_--P_Cy_-=-2\phi_2\in\neg C.
	\end{equation*}
	The last inclusion follows directly because the residual has non-positive
	inner product with both generators of $C$.  Thus $P_Cy_-$ and
	$y_--P_Cy_-$ form the orthogonal cone--polar decomposition later formalized
	by the Moreau theorem in \sectref{sect:moreau-decomposition}. The neuron output has the opposite
	sign,
	\begin{equation*}
	 z=\myvec{w}^T\myvec{x}-y_-=2\phi_2,
	\end{equation*}
	so it is the reflection of the residual.
\end{proof}

\subsection{Plasticity}
The neuron updates the excitatory weights $\myvec{w}$ using a built-in {adaptive combiner}. Such a neuron model, ADALINE for ADaptive LInear NEuron, was first proposed by \citet{Widrow.Hoff.1960asc} based on an adaptive {\em linear} combiner using the {\em LMS learning rule} \citep{Widrow.Hoff.1960asc}
\begin{equation}
	\label{eq:lms-learning-rule}
	w_k := w_k - \varepsilon x_k z,
\end{equation}
where $\varepsilon > 0$ is the learning rate. The name is appropriate because the adaptive linear combiner attempts to adjust a linear combination of input signals to match a desired response signal. At the time, it was difficult to explain mechanistically how the neuron could implement such a mechanism because little was known about the internals of neurons. It was assumed that the mechanism used external feedback, but such feedback would be too slow for biological neurons. ADALINE was abandoned as a neuron model, but under the abbreviation's new interpretation ``ADaptive LINear Element'' became the basis for the explosive and highly successful development of adaptive filters in signal processing
\citep{Widrow.Stearns.1985asp,Haykin.2002aft,Sayed.2003foa}.

However, it was recently shown for the present mechanistic neuron model
that sufficiently fast internal feedback does exist \citep{Nilsson.2026meo},
and that neurons implement an adaptive {\em conical} combiner
using the modified LMS learning rule
\begin{equation}
	\label{eq:conical-lms-rule}
	w_k := \max(0, w_k - \varepsilon x_k z),
\end{equation}
which ensures the non-negativity of synaptic weights.

\subsubsection{Relation to non-negative least squares}
\label{sect:nnls-interpretation}
When $z$ is the instantaneous residual---in particular, when $\lambda=1$ in
\eqref{eq:defining-equation}---the modified rule is precisely a projected
stochastic-gradient method for non-negative least squares (NNLS), rather than a
new least-squares objective. To make this statement exact, collect $N$ sampled
input vectors as the columns of
$X=[\myvec{x}^{(1)},\ldots,\myvec{x}^{(N)}]\in\mathbb{R}^{n\times N}$ and the
corresponding inhibitory samples in
$\myvec{y}=(y^{(1)},\ldots,y^{(N)})^T$. The batch problem is
\begin{equation}
 \min_{\myvec{w}\geq0} F(\myvec{w}),
 \qquad
 F(\myvec{w})=\frac12\|X^T\myvec{w}-\myvec{y}\|_2^2,
 \label{eq:nnls-objective}
\end{equation}
which is the classical NNLS problem \citep{Lawson.Hanson.1995sls}. For sample
$p$, put
\begin{equation}
 z^{(p)}(\myvec{w})=\myvec{w}^T\myvec{x}^{(p)}-y^{(p)},
 \qquad
 \ell_p(\myvec{w})=\frac12\bigl(z^{(p)}(\myvec{w})\bigr)^2,
\end{equation}
where $\ell_p(\myvec{w})$ is the contribution of observation $p$ to the
batch loss, so that $F(\myvec{w})=\sum_{p=1}^N\ell_p(\myvec{w})$. Then
$\nabla\ell_p(\myvec{w})=\myvec{x}^{(p)}z^{(p)}(\myvec{w})$, and the vector form of
\eqref{eq:conical-lms-rule} is
\begin{equation}
 \myvec{w}^{(p+1)}=
 P_{\mathbb{R}_+^n}\!\left(
 \myvec{w}^{(p)}-\varepsilon_p\nabla\ell_p(\myvec{w}^{(p)})
 \right),
 \label{eq:projected-stochastic-nnls}
\end{equation}
where $\varepsilon_p>0$ is the learning rate used for observation $p$,
$\mathbb{R}_+^n=\{\myvec{w}\in\mathbb{R}^n\mid w_k\geq0\}$ is the non-negative
orthant, and $P_{\mathbb{R}_+^n}$ denotes Euclidean projection onto it.  This
projection is the componentwise operation $\max(0,\cdot)$. Thus each
application of the rule is exactly a projected
stochastic-gradient step \citep{Bertsekas.1976otg,Geiersbach.Pflug.2019psg}; it
is not, by itself, a one-step closed-form NNLS solver.

There is also a short proof that the corresponding full-batch fixed points are
exactly the NNLS solutions. The batch gradient is
\begin{equation}
 \myvec{g}(\myvec{w})=\nabla F(\myvec{w})
 =X(X^T\myvec{w}-\myvec{y}).
\end{equation}
For any $\varepsilon>0$, the fixed-point relation
\begin{equation}
 \myvec{w}=P_{\mathbb{R}_+^n}
 (\myvec{w}-\varepsilon\myvec{g}(\myvec{w}))
 \label{eq:nnls-fixed-point}
\end{equation}
holds if and only if, componentwise,
\begin{equation}
 \myvec{w}\geq0,
 \qquad \myvec{g}(\myvec{w})\geq0,
 \qquad w_k g_k(\myvec{w})=0,
 \label{eq:nnls-kkt}
\end{equation}
where $g_k(\myvec{w})$ is component $k$ of $\myvec{g}(\myvec{w})$. Indeed, a
positive component is unchanged only when its gradient component is
zero, whereas a zero component remains clipped to zero exactly when its
gradient component is non-negative. Conditions \eqref{eq:nnls-kkt} are the
Karush--Kuhn--Tucker conditions for \eqref{eq:nnls-objective}; since the
objective and feasible set are convex, they are necessary and sufficient.
Consequently, the fixed points of the batch projected update are precisely the
NNLS minimizers. For $0<\varepsilon<2/\|X\|_2^2$, where $\|X\|_2$ denotes the
matrix spectral norm, the deterministic batch iteration converges to such a
minimizer; it is unique when $X^T$ has full column rank.

For the instantaneous LMS form, the conclusion needs the usual stochastic
qualification. Under unbiased sampling, suitable moment bounds, and diminishing
learning rates satisfying
$\sum_{p=1}^{\infty}\varepsilon_p=\infty$ and
$\sum_{p=1}^{\infty}\varepsilon_p^2<\infty$, projected stochastic-gradient convergence to
the NNLS solution set follows under standard assumptions
\citep{Geiersbach.Pflug.2019psg}. A constant small learning rate, as commonly
used by an adaptive filter, generally leaves a noise-dependent steady-state
misadjustment and should be said to track or approximate the NNLS solution, not
to compute it exactly. Likewise, if $z$ contains the low-pass memory represented
by $\lambda<1$, the clipping remains an exact projection, but $x_kz$ is an exact
gradient only for a correspondingly filtered objective with matched filtering.
Without that additional identification, NNLS describes the equilibrium target,
not literally every transient update.

The same interpretation applies before temporal sampling. For wavelets
$x_k,y\in L_2$, the finite coefficient problem is
\begin{equation}
 \min_{\myvec{w}\geq0}
 \frac12\left\|\sum_k w_kx_k-y\right\|_{L_2}^2,
 \qquad
 g_k=\left\langle x_k,\sum_jw_jx_j-y\right\rangle_{L_2},
 \label{eq:hilbert-nnls}
\end{equation}
where $g_k$ is component $k$ of the Hilbert-space gradient and $j$ indexes the
input channels. Thus the signal space may be infinite-dimensional while the NNLS variable
$\myvec{w}$ remains finite-dimensional. When $z(t)$ is the unfiltered
residual, the product $x_k(t)z(t)$ used by the online rule is the
instantaneous contribution to the Hilbert-space gradient; its temporal
integral gives $g_k$. Under an appropriate random temporal-sampling
model, it may equivalently be interpreted, with the corresponding
normalization, as an unbiased stochastic estimate of $g_k$.

Although the input-filter responses are neither orthogonal nor linearly
independent, their different time constants generate a rich family of
temporal functions. If the closed conic hull of these responses contains
the target response $y$, then conical combinations of sufficiently many
input-filter responses can approximate $y$ arbitrarily closely in the
$L_2$ norm. Thus, subject to this richness condition, the neuron model
can in principle approximate any linear filter characteristic belonging
to the closed cone generated by its excitatory inputs.

The adaptive combiner can be viewed as a device performing a wavelet transform \citep{Doroslovacki.Fan.1996wbl,Mallat.2009awt}. Its adjusted weights express the inhibitory input as a conical combination of the excitatory inputs.

\begin{proof}[{\qreff{Plasticity}
		{ex:plasticity}}]
	Continue example \ref{ex:hilbert-space-inputs} with
	$\myvec{x}=(\phi_1,\phi_2)^T$ and
	\begin{equation}
	 y_+=\phi_1+2\phi_2+3\phi_3.
	\end{equation}
	The neuron adjusts $\myvec{w}\geq0$ to minimize
	$\|\myvec{w}^T\myvec{x}-y_+\|_{L_2}$. With the cone $C$ from example
	\ref{ex:non-negative-weights}, orthonormality gives
	\begin{equation}
	 P_Cy_+=\phi_1+2\phi_2,
	 \qquad y_+-P_Cy_+=3\phi_3,
	\end{equation}
	so the adjusted weights are again $\myvec{w}=(1,2)^T$. Because the output is
	defined as approximation minus inhibitory target, its sign is
	\begin{equation}
	 z=\myvec{w}^T\myvec{x}-y_+=-3\phi_3.
	\end{equation}
	The positive polar residual is $3\phi_3$; the neuronal output is its
	reflection. More generally, the entries of $\myvec{w}$ are non-negative
	wavelet coefficients of the conic projection, and need not be unique for an
	overcomplete dictionary.
\end{proof}

\subsection{Abstract model of populations}
The generalization to a population of $m$ neurons is straightforward.
We have $m$-dimen\-sion\-al vectors $\myvec{y}$
and $\myvec{z}$ of inhibitory inputs and outputs, respectively. As before,
we have an $n$-dimensional vector $\myvec{x}$ of excitatory inputs.

\begin{proof}[{\qreff{Populations}
		{ex:populations}}]
	Expand the preceding examples to two input channels and two output neurons.
	Consider the two dense input messages
	\begin{equation}
	 \myvec{x}^{(1)}=(\phi_1,\phi_2)^T,
	 \qquad
	 \myvec{x}^{(2)}=(\phi_2,\phi_1)^T
	 \quad\text{in }L_2^2,
	\end{equation}
	and the non-negative linear map
	\begin{equation}
	 A=\begin{pmatrix}1&2\\2&1\end{pmatrix}=W^T.
	\end{equation}
	Its predicted output messages are
	\begin{align}
	 \widehat{\myvec{y}}^{(1)}
	 &=A\myvec{x}^{(1)}
	 =(\phi_1+2\phi_2,\,2\phi_1+\phi_2)^T,\\
	 \widehat{\myvec{y}}^{(2)}
	 &=A\myvec{x}^{(2)}
	 =(2\phi_1+\phi_2,\,\phi_1+2\phi_2)^T.
	\end{align}
	Now define the output-space prediction-error direction
	$\myvec{e}=(\phi_3,0)^T$ and the two
	inhibitory target messages
	\begin{equation}
	 \myvec{y}^{(1)}=\widehat{\myvec{y}}^{(1)}+3\myvec{e},
	 \qquad
	 \myvec{y}^{(2)}=\widehat{\myvec{y}}^{(2)}.
	\end{equation}
	Because $\phi_3$ is orthogonal to $\phi_1$ and $\phi_2$, the added component
	cannot be synthesized from these excitatory messages and does not change the
	least-squares map $W^T=A$. The population outputs are therefore
	\begin{equation}
	 \myvec{z}^{(1)}=W^T\myvec{x}^{(1)}-\myvec{y}^{(1)}=-3\myvec{e},
	 \qquad
	 \myvec{z}^{(2)}=W^T\myvec{x}^{(2)}-\myvec{y}^{(2)}=\myvec{0}.
	\end{equation}
	The same $W$ maps every conical combination
	$\alpha\myvec{x}^{(1)}+\beta\myvec{x}^{(2)}$, $\alpha,\beta\geq0$, to the
	corresponding conical combination of predicted outputs. The matrix $A$ here
	denotes the learned computational map; later, $M$ will be reserved for an
	anatomical transmission map between populations.
\end{proof}

The population output $\myvec{z}$ is updated as (cf. \eqref{eq:defining-equation})
\begin{equation}
    \label{eq:adaptive.linear-combiner-update-formula}
	\myvec{z} := \lambda \left(W^T \myvec{x} - \myvec{y}\right) + (1 - \lambda) \myvec{z} ,
\end{equation}
where the update rule for the weight matrix $W$ in the time domain is
\begin{equation}
	W := 
	W - \varepsilon \myvec{x} \myvec{z}^T
\end{equation}
for the linear combiner and
\begin{equation}
	W := 
	\max\left(0, W - \varepsilon \myvec{x} \myvec{z}^T \right)
\end{equation}
for the conical combiner, where $\max(\cdot)$ is applied component-wise.

For the subspace case with $\lambda=1$, suppose that the components
of $\myvec{y}$ in
\eqref{eq:adaptive.linear-combiner-update-formula}
are independent, identically distributed zero-mean Gaussian random
variables and that
\begin{equation}
	\myvec{x}=B\myvec{y},
\end{equation}
where $B\in\mathbb{R}^{n\times m}$ is fixed. The mean squared error
\begin{equation}
	\mathbb{E}\!\left[
	\|W^T\myvec{x}-\myvec{y}\|_2^2
	\right]
\end{equation}
is minimized by every matrix of the form
\begin{equation}
	W^T
	=
	B^+ + \Phi(I-BB^+),
\end{equation}
where $\Phi\in\mathbb{R}^{m\times n}$ is arbitrary. Thus, the
least-squares problem does not generally determine a unique weight
matrix. Nevertheless, all minimizing matrices have the same action
on the input, because
\begin{equation}
	\left[B^+ + \Phi(I-BB^+)\right]B\myvec{y}
	=
	B^+B\myvec{y}.
\end{equation}
Consequently, the population output at any minimum is
\begin{equation}
	W^T\myvec{x}-\myvec{y}
	=
	(B^+B-I)\myvec{y}.
\end{equation}
Under conditions ensuring convergence of the adaptive learning rule to
the least-squares minimizing set, its input--output mapping therefore
converges to this uniquely determined relation, even though the weight
matrix itself need not converge to a uniquely determined matrix.

Since $B^+B$ is the orthogonal projector onto
$\Col(B^T)$, the vector
\begin{equation}
	(I-B^+B)\myvec{y}
\end{equation}
is the part of $\myvec{y}$ that cannot be represented by
$\myvec{x}=B\myvec{y}$. If $\myvec{x}$, $\myvec{y}$, and
$\myvec{z}$ are regarded as samples from the subspaces $q$, $p$, and
$r$, respectively, this relation represents the rejection of $q$
from $p$, with the learned alignment of their population coordinate
systems understood:
\begin{equation}
	\label{eq:description-of-rejection}
	r=p\neg q.
\end{equation}

Unlike the unconstrained pseudoinverse formula above, there is no analogous
general closed-form expression for an optimal $W$ in the conical case; it is an
NNLS problem. As we shall see in \sectref{sect:population-as-operator},
there is a generalization of \eqref{eq:description-of-rejection}
to convex cones.

\section{The algebra of convex cones}
\label{sect:algebra-of-cones}
In this section, we investigate the basic mathematical properties of convex cones and the operations in an algebra of such objects. We postpone the issue of implementation by biological neurons until \sectref{sect:neuronal-implementation}.

\subsection{Finitely generated convex cones}

A {\em frame}
$F_a=\{\myvec{v}_1,\ldots,\myvec{v}_m\}\subset H$
for a cone $a$ is a finite set whose non-negative linear combinations
generate $a$. A cone possessing such a frame is called {\em finitely
	generated}. Unless stated otherwise, a {\em cone} in this paper means a
finitely generated convex cone in a real Hilbert space $H$.

More generally, for any subset $c$ of $H$, let $\Conic(c)$ denote the
smallest closed convex cone containing $c$. More precisely,
\begin{equation}
	\Conic(c)
	\triangleq
	\overline{
		\left\{
		\sum_{i=1}^{n}\lambda_i\myvec{x}_i
		\,\middle|\,
		n\in\mathbb{N},\,
		\myvec{x}_i\in c,\,
		\lambda_i\geq 0
		\right\}},
\end{equation}
where the overbar denotes closure in the norm topology of $H$. In
particular, a cone with frame $F_a$ can be written as
\begin{equation}
	a=\Conic(F_a)
	=\left\{\sum_{k=1}^{m}\lambda_k\myvec{v}_k
	\ \middle|\ \lambda_k\geq0,\ \myvec{v}_k\in F_a\right\},
\end{equation}
where the non-negative scalars $\lambda_k$ are the conic-combination
coefficients. Since a frame contains finitely many vectors, its conic
hull is polyhedral within the finite-dimensional space
$\operatorname{span}(F_a)$ and is consequently closed in $H$. The
overbar therefore has no effect in this case. This remains true when
$H$ itself is infinite-dimensional. For example, the frame vectors may
be vector-valued wavelets in $L_2([0,T];\mathbb{R}^n)$, while their span
remains finite-dimensional and closed. The term ``finite cone'' is
sometimes used for this construction, but ``finitely generated'' states
the intended property more precisely.

The frame vectors can be visualized as the ribs of an umbrella with the cap
at the origin, or as the flowers in a bouquet.  Although more general convex
cones are also possible invariants, finite generation is convenient for
representation and computer implementation and permits induction over
frames.  When a result is stated for a cone that is not assumed to be
finitely generated, we require it to be a {\em closed convex
cone}; closedness is the hypothesis that guarantees existence of its metric
projection.

A {\em face} $F$ of a convex cone $a$ is a convex cone
$F\subseteq a$ such that
\begin{equation}
	\myvec{x},\myvec{y}\in a,
	\qquad
	\myvec{x}+\myvec{y}\in F
	\quad\Longrightarrow\quad
	\myvec{x},\myvec{y}\in F.
\end{equation}
Equivalently, if an interior point of a line segment between two
points of $a$ belongs to $F$, then both endpoints belong to $F$.
Geometrically, faces describe the boundary pieces of a cone; the
whole cone $a$ is also regarded as a face of itself.

\subsection{The projection and Moreau decomposition theorems}
\label{sect:moreau-decomposition}

Moreau decomposition is the fundamental result that decomposes every
vector into two orthogonal components: its projection onto a closed
convex cone and the residual from that projection.

Let $a$ be a closed convex cone in the real Hilbert space $H$. The
{\em projection theorem}
\citep{Luenberger.1969obv}
for closed convex sets guarantees that every
$\myvec{x}\in H$ has a unique closest point in $a$. We denote this
point by $P_a\myvec{x}$ and define
\begin{equation}
	P_a\myvec{x}
	\triangleq
	\argmin_{\myvec{y}\in a}
	\|\myvec{x}-\myvec{y}\|.
	\label{eq:pointwise-cone-projection}
\end{equation}
Thus, $P_a:H\rightarrow a$ is the pointwise metric, or orthogonal,
projection onto $a$. Unlike orthogonal projection onto a subspace,
$P_a$ is generally not linear. Conic projection of one cone onto
another will be defined by applying this pointwise projection to all
points of the input cone and taking the closed conic hull of the
result.

The {\em polar} of $a$ is the closed convex cone defined as
\begin{equation}
	\neg a
	\triangleq
	\left\{
	\myvec{z}\in H
	\,\middle|\,
	\left\langle\myvec{z},\myvec{y}\right\rangle\leq0
	\text{ for every }\myvec{y}\in a
	\right\}.
	\label{eq:polar-cone-definition}
\end{equation}

The \emph{Moreau decomposition theorem}
is fundamental for convex cones
\citep{Moreau.1962dod}; see also
\citet[Lemma~2(iii), p.~345]{Ingram.Marsh.1991poc}.
It says that for every closed convex cone $a$ and every $\myvec{x}\in H$,
\begin{equation}
	\myvec{x}
	=
	P_a\myvec{x}
	+
	P_{\neg a}\myvec{x},
	\qquad
	\left\langle
	P_a\myvec{x},
	P_{\neg a}\myvec{x}
	\right\rangle
	=
	0.
	\label{eq:moreau-decomposition}
\end{equation}
Equivalently, if
$\myvec{x}=\myvec{y}+\myvec{z}$, where
$\myvec{y}\in a$, $\myvec{z}\in\neg a$, and
$\langle\myvec{y},\myvec{z}\rangle=0$, then
\begin{equation}
	\myvec{y}=P_a\myvec{x},
	\qquad
	\myvec{z}=P_{\neg a}\myvec{x}.
\end{equation}
The two components of the decomposition are therefore unique.

For later use, a point $\myvec{y}\in a$ is the projection of
$\myvec{x}$ onto $a$ if and only if
\begin{equation}
	\left\langle
	\myvec{x}-\myvec{y},
	\myvec{w}-\myvec{y}
	\right\rangle
	\leq0
	\qquad
	\text{for every }\myvec{w}\in a.
	\label{eq:projection-characterization}
\end{equation}
Because $a$ is a cone, this is equivalent to
\begin{equation}
	\myvec{x}-\myvec{y}\in\neg a,
	\qquad
	\left\langle
	\myvec{y},
	\myvec{x}-\myvec{y}
	\right\rangle
	=
	0.
	\label{eq:cone-projection-characterization}
\end{equation}

Moreau decomposition also gives the useful residual identity
\begin{equation}
	P_{\neg a}\myvec{x}
	=
	\myvec{x}-P_a\myvec{x}.
	\label{eq:moreau-residual}
\end{equation}
Consequently, projection onto the polar can be obtained as the residual
after projection onto the original cone. This identity will allow
conic rejection to be described without explicitly constructing the
polar cone.

\subsection{Basic operations on convex cones}
\label{sect:definitions-of-basic-operations-on-cones}
\begin{figure}[!hbt] 
	\includegraphics[width=\textwidth]{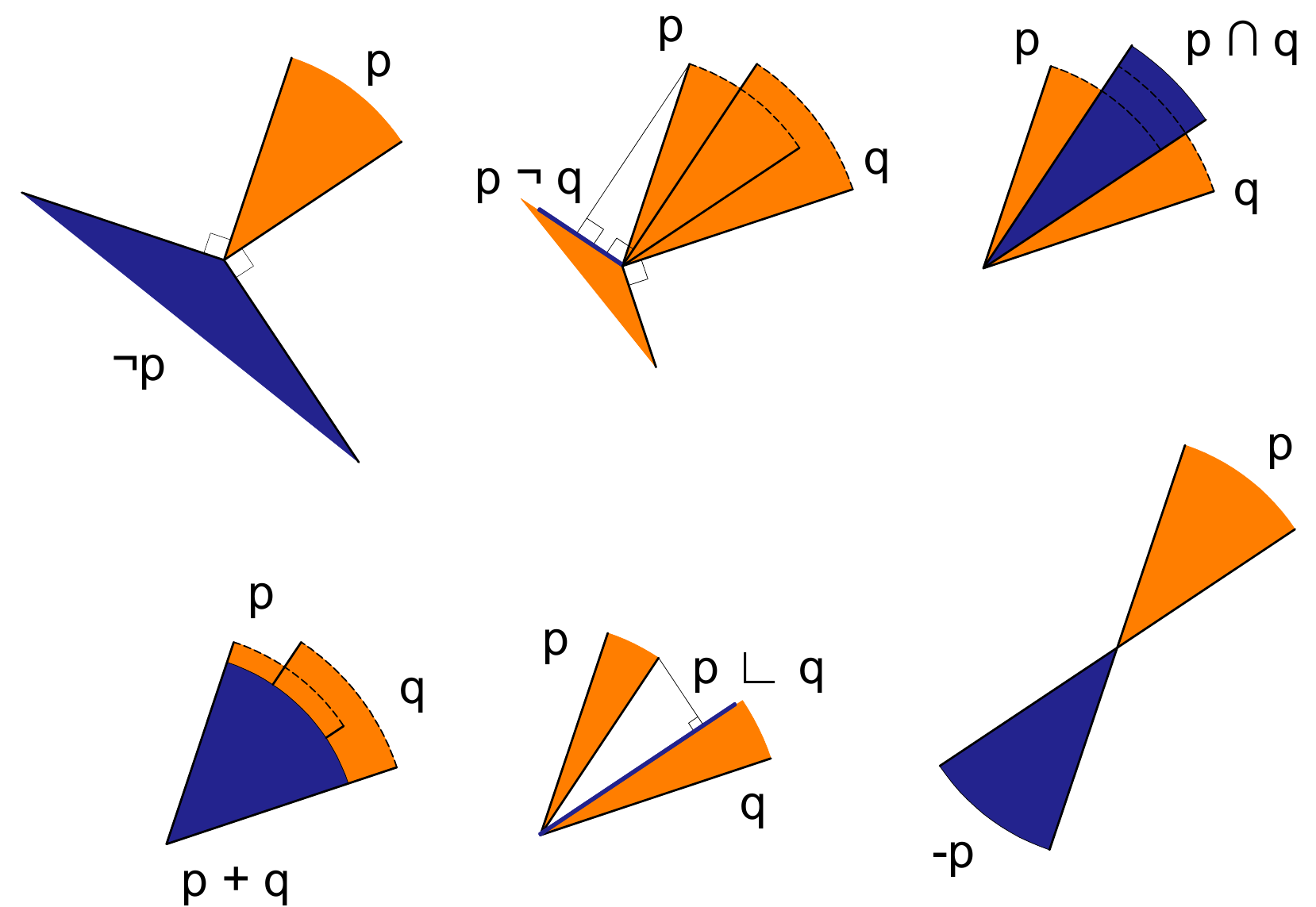}
	\caption{{\em Basic operations on convex cones.}
		Because cones arise from non-negative linear combinations,
		the {\em reflection} operation is needed. The other operations are
		generalizations of subspace operations. 	
		\label{fig:mathematical-cone-operations}}
\end{figure}

We will generalize the five basic operations {\em complement}, {\em sum}, {\em intersection}, {\em projection}, and {\em rejection} from the algebra of subspaces in \sectref{sect:subspace-operations} (\figref{fig:mathematical-cone-operations}). The non-negativity of weights mandates one new operation, {\em reflection} (unary `$-$'), which we give the same precedence order as `$+$'.

\subsubsection{Polar}
The {\em polar} $\neg a$, defined in
\eqref{eq:polar-cone-definition}, generalizes the {\em complement}
operation.

The polar is needed below to define and analyse conic rejection, but it
is not included as a basic operation of the cone algebra. In addition
to lacking an evident biological implementation, polarity does not
preserve finite generation in an infinite-dimensional Hilbert space.
Indeed, if a finitely generated cone $a$ is contained in the
finite-dimensional subspace $S=\operatorname{span}(a)$, then
$S^\perp\subseteq \neg a$. Thus, $\neg a$ generally contains an
infinite-dimensional subspace and cannot be finitely generated.
Polarity is therefore used only as an auxiliary mathematical
construction.

\subsubsection{Sum}
{\em Sum} is the closed conic hull of two convex cones. We write the
sum of two cones $a$ and $b$ as
\begin{equation}
	a+b
	\triangleq
	\overline{
		\left\{
		\myvec{u}+\myvec{w}
		\,\middle|\,
		\myvec{u}\in a,\ \myvec{w}\in b
		\right\}}
	=
	\Conic(a\cup b).
\end{equation}
For finitely generated cones, this Minkowski sum is already closed, and
the overbar has no effect.

\subsubsection{Conic projection}
The orthogonal projection of a point $x$ on a closed convex cone is the point in the cone nearest to~$x$.
Such a point always exists and is unique according to the projection theorem (\sectref{sect:moreau-decomposition}).
The orthogonal projection of a cone $a$ on another cone $b$ is a cone but not necessarily convex.
An example is when $a$ and $b$ are disjoint except for the apex, and the projection of $a$ on $b$ is limited to the boundary of~$b$.
We extend the projection to its conic hull and name it {\em conic projection} to guarantee a convex result,
\begin{equation}
	a\llcorner b
	\triangleq
	\Conic
	\left\{
	P_b\myvec{x}
	\,\middle|\,
	\myvec{x}\in a
	\right\}.
\end{equation}
Unless stated otherwise, when we write {\em projection} of a cone below, we mean {\em conic projection}.
When referring to the original, non-extended projection, we prefix by ``orthogonal''.

\subsubsection{Conic rejection}
We define {\em conic rejection} of $b$ from $a$ as the {\em conic
	projection} of $a$ onto the {\em polar} of~$b$,
\begin{equation}
		\label{eq:rejection-definition}
	a\,\neg\,b
	\triangleq
	\Conic
	\left\{
	\myvec{x}-P_b\myvec{x}
	\,\middle|\,
	\myvec{x}\in a
	\right\}
		=
	a \, \llcorner \, (\neg b).
\end{equation}
{\em Conic projection} can, in turn, be expressed as two successive
{\em conic rejections} (proved in
\sectref{sect:projection-by-rejection}),
\begin{equation}
	\label{eq:projection-by-rejection}
	a\,\llcorner\,b
	=
	a\,\neg\,(a\,\neg\,b).
\end{equation}

\subsubsection{Intersection}
The {\em intersection} of two convex cones is trivially
\begin{equation}
	a \, \cap \, b \triangleq \{\myvec{v}\in a \text{ and }\myvec{v}\in  b\} .
\end{equation}
It can be expressed as a combination of three {\em conic rejections} and two {\em sums}
(Proof in \sectref{sect:intersection-by-rejection-and-span}):
\begin{align}
		a \cap b 
		& = a \, \neg \, (b \, \neg \, a \,+\, a \, \neg \, b) \\
		& = b \, \neg \, (b \, \neg \, a \,+\, a \, \neg \, b) \\
		& = (a + b) \, \neg \, (b \, \neg \, a \,+\, a \, \neg \, b) .
    	\label{eq:intersection-by-rejection}
\end{align}

\subsubsection{Reflection} 
{\em Reflection} mirrors a cone through the origin.
This operation is trivial for subspaces because a subspace containing a vector $\myvec{v}$ always
contains $-\myvec{v}$, but it is a meaningful operation for cones. It is defined simply as
\begin{equation}
	-a \triangleq \left\{-\myvec{v} \, | \, \myvec{v} \in a \right\} .
\end{equation}
The {\em reflection} of the {\em\polar} $-(\neg a)$ is the {\em dual} cone. The linear hull of a convex cone $a$ equals the sum of $a$ and its reflection,
\begin{equation}
\operatorname{span}(a) = a+(-a).
\end{equation}

\begin{proof}[{\qreff{Operations on cones}
		{ex:operations-on-cones}}]
	Continue example \ref{ex:populations} and define
	\begin{equation}
	 X=\Conic\{\myvec{x}^{(1)},\myvec{x}^{(2)}\},
	 \qquad
	 \widehat{Y}=AX
	 =\Conic\{\widehat{\myvec{y}}^{(1)},
	             \widehat{\myvec{y}}^{(2)}\},
	\end{equation}
	and
	\begin{equation}
	 Y=\Conic\{\myvec{y}^{(1)},\myvec{y}^{(2)}\}.
	\end{equation}
	The comparison relevant to rejection is between $Y$ and $\widehat{Y}$ in the
	same output Hilbert space, not directly between the input cone $X$ and the
	output cone $Y$. By construction, $\myvec{e}\perp\widehat{Y}$. Hence, for
	$\alpha,\beta\geq0$,
	\begin{equation}
	 \alpha\myvec{y}^{(1)}+\beta\myvec{y}^{(2)}
	 =\left(\alpha\widehat{\myvec{y}}^{(1)}
	       +\beta\widehat{\myvec{y}}^{(2)}\right)
	  +3\alpha\myvec{e}
	\end{equation}
	is an orthogonal cone--polar decomposition of that point relative to
	$\widehat{Y}$: the first term belongs to $\widehat{Y}$ and the second belongs
	to its polar and is orthogonal to it. This is the decomposition later
	formalized by the Moreau theorem. It follows that
	\begin{equation}
	 Y\,\neg\,\widehat{Y}=\Conic\{\myvec{e}\}.
	\end{equation}
	The neuronal sign convention reflects this residual, so the output cone is
	\begin{equation}
	 Z=\Conic\{\myvec{z}^{(1)},\myvec{z}^{(2)}\}
	 =\Conic\{-\myvec{e}\}
	 =-\bigl(Y\,\neg\,\widehat{Y}\bigr).
	\end{equation}
	This makes both the space in which rejection is performed and the final sign
	explicit.
\end{proof}

\subsection{Comparing convex cones}
\label{sect:cone-relations}

The comparison of convex cones parallels the comparison of subspaces
described in \sectref{sect:emptiness-and-normality}. In both cases,
inclusion defines a partial order, while projection and rejection
provide graded measures of similarity. Cone inclusion is more
informative, however, because two cones may have the same linear hull
and still differ in their extent or orientation.

We use the same definition of {\em emptiness} as for subspaces in
\sectref{sect:emptiness-and-normality}, namely, being equal to
$\{\myvec{0}\}$.

Convex cones have a key advantage over subspaces: they can be ordered
even when they have the same linear hull. We define $p\preceq q$ by
$p\subseteq q$. Conic rejection tests this relation: a convex cone $p$
is a subset of another convex cone $q$ exactly when $p\neg q$ is
empty.

Cones $a$ and $b$ can therefore be compared by computing the
rejections $a\neg b$ and $b\neg a$. If $a\neg b$ is empty, then
$a\subseteq b$; if $b\neg a$ is empty, then $b\subseteq a$; if both
are empty, then $a=b$.

For a point $\myvec{x}$ and a closed convex cone $c$, let
$P_c\myvec{x}$ denote the orthogonal projection of $\myvec{x}$ onto
$c$. For two cones $a$ and $b$ that each contain a nonzero vector,
define the unit sphere
\begin{equation}
	\mathbb{S}_H
	=
	\{\myvec{x}\in H\mid\|\myvec{x}\|=1\}
\end{equation}
and the symmetric worst-case similarity
\begin{align}
	s(a,b)
	&=
	\min\left\{
	\inf_{\myvec{x}\in a\cap\mathbb{S}_H}
	\|P_b\myvec{x}\|,
	\inf_{\myvec{x}\in b\cap\mathbb{S}_H}
	\|P_a\myvec{x}\|
	\right\},
	\label{eq:cone-worst-case-similarity}\\
	\theta(a,b)
	&=
	\arccos s(a,b).
	\label{eq:cone-comparison-angle}
\end{align}
Here, $0\leq s(a,b)\leq1$. Identical cones have similarity one.
The similarity is zero if either cone contains a direction in the
polar of the other. Thus, $\theta(a,b)$ measures the largest
directional mismatch between the cones, considering both directions.
The empty cone $\{\myvec{0}\}$ is handled separately by the emptiness
test.

The exact infima range over all directions in the cones. A finite
frame gives a computable proxy. Let
\begin{equation}
	\widehat F_c
	=
	\left\{
	\frac{\myvec{v}}{\|\myvec{v}\|}
	\;\middle|\;
	\myvec{v}\in F_c,\ \myvec{v}\ne\myvec{0}
	\right\}
\end{equation}
be the normalized frame of $c$. We then define
\begin{equation}
	\widehat{s}(a,b)
	=
	\min\left\{
	\min_{\myvec{u}\in\widehat F_a}\|P_b\myvec{u}\|,
	\min_{\myvec{v}\in\widehat F_b}\|P_a\myvec{v}\|
	\right\}.
	\label{eq:cone-frame-similarity}
\end{equation}
Each projection is an NNLS problem. The proxy is generally optimistic
because an interior direction can differ more than every frame vector.
Suppose that, for some $\gamma\geq0$, every unit direction in each
cone lies within distance $\gamma$ of a normalized frame vector.
Nonexpansiveness of metric projection then gives
\begin{equation}
	0
	\leq
	\widehat{s}(a,b)-s(a,b)
	\leq
	\gamma.
\end{equation}
Without such directional coverage, no uniform approximation guarantee
follows from the frame alone.

Thus, cone comparison retains the partial-order and graded-comparison
structure used for subspaces, while distinguishing cones that would
become identical if represented only by their linear hulls.

\subsection{Properties of the algebra of convex cones}
This section is the mathematical core of the paper. It contains two important but
fairly long proofs and may therefore feel demanding. Readers mainly interested in
the biological argument may skip the proofs on a first reading and return to them later.

Although the neuronal representations considered in this paper are
finitely generated convex cones, the proofs in this subsection do not
depend on finite generation. They hold more generally for closed convex
cones in a real Hilbert space, using the projection theorem,
Moreau's decomposition, and the bipolar identity
$\neg(\neg a)=a$. In this general setting, conic hulls and sums are
understood to be topologically closed whenever necessary.

\begin{figure}[!hbt] 
    \centering
	\includegraphics[scale=1.0]{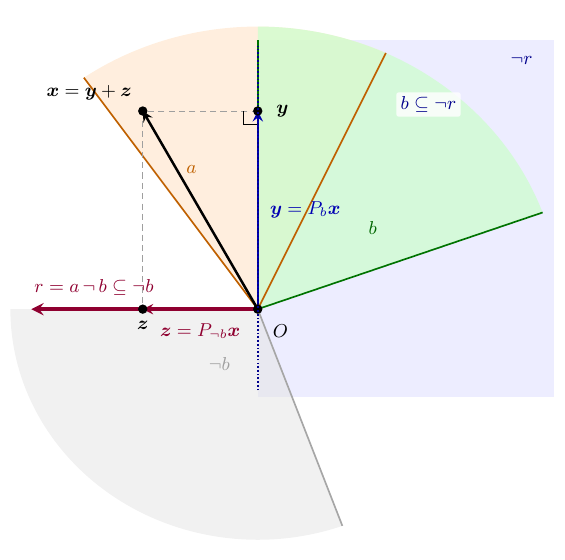}
	\caption{{\em Moreau decomposition underlying double rejection.}
	For $\myvec{x}\in a$, let
	$\myvec{y}=P_b\myvec{x}$ and
	$\myvec{z}=P_{\neg b}\myvec{x}$.
	Moreau decomposition gives
	$\myvec{x}=\myvec{y}+\myvec{z}$ with
	$\langle\myvec{y},\myvec{z}\rangle=0$.
	Since $\myvec{z}\in r=a\,\neg\,b$ and
	$r\subseteq\neg b$, polarity implies
	$b\subseteq\neg r$; consequently,
	$\myvec{y}=P_{\neg r}\myvec{x}$.
	The cones are shown schematically in a two-dimensional cross-section.
	\label{fig:cone-formulas}}
\end{figure}

\subsubsection{Some well-known properties of convex cones}
\label{sect:cone-properties}
Many useful properties of convex cones exist
\citep{Greer.1984ato,Ingram.Marsh.1991poc},
\begin{align}
& a\, \llcorner \, b \subseteq b , \\
& a \subseteq b \Leftrightarrow \neg \, a \supseteq \, \neg \, b , \\
& a = \neg \, \neg \, a ,\\
& a \, \llcorner \, (a \, \llcorner \, b) = a \, \llcorner \, b \\
& a \, \cap \, b = \neg \, (\neg \, a + \, \neg \, b) .
\label{eq:sum-intersection}
\end{align}
Below, we present two additional essential relations.

\subsubsection{Conic projection by conic rejections}
\label{sect:projection-by-rejection}
In this subsection, we prove that
\begin{equation}
 a\llcorner b=a\neg(a\neg b)
 \label{eq:projection-by-double-rejection-goal}
\end{equation}
for all closed convex cones $a$ and $b$. Thus conic projection, although
defined as a separate operation, can be expressed by applying conic rejection
twice.

Let
\begin{equation}
 r=a\,\neg\,b
   =\Conic\{P_{\neg b}\myvec{x}\mid\myvec{x}\in a\}.
 \label{eq:projection-rejection-r}
\end{equation}
We show that projecting a point of $a$ on $b$ gives the same result as
projecting it on $\neg r$. This is not a consequence merely of the inclusion
$b\subseteq\neg r$: projection on a larger convex set is generally different.
The equality follows from the particular orthogonal decomposition supplied by
the Moreau theorem.

Let $\myvec{x}\in a$ and put
\begin{equation}
 \myvec{y}=P_b\myvec{x},
 \qquad
 \myvec{z}=P_{\neg b}\myvec{x}.
\end{equation}
By \eqref{eq:moreau-decomposition},
$\myvec{x}=\myvec{y}+\myvec{z}$ and
$\langle\myvec{y},\myvec{z}\rangle=0$. By the definition of $r$,
$\myvec{z}\in r$: indeed, $\myvec{z}=P_{\neg b}\myvec{x}$ is one of the
vectors whose conic hull defines $r$. Moreover, $r\subseteq\neg b$. To see
this explicitly, every generator $P_{\neg b}\myvec{x}$ belongs to
$\neg b$, and a non-negative linear combination of vectors in the convex
cone $\neg b$ again belongs to $\neg b$.

Polarity reverses inclusion. In fact, if $c_1\subseteq c_2$, then a vector
having non-positive inner product with every vector of $c_2$ necessarily has
non-positive inner product with every vector of $c_1$; hence
$\neg c_2\subseteq\neg c_1$. Applying this observation to
$r\subseteq\neg b$ gives
\begin{equation}
 b=\neg\neg b\subseteq\neg r.
\end{equation}
Here we used the bipolar identity $\neg\neg b=b$, which holds because $b$ is
a closed convex cone.

We now apply Moreau decomposition a second time, this time to the closed
convex cone $\neg r$ and to the same point $\myvec{x}$.  The polar cone paired
with $\neg r$ is $\neg\neg r$.  We already know that
\begin{equation}
 \myvec{x}=\myvec{y}+\myvec{z},
 \qquad
 \myvec{y}\in\neg r,
 \qquad
 \myvec{z}\in r\subseteq\neg\neg r,
 \qquad
 \langle\myvec{y},\myvec{z}\rangle=0.
\end{equation}
Thus $\myvec{y}$ and $\myvec{z}$ satisfy exactly the membership, sum, and
orthogonality conditions of the Moreau decomposition for the cone $\neg r$.
The uniqueness part of the theorem therefore identifies $\myvec{y}$ as the
projection of $\myvec{x}$ on $\neg r$. Since
$\myvec{y}=P_b\myvec{x}$ by its original definition, we have proved the
pointwise identity
\begin{equation}
 P_b\myvec{x}=\myvec{y}=P_{\neg r}\myvec{x}
 \qquad\text{for every }\myvec{x}\in a.
 \label{eq:point-double-rejection}
\end{equation}

We next pass carefully from the pointwise identity to an identity between
cones. By definition,
\begin{align}
 a\llcorner b
 &=\Conic\{P_b\myvec{x}\mid\myvec{x}\in a\},\label{eq:double-rejection-hull-b}\\
 a\llcorner\neg r
 &=\Conic\{P_{\neg r}\myvec{x}\mid\myvec{x}\in a\}.
 \label{eq:double-rejection-hull-r}
\end{align}
Equation \eqref{eq:point-double-rejection} says that the set of generators on
the right-hand side of \eqref{eq:double-rejection-hull-b} is identical to the
set of generators on the right-hand side of
\eqref{eq:double-rejection-hull-r}: for each $\myvec{x}\in a$, the two
projected vectors are the same.  Hence their conic hulls are the same, and
\begin{equation}
 a\llcorner b=a\llcorner\neg r.
\end{equation}
Finally, $r=a\neg b$, and the definition of rejection gives
\begin{equation}
 a\neg r=a\llcorner\neg r.
\end{equation}
Substituting these two identities yields
\begin{equation}
	\label{eq:double-rejection}
		a \, \llcorner \, b 
		 = a \, \llcorner \, \neg \, (a \, \neg \, b)
		 = a \, \neg \, (a \, \neg \, b) .
\end{equation}
\myqed

\subsubsection{Intersection by conic rejections and sums}
\label{sect:intersection-by-rejection-and-span}
In this subsection, we prove that intersection can be expressed using only
conic rejections and a sum. More precisely, for all closed convex cones $a$
and $b$, we prove the three equivalent formulas
\begin{align}
 a\cap b
 &=a\neg(b\neg a+a\neg b)\notag\\
 &=b\neg(b\neg a+a\neg b)\notag\\
 &=(a+b)\neg(b\neg a+a\neg b).
 \label{eq:intersection-rejection-goal}
\end{align}
The proof is longer than the corresponding identity using polar cones because
the expression above deliberately eliminates explicit polar operations.

Let
\begin{equation}
 c=a\cap b,\qquad
 r_a=a\,\neg\,b,\qquad
 r_b=b\,\neg\,a,
 \qquad
 r=r_a+r_b,\qquad
 k=\neg r.
 \label{eq:intersection-auxiliary-cones}
\end{equation}
By \eqref{eq:sum-intersection},
\begin{equation}
	 k=\neg(r_a+r_b)=\neg r_a\cap\neg r_b.
 \label{eq:intersection-k}
\end{equation}
Since $r_a\subseteq\neg b$ and $r_b\subseteq\neg a$, polarity gives
$b\subseteq\neg r_a$ and $a\subseteq\neg r_b$. It follows that
\begin{equation}
	 c=a\cap b\subseteq k.
 \label{eq:intersection-contained-in-k}
\end{equation}

Let us spell out why \eqref{eq:intersection-contained-in-k} supplies one of
the two inclusions that will be needed. If $\myvec{x}\in c$, then
$\myvec{x}$ belongs to $a$, to $b$, and to $a+b$; it also belongs to $k$.
A point already in a closed convex cone is its own projection on that cone,
so
\begin{equation}
 P_k\myvec{x}=\myvec{x}\qquad(\myvec{x}\in c).
 \label{eq:intersection-fixed-points}
\end{equation}
Consequently, projecting $a$, $b$, or $a+b$ on $k$ produces a set containing
$c$. The non-obvious reverse inclusion is that no point outside $c$ can be
produced by these projections. We first establish the geometric fact needed
for that conclusion.

\paragraph{Lemma.}
For every $\myvec{q}\in k$,
\begin{equation}
 P_a\myvec{q}=P_b\myvec{q}=P_c\myvec{q}.
 \label{eq:intersection-projection-lemma}
\end{equation}
Furthermore, if $\myvec{g}=P_c\myvec{q}$ and
$\myvec{t}=\myvec{q}-\myvec{g}$, then
\begin{equation}
 \myvec{t}\in\neg(a+b),
 \qquad
 \langle\myvec{g},\myvec{t}\rangle=0.
 \label{eq:intersection-lemma-remainder}
\end{equation}

\emph{Proof.}
We apply Moreau decomposition four times.  The first application decomposes
$\myvec{q}$ relative to $a$ and its polar $\neg a$; the second decomposes the
same point relative to $b$ and $\neg b$.  The third decomposes the first
projection $\myvec{u}$ relative to $b$ and $\neg b$, while the fourth
decomposes the second projection $\myvec{v}$ relative to $a$ and $\neg a$:
\begin{align}
 \myvec{q}&=\myvec{u}+\myvec{u}^{\circ},
 &\myvec{u}&=P_a\myvec{q},
 &\myvec{u}^{\circ}&=P_{\neg a}\myvec{q},
 \label{eq:intersection-moreau-a}\\
 \myvec{q}&=\myvec{v}+\myvec{v}^{\circ},
 &\myvec{v}&=P_b\myvec{q},
 &\myvec{v}^{\circ}&=P_{\neg b}\myvec{q},
 \label{eq:intersection-moreau-b}\\
 \myvec{u}&=\myvec{v}_u+\myvec{r}_u,
 &\myvec{v}_u&=P_b\myvec{u},
 &\myvec{r}_u&=P_{\neg b}\myvec{u},
 \label{eq:intersection-moreau-ab}\\
 \myvec{v}&=\myvec{u}_v+\myvec{r}_v,
 &\myvec{u}_v&=P_a\myvec{v},
 &\myvec{r}_v&=P_{\neg a}\myvec{v}.
 \label{eq:intersection-moreau-ba}
\end{align}
Moreau decomposition supplies three facts in each row: the two displayed
components add to the vector on the left, the first component belongs to the
named cone and the second to its polar, and the two components are
orthogonal. Thus, in addition to the memberships displayed above,
$\myvec{u}\perp\myvec{u}^{\circ}$,
$\myvec{v}\perp\myvec{v}^{\circ}$,
$\myvec{v}_u\perp\myvec{r}_u$, and
$\myvec{u}_v\perp\myvec{r}_v$.

Because $\myvec{u}\in a$, the residual
$\myvec{r}_u=P_{\neg b}\myvec{u}$ belongs to $r_a=a\neg b$.
Similarly, $\myvec{r}_v\in r_b=b\neg a$. Since
$\myvec{q}\in k=\neg r_a\cap\neg r_b$,
\begin{equation}
 \langle\myvec{q},\myvec{r}_u\rangle\leq0,
 \qquad
 \langle\myvec{q},\myvec{r}_v\rangle\leq0.
 \label{eq:intersection-upper-bound}
\end{equation}
We next derive a lower bound for the sum of these inner products. From
\eqref{eq:intersection-moreau-a} and
\eqref{eq:intersection-moreau-ab},
we have
\begin{align}
 \langle\myvec{q},\myvec{r}_u\rangle
 &=\langle\myvec{u}+\myvec{u}^{\circ},\myvec{r}_u\rangle\\
 &=\langle\myvec{v}_u+\myvec{r}_u,\myvec{r}_u\rangle
   +\langle\myvec{u}^{\circ},\myvec{r}_u\rangle\\
 &=\|\myvec{r}_u\|^2
   +\langle\myvec{u}^{\circ},\myvec{r}_u\rangle.
\end{align}
Because $\myvec{u}^{\circ}\perp\myvec{u}$ and
$\myvec{r}_u=\myvec{u}-\myvec{v}_u$,
\begin{equation}
 \langle\myvec{u}^{\circ},\myvec{r}_u\rangle
 =\langle\myvec{u}^{\circ},\myvec{u}-\myvec{v}_u\rangle
 =-\langle\myvec{u}^{\circ},\myvec{v}_u\rangle.
\end{equation}
Therefore
\begin{equation}
 \langle\myvec{q},\myvec{r}_u\rangle
 =\|\myvec{r}_u\|^2
  -\langle\myvec{u}^{\circ},\myvec{v}_u\rangle.
 \label{eq:intersection-inner-product-a}
\end{equation}
Repeating the same calculation after interchanging $a$ and $b$ gives
\begin{equation}
 \langle\myvec{q},\myvec{r}_v\rangle
 =\|\myvec{r}_v\|^2
  -\langle\myvec{v}^{\circ},\myvec{u}_v\rangle.
 \label{eq:intersection-inner-product-b}
\end{equation}
Now $\myvec{u}^{\circ}\in\neg a$ and $\myvec{u}_v\in a$, while
$\myvec{v}^{\circ}\in\neg b$ and $\myvec{v}_u\in b$. Hence
\begin{equation}
 \langle\myvec{u}^{\circ},\myvec{u}_v\rangle\leq0,
 \qquad
 \langle\myvec{v}^{\circ},\myvec{v}_u\rangle\leq0.
 \label{eq:intersection-polar-cross-signs}
\end{equation}
The two quantities
\begin{equation}
-\langle\myvec{u}^{\circ},\myvec{u}_v\rangle
\qquad\hbox{and}\qquad
-\langle\myvec{v}^{\circ},\myvec{v}_u\rangle
\end{equation}
are non-negative.  By adding them to the expression
\begin{equation}
\langle\myvec{u}^{\circ},\myvec{v}_u\rangle
+\langle\myvec{v}^{\circ},\myvec{u}_v\rangle,
\end{equation}
we obtain
\begin{align}
 &\langle\myvec{u}^{\circ},\myvec{v}_u\rangle
  +\langle\myvec{v}^{\circ},\myvec{u}_v\rangle \notag\\
 &\quad\leq
 \left\langle
   \myvec{u}^{\circ}-\myvec{v}^{\circ},
   \myvec{v}_u-\myvec{u}_v
 \right\rangle.
 \label{eq:intersection-cross-bound}
\end{align}
Put $\myvec{h}=\myvec{v}_u-\myvec{u}_v$. From
\eqref{eq:intersection-moreau-a}--\eqref{eq:intersection-moreau-ba},
\begin{equation}
 \myvec{u}^{\circ}-\myvec{v}^{\circ}
 =\myvec{v}-\myvec{u}
 =-\myvec{h}-(\myvec{r}_u-\myvec{r}_v).
\end{equation}
It follows that
\begin{equation}
 \left\langle
   \myvec{u}^{\circ}-\myvec{v}^{\circ},\myvec{h}
 \right\rangle
 =-\|\myvec{h}\|^2
  -\langle\myvec{r}_u-\myvec{r}_v,\myvec{h}\rangle.
 \label{eq:intersection-square-expression}
\end{equation}

Here we can use the identity
\begin{equation}
 -\|\myvec{h}\|^2
 -\langle\myvec{r}_u-\myvec{r}_v,\myvec{h}\rangle
 =\frac14\|\myvec{r}_u-\myvec{r}_v\|^2
 -\left\|\myvec{h}+\frac12(\myvec{r}_u-\myvec{r}_v)\right\|^2,
 \label{eq:intersection-completing-square-identity}
\end{equation}
which is obtained by expanding the squared norm on the right. Since that
squared norm is non-negative, equations
\eqref{eq:intersection-square-expression} and
\eqref{eq:intersection-completing-square-identity} give
\begin{align}
 \left\langle
   \myvec{u}^{\circ}-\myvec{v}^{\circ},\myvec{h}
 \right\rangle
 &\leq \frac{1}{4}\|\myvec{r}_u-\myvec{r}_v\|^2 \\
 &\leq \frac{1}{2}
   \left(\|\myvec{r}_u\|^2+\|\myvec{r}_v\|^2\right).
 \label{eq:intersection-square-bound}
\end{align}
Adding \eqref{eq:intersection-inner-product-a} and
\eqref{eq:intersection-inner-product-b}, and then applying
\eqref{eq:intersection-cross-bound} and
\eqref{eq:intersection-square-bound}, gives the following explicit chain:
\begin{align}
 \left\langle\myvec{q},\myvec{r}_u+\myvec{r}_v\right\rangle
 &=\|\myvec{r}_u\|^2+\|\myvec{r}_v\|^2
 -\left(
   \langle\myvec{u}^{\circ},\myvec{v}_u\rangle
   +\langle\myvec{v}^{\circ},\myvec{u}_v\rangle
  \right)\\
 &\geq\|\myvec{r}_u\|^2+\|\myvec{r}_v\|^2
 -\left\langle
   \myvec{u}^{\circ}-\myvec{v}^{\circ},
   \myvec{v}_u-\myvec{u}_v
  \right\rangle\\
 &\geq\frac{1}{2}
 \left(\|\myvec{r}_u\|^2+\|\myvec{r}_v\|^2\right).
 \label{eq:intersection-lower-bound}
\end{align}
The left-hand side is non-positive by
\eqref{eq:intersection-upper-bound}, because each of its two summands is
non-positive. The right-hand side is a sum of squared norms and is therefore
non-negative. The same number can satisfy both bounds only if the right-hand
side is zero. Consequently,
\begin{equation}
 \myvec{r}_u=\myvec{r}_v=\myvec{0}.
\end{equation}
The equation $\myvec{r}_u=\myvec{0}$ changes
$\myvec{u}=\myvec{v}_u+\myvec{r}_u$ into
$\myvec{u}=\myvec{v}_u\in b$. Since $\myvec{u}\in a$ already, this shows
$\myvec{u}\in c=a\cap b$. Similarly,
$\myvec{r}_v=\myvec{0}$ gives $\myvec{v}=\myvec{u}_v\in a$, and since
$\myvec{v}\in b$, we have $\myvec{v}\in c$.

We now identify these vectors as projections on $c$. The first Moreau
decomposition above reads
\begin{equation}
 \myvec{q}=\myvec{u}+\myvec{u}^{\circ}.
\end{equation}
We have just proved $\myvec{u}\in c$. Moreover,
$\myvec{q}-\myvec{u}=\myvec{u}^{\circ}\in\neg a\subseteq\neg c$ and
$\myvec{u}\perp\myvec{u}^{\circ}$. Thus this same equation is an orthogonal
decomposition of $\myvec{q}$ into a vector of $c$ and a vector of its polar
$\neg c$. By the uniqueness part of Moreau decomposition for the cone $c$,
the first component must be $P_c\myvec{q}$. Hence
$\myvec{u}=P_c\myvec{q}$. The same argument applied to
$\myvec{q}=\myvec{v}+\myvec{v}^{\circ}$ gives
$\myvec{v}=P_c\myvec{q}$. This proves
\eqref{eq:intersection-projection-lemma}.

Let $\myvec{g}=P_c\myvec{q}=\myvec{u}=\myvec{v}$ and put
$\myvec{t}=\myvec{q}-\myvec{g}$. Because
$\myvec{t}=\myvec{q}-\myvec{u}=\myvec{u}^{\circ}\in\neg a$ and, since
$\myvec{u}=\myvec{v}$,
$\myvec{t}=\myvec{q}-\myvec{v}=\myvec{v}^{\circ}\in\neg b$, we have
\begin{equation}
 \myvec{t}\in\neg a\cap\neg b=\neg(a+b).
\end{equation}
Here the last equality $\neg a\cap\neg b=\neg(a+b)$ follows directly from the
definition of the polar: a vector has non-positive inner product with every
sum $\myvec{a}+\myvec{b}$ precisely when it has non-positive inner product
with every $\myvec{a}\in a$ and every $\myvec{b}\in b$ separately.
Finally, $\myvec{g}\perp\myvec{t}$ follows from the Moreau decomposition of
$\myvec{q}$ relative to $c$. This proves
\eqref{eq:intersection-lemma-remainder}.
\myqed

We can now complete the proof of the intersection identities. Let
$\myvec{x}\in a+b$ and put
\begin{equation}
 \myvec{q}=P_k\myvec{x},
 \qquad
 \myvec{p}=\myvec{x}-\myvec{q}.
\end{equation}
Apply Moreau decomposition to the point $\myvec{x}$ and the closed convex
cone $k$. Since $\myvec{q}=P_k\myvec{x}$, its other component is
$P_{\neg k}\myvec{x}=\myvec{x}-\myvec{q}=\myvec{p}$. Therefore
\begin{equation}
 \myvec{x}=\myvec{q}+\myvec{p},
 \qquad
 \myvec{q}\in k,
 \qquad
 \myvec{p}\in\neg k,
 \qquad
 \myvec{p}\perp\myvec{q}.
 \label{eq:intersection-moreau-k}
\end{equation}
We may apply the lemma because $\myvec{q}\in k$. It gives
\begin{equation}
 \myvec{q}=\myvec{g}+\myvec{t},
 \qquad
 \myvec{g}\in c,
 \qquad
 \myvec{t}\in\neg(a+b),
 \qquad
 \myvec{g}\perp\myvec{t}.
\end{equation}
Since $c\subseteq k$, $\myvec{g}\in k$, and therefore
$\langle\myvec{p},\myvec{g}\rangle\leq0$. Together with
$\myvec{p}\perp\myvec{q}$, this yields
\begin{equation}
 \langle\myvec{p},\myvec{t}\rangle
 =\langle\myvec{p},\myvec{q}-\myvec{g}\rangle
 =-\langle\myvec{p},\myvec{g}\rangle\geq0.
\end{equation}
On the other hand, $\myvec{x}\in a+b$ and
$\myvec{t}\in\neg(a+b)$, so $\langle\myvec{t},\myvec{x}\rangle\leq0$.
Using $\myvec{x}=\myvec{q}+\myvec{p}=\myvec{g}+\myvec{t}+\myvec{p}$,
$\myvec{g}\perp\myvec{t}$, and
$\langle\myvec{t},\myvec{p}\rangle\geq0$, we also obtain
\begin{align}
 \langle\myvec{t},\myvec{x}\rangle
 &=\langle\myvec{t},\myvec{g}+\myvec{t}+\myvec{p}\rangle \\
 &=\|\myvec{t}\|^2+\langle\myvec{t},\myvec{p}\rangle
 \geq\|\myvec{t}\|^2.
\end{align}
Combining the two bounds gives
\begin{equation}
 0\geq\langle\myvec{t},\myvec{x}\rangle
 \geq\|\myvec{t}\|^2\geq0.
\end{equation}
Hence $\|\myvec{t}\|=0$, so $\myvec{t}=\myvec{0}$, and therefore
$P_k\myvec{x}=\myvec{q}=\myvec{g}\in c$. We have shown that
\begin{equation}
 \{P_k\myvec{x}\mid\myvec{x}\in a+b\}\subseteq c.
\end{equation}
Conversely, if $\myvec{g}\in c$, then, as shown before the lemma,
$\myvec{g}\in a+b$ and $\myvec{g}\in k$. A point of $k$ projects on itself,
so $P_k\myvec{g}=\myvec{g}$. Thus every point of $c$ occurs among the
projections of points of $a+b$, proving the reverse inclusion. Consequently,
\begin{equation}
 \{P_k\myvec{x}\mid\myvec{x}\in a+b\}=c.
 \label{eq:intersection-point-projection}
\end{equation}
The same two-inclusion argument applies when the source is restricted to
$a$ or to $b$. Since $a,b\subseteq a+b$, the projections of their points on
$k$ lie in $c$ by \eqref{eq:intersection-point-projection}. Conversely,
$c\subseteq a$, $c\subseteq b$, and every point of $c\subseteq k$ projects
on itself. Hence
\begin{equation}
 \{P_k\myvec{x}\mid\myvec{x}\in a\}
 =\{P_k\myvec{x}\mid\myvec{x}\in b\}=c.
 \label{eq:intersection-point-projection-ab}
\end{equation}

It remains only to translate these pointwise projection identities into the
notation of conic rejection. Since $k=\neg r$ and
$r=r_a+r_b=b\neg a+a\neg b$, the definition of rejection gives
\begin{align}
 a\neg r
 &=\Conic\{P_{\neg r}\myvec{x}\mid\myvec{x}\in a\}
 =\Conic\{P_k\myvec{x}\mid\myvec{x}\in a\},\\
 b\neg r
 &=\Conic\{P_k\myvec{x}\mid\myvec{x}\in b\},\\
 (a+b)\neg r
 &=\Conic\{P_k\myvec{x}\mid\myvec{x}\in a+b\}.
\end{align}
Equations \eqref{eq:intersection-point-projection} and
\eqref{eq:intersection-point-projection-ab} say that each set inside braces
is exactly $c$. Because $c$ is already a closed convex cone,
$\Conic(c)=c$. Therefore
\begin{align}
 a\cap b
 &=a\,\neg\,(b\,\neg\,a+a\,\neg\,b),
 \label{eq:intersection-rejection-a}\\
 &=b\,\neg\,(b\,\neg\,a+a\,\neg\,b),
 \label{eq:intersection-rejection-b}\\
 &=(a+b)\,\neg\,(b\,\neg\,a+a\,\neg\,b).
 \label{eq:intersection-rejection-sum}
\end{align}
\myqed

\subsection{Approximate invariance of the algebra of convex cones}
\label{sect:approximate-cone-invariance}
Here, \emph{invariance} refers to invariance of meaning, not to a cone remaining
fixed geometrically.  A message may acquire new coordinates and a different
geometric representation as it passes from one neuron population to another.
Its meaning is invariant when the cone operations commute, exactly or to a
controlled approximation, with this transformation.  In mathematical
terminology this property is also called \emph{equivariance}.

Exact equivariance of the complete cone algebra imposes severe conditions: a
transmission map would have to preserve the relevant inner products, and hence
angles and projections, up to one common scale factor.  Such a requirement is
not credible on the entire, very high-dimensional space of all possible
population activities.  It is also much stronger than semantic invariance
requires.  Neural messages are sparse or have low intrinsic dimensionality,
cone calculations use only a restricted collection of directions, and
biological decoding can tolerate perturbations that remain smaller than the
separation between meanings.  We therefore proceed from the exact reference
case to a restricted approximate condition.  The resulting condition allows
sparse rectangular maps, unequal population sizes, axonal branching, and even
reduction of ambient dimensionality.  The purpose of this section is to show
that the globally severe requirement can thereby be relaxed to conditions
that are reasonable for biological transmission.

Let $M:H\longrightarrow H'$ be the linear transformation between the message
spaces of two populations, and let
\begin{equation}
 Ma=\{M\myvec{x}\mid \myvec{x}\in a\}.
 \label{eq:linear-image-cone}
\end{equation}
The image $Ma$ is a convex cone for every linear $M$; entrywise
non-negativity of $M$ is not required for this fact.  Non-negativity is,
however, biologically relevant when the entries describe excitatory synaptic
transmission.  It must not be confused with preservation of angles or
projections.

Linearity preserves reflection and sum exactly for finitely generated convex cones:
\begin{equation}
 M(-a)=-Ma,
 \qquad
 M(a+b)=Ma+Mb.
 \label{eq:linear-exact-equivariance}
\end{equation}
Intersection is also preserved when $M$ is injective on
$\operatorname{span}(a\cup b)$; otherwise distinct messages can collide.
Projection, rejection, and polarity depend on the inner product. Within a
relevant subspace $S\subseteq H$, their exact preservation is guaranteed by the
reference condition
\begin{equation}
 \langle M\myvec{x},M\myvec{y}\rangle
 =\alpha\langle\myvec{x},\myvec{y}\rangle,
 \qquad \myvec{x},\myvec{y}\in S,
 \qquad \alpha>0,
 \label{eq:exact-restricted-conformality}
\end{equation}
where $\alpha$ is a positive scale factor. Thus, $M$ must be a scaled isometry
on $S$. Under this condition, metric projection and rejection commute with
$M$, and polarity is preserved within the embedded subspace $M(S)$. This is a
useful benchmark, not a biological requirement. Neural transmission only needs
to preserve the operations approximately on the sparse or low-dimensional
message model that is actually active. We therefore use the following
quantitative relaxation.

\subsubsection{Restricted approximate isometry}
Exact scaled isometry implies
$\|M\myvec{u}\|^2=\alpha\|\myvec{u}\|^2$.  Its natural quantitative
relaxation is to allow a relative geometric distortion $\rho$ in this norm
identity.  On a relevant subspace $S\subseteq H$, the resulting condition is
\begin{equation}
 (1-\rho)\alpha\|\myvec{u}\|^2
 \leq \|M\myvec{u}\|^2
 \leq (1+\rho)\alpha\|\myvec{u}\|^2,
 \qquad \myvec{u}\in S,
 \label{eq:restricted-scaled-isometry}
\end{equation}
where $0<\rho<1$ is the geometric distortion.  This is the standard scaled subspace-embedding
condition; on a union of sparse coordinate subspaces it is the restricted
isometry property (RIP) \mbox{\citep{Candes.Tao.2005dlp}}.  The condition is
deterministic; if $M$ is drawn from a random ensemble, one requires the
simultaneous inequalities to hold with probability at least $1-\delta$, where
$0<\delta<1$ is the allowed failure probability.

The subspace or sparse model must contain not only the observed frame vectors
but also all vectors produced while the cone operation is performed: conic
combinations, sums, differences, projected vectors, residuals, and the
relevant tangent or normal directions.  In the sparse case, both the sum and
the difference of two $k$-sparse vectors can have the union of their supports
and hence be $2k$-sparse.  A sum operation gives the most immediate example of
this temporary support growth.  An RIP of at least order $2k$ is therefore
normally needed even when the incoming and outgoing messages are intended to
be $k$-sparse.  An RIP stated only for the original frame vectors is not
sufficient unless all intermediate projections and residuals remain in the
same sparse model.

Biological activation functions help prevent this support growth from accumulating
over successive operations.  In the rectifying range, the activation
function suppresses weak components and can return an expanded representation
toward a chosen sparsity level (\sectref{sect:activation-function}).  For
non-negative activities, the map
$u_i\mapsto\max(0,u_i-\tau)$, where $\tau\geq0$ is the threshold, is precisely
non-negative soft thresholding, the proximal operation associated with an
$\ell_1$ penalty, with $\|\myvec{u}\|_1=\sum_i|u_i|$, under a non-negativity
constraint. Thus the activation
function provides a biological analogue of $\ell_1$ regularization and
shrinkage \citep{Donoho.1995dnb}.  It does not remove the need for a
$2k$-order guarantee during the operation itself, but it can restore sparsity
before the result is transmitted to the next algebraic stage.

On a subspace $S$, polarization of
\eqref{eq:restricted-scaled-isometry} gives
\begin{equation}
 \left|
 \langle M\myvec{u},M\myvec{v}\rangle
 -\alpha\langle\myvec{u},\myvec{v}\rangle
 \right|
 \leq \rho\alpha
 \|\myvec{u}\|\,\|\myvec{v}\|,
 \qquad \myvec{u},\myvec{v}\in S.
 \label{eq:restricted-inner-product-bound}
\end{equation}
For $k$-sparse $\myvec{u}$ and $\myvec{v}$, the same conclusion follows when
the RIP applies to their joint support, of size at most $2k$.
This is the central condition for approximate semantic invariance.  It says
that angles, and not merely the individual message norms, are approximately
preserved.  
Random Johnson–Lindenstrauss projections preserve finite message families with high probability
\citep{Johnson.Lindenstrauss.1984eol,Dasgupta.Gupta.2002aep}. A complete cone calculation requires a stronger guarantee: the map must also preserve the relevant subspaces, sparse-subspace unions, or tangent and normal directions.

The qualification ``approximate'' is indispensable near a polar boundary.
Equation \eqref{eq:restricted-inner-product-bound} preserves the sign of an
inner product whenever it has the angular margin
\begin{equation}
 |\langle\myvec{u},\myvec{v}\rangle|
 >\rho\|\myvec{u}\|\,\|\myvec{v}\|.
 \label{eq:angular-margin}
\end{equation}
No approximate isometry can uniformly preserve the sign of inner products
that are zero or arbitrarily close to zero.  Thus, points strictly inside a
polar cone are stable when their margin exceeds the distortion, whereas the
classification of points on its boundary can change.  The same phenomenon
occurs when an intersection is nearly tangential or when a small perturbation
changes the active face selected by a projection.  Semantic decoding must
therefore tolerate a small angular error, or the represented cones must be
separated by an adequate margin.

This limitation is biologically natural rather than pathological. Many learned
and perceptual categories have graded or context-dependent boundaries
\citep{Rosch.Mervis.1975frs}. Small changes in the classification of points near
such a boundary need not represent a loss of meaning. Robust semantic transmission
requires stability of well-separated message families, not invariant classification
of every boundary case.

There is also a direct projection-error bound. Let $S\subseteq H$ be a
closed subspace containing $\myvec{x}$ and $b$, and assume that
\eqref{eq:restricted-scaled-isometry} holds on $S$. Its lower bound makes
the restriction of $M$ to $S$ injective and bounded below. Since $b$ is
closed, it follows that $Mb$ is also closed. Consequently,
$P_{Mb}(M\myvec{x})$ exists and belongs to $Mb$.

Let $\myvec{p}=P_b\myvec{x}$. There is then a unique $\myvec{q}\in b$
such that
\begin{equation}
	M\myvec{q}=P_{Mb}(M\myvec{x}).
\end{equation}
Optimality of $\myvec{q}$ in the transformed space gives
\begin{equation}
 \|\myvec{x}-\myvec{q}\|^2
 \leq \frac{1+\rho}{1-\rho}
 \|\myvec{x}-\myvec{p}\|^2.
\end{equation}
The projection variational inequality also gives
\begin{equation}
 \|\myvec{x}-\myvec{q}\|^2
 \geq \|\myvec{x}-\myvec{p}\|^2
       +\|\myvec{q}-\myvec{p}\|^2.
\end{equation}
Combining these inequalities yields
\begin{equation}
 \|\myvec{q}-\myvec{p}\|
 \leq
 \sqrt{\frac{2\rho}{1-\rho}}\,
 \operatorname{dist}(\myvec{x},b),
 \label{eq:approximate-projection-error}
\end{equation}
where
$\operatorname{dist}(\myvec{x},b)
\triangleq\inf_{\myvec{y}\in b}\|\myvec{x}-\myvec{y}\|
=\|\myvec{x}-\myvec{p}\|$
is the distance from the point $\myvec{x}$ to the cone $b$.  In the transmitted
space, the equivariance error is consequently bounded by
\begin{align*}
 &\|P_{Mb}(M\myvec{x})-M(P_b\myvec{x})\| \\
 &\qquad\leq
 \sqrt{\frac{2\rho(1+\rho)\alpha}{1-\rho}}\,
 \operatorname{dist}(\myvec{x},b).
\end{align*}
Hence projection, and by Moreau decomposition rejection, commute with $M$ up
to a controlled error.  This pointwise statement extends to a finite frame of
messages by applying the bound to every frame vector.  The resulting output
cones retain their semantic identity whenever this error is smaller than the
angular or decoding margin that distinguishes them from competing cones.

\subsubsection{Rectangular maps and dimensional reduction}
Two communicating populations will generally contain different numbers of
neurons.  If the source and target populations have $n$ and $m$ spatial
coordinates, respectively, their effective transmission map is therefore
rectangular, $M\in\mathbb{R}^{m\times n}$.  In particular, when $m<n$, the map
has a non-trivial kernel and cannot be injective, let alone isometric, on all
of $\mathbb{R}^n$.  Global exact equivariance is then impossible.

This loss of ambient dimensions is not in itself an obstacle.  Suppose all
vectors needed for a cone calculation lie in an intrinsic subspace $S$ of
dimension $d$.  If $d\leq m$ and
\eqref{eq:restricted-scaled-isometry} holds on $S$, its lower bound makes
$M|_S$ injective and the calculation can be preserved even though $M$ is
non-injective on the ambient space.  The Johnson--Lindenstrauss lemma makes
the dimensional advantage explicit: $N$ relevant points can be embedded with
distortion $\rho$ and failure probability $\delta$ in
\begin{equation}
 m=O\!\left(\rho^{-2}\log\frac{N}{\delta}\right)
\end{equation}
dimensions.  A uniform embedding of a fixed $d$-dimensional subspace instead
requires, for standard random constructions,
$m=O(\rho^{-2}[d+\log(1/\delta)])$
\citep{Johnson.Lindenstrauss.1984eol,Dasgupta.Gupta.2002aep}.  Thus the
decisive quantities are the number of relevant messages, their intrinsic
dimension, or their sparsity---not the full size $n$ of the source
population.  Dimensional reduction is compatible with semantic invariance
when the represented structure has dimension
$d\ll\min(m,n)$; it fails if $S\cap\ker M$ contains a nonzero semantically
relevant direction.

The Johnson--Lindenstrauss result establishes the geometric possibility of
such compression; it is not by itself an anatomical model, since the usual
random matrices need not be sparse or non-negative.  We next give a separate
sufficient condition adapted to sparse axonal branching.

\subsubsection{A sparse collateral-branching model}
An ideal one-to-one routing between equally large populations can be
represented by a permutation matrix and is an exact isometry.  This is useful
as a limiting reference case, not as a realistic anatomical hypothesis:
communicating populations will rarely have equal size, and axons normally
branch and converge.  The relevant biological model is therefore a sparse
rectangular matrix.  Column $\myvec{m}_i$ of $M$ describes the effective
outputs originating from input axon $i$.

Collateral branching by itself need not impair invariance.  If different
input axons terminate on disjoint output coordinates and the columns have
equal squared norm $\alpha$, then
\begin{equation}
 M^TM=\alpha I,
\end{equation}
so the branching map is an exact scaled isometric embedding.  This disjoint
construction requires at least as many output as input coordinates.  When
the target population is smaller, exact orthogonality of all columns is
impossible, but restricted approximate isometry on the active message model
can still hold as described above.

Two biologically meaningful deviations determine the approximate error:
unequal total transmission strength and overlap between the collateral trees
of different axons.  After choosing a common scale $\alpha$, suppose
\begin{equation}
 \left|\frac{\|\myvec{m}_i\|^2}{\alpha}-1\right|\leq\eta,
 \qquad
 \frac{|\langle\myvec{m}_i,\myvec{m}_j\rangle|}{\alpha}\leq\mu
 \quad(i\ne j),
 \label{eq:biological-gain-coherence}
\end{equation}
where $\eta$ measures variation in total effective squared gain and $\mu$
measures the coherence caused by shared targets. The gain is naturally
interpreted through pulse-frequency modulation (PFM), the description of
rate-coded neuronal communication adopted here. Internal potential modulates
the output spike frequency, and the
receiving synapses and dendrites low-pass filter the pulse train to recover a
graded signal.  Channel-current fluctuations introduce high-frequency spike
timing variability that is removed by this filtering but improves the
resolution of low-intensity PFM signals
\citep{Nilsson.Jorntell.2021ccf}.  Thus $\|\myvec{m}_i\|^2$ represents the
effective gain from an input firing-rate waveform to the collection of
filtered responses at its collateral targets; it is not merely a count of
synapses or the gain of a single spike.

For a message supported on a set $I$ of at most $k$ input axons, let $M_I$ be
the submatrix formed by the active columns of $M$. Its Gram matrix is
$G_I=M_I^TM_I$, with entries
$(G_I)_{ij}=\langle\myvec{m}_i,\myvec{m}_j\rangle$. If $\myvec{x}_I$ contains
the active coordinates, then
$\|M\myvec{x}\|^2=\myvec{x}_I^TG_I\myvec{x}_I$. The bounds in
\eqref{eq:biological-gain-coherence} place the diagonal entries of
$G_I/\alpha$ within $\eta$ of one and every off-diagonal entry within $\mu$ of
zero. The Gershgorin circle theorem
\citep{Horn.Johnson.2012ma} therefore places every eigenvalue
of $G_I/\alpha$ in
$[1-\eta-(k-1)\mu,\,1+\eta+(k-1)\mu]$. It follows that
\begin{equation}
 (1-\rho_k)\alpha\|\myvec{x}\|^2
 \leq\|M\myvec{x}\|^2
 \leq(1+\rho_k)\alpha\|\myvec{x}\|^2,
 \qquad
 \rho_k\leq\eta+(k-1)\mu,
 \label{eq:collateral-rip-bound}
\end{equation}
where $\rho_k$ is the resulting order-$k$ geometric distortion. Whenever
$\rho_k<1$, this is an RIP bound of order $k$.  The relation
makes the biological trade-off explicit.  Sparse messages
(small $k$), homeostatic normalization of each axon's total effective output
(small $\eta$), and little overlap between collateral target sets (small
$\mu$) together imply approximate invariance.

For example, suppose every axon makes $d$ equally weighted effective branches
and any two axons share at most $r$ output targets.  After normalizing each
column, $\eta=0$ and
\begin{equation}
 \mu\leq\frac{r}{d},
 \qquad
 \rho_k\leq(k-1)\frac{r}{d}.
 \label{eq:branch-overlap-bound}
\end{equation}
Consequently, additional branches are not intrinsically harmful: they can
improve robustness if they provide redundant, largely disjoint transmission.
The distortion grows instead with the fraction of shared targets and with the
number of simultaneously active input axons.  Entrywise non-negativity makes
overlap errors add rather than cancel.  Consequently, sparse, approximately
disjoint branching is especially important for a purely excitatory map.  Signed or
inhibitory contributions could produce cancellation, but they require a model
that keeps excitation and inhibition, or their effective signs, explicit.

The phrase ``simultaneously active input axons'' should not be interpreted as
if each axon supplied one static scalar to a classical weighted-sum neuron.
Each axon carries a time-varying PFM signal, and each synaptic and dendritic
path has its own low-pass characteristics
(\sectref{sect:mathematical-model}).  More generally, let $\mathcal{M}$ denote
the full spatiotemporal transmission operator, let $g_{ji}$ be the effective
non-negative gain from input axon $i$ to output neuron $j$, let $h_{ji}$ be the
impulse response of that synaptic and dendritic path, and let $*$ denote
temporal convolution. The transmitted signal at output neuron $j$ then has the form
\begin{equation}
 (\mathcal{M}\myvec{x})_j(t)
 =\sum_i g_{ji}\,(h_{ji}*x_i)(t).
\end{equation}
Restricting the signals to a finite temporal
wavelet dictionary turns the operator $\mathcal{M}$ into a finite rectangular
block matrix, which is the effective $M$ used above.

This spatiotemporal description is richer than anatomical target overlap
alone.  Signals from two axons converging on the same neuron need not become
identical, because different path filters can place their responses in
different temporal directions.  The coherence $\mu$ should therefore be
computed between the complete filtered spatiotemporal responses, not merely
from the number of shared target neurons.  Filter diversity may reduce the
effective coherence of converging routes, although excessive gain or phase
variation can also increase distortion.  Equations
\eqref{eq:biological-gain-coherence} and
\eqref{eq:collateral-rip-bound} are accordingly conditions on this full
effective operator.  The subsequent activation function can then suppress
weak output components and re-establish a suitable sparsity level.

Hence projection, and by Moreau decomposition rejection, commute with
$M$ up to a controlled pointwise error. If
\eqref{eq:restricted-scaled-isometry} holds uniformly on a common
subspace containing $a$ and $b$, the bound applies to every
$\myvec{x}\in a$ and therefore controls the complete conic projection,
rather than only the projections of a chosen frame. In the sparse case,
the RIP must instead cover all conic combinations, projections, and
residuals required by the operation, as specified above. The resulting
output cones retain their semantic identity whenever the error is
smaller than the angular or decoding margin that distinguishes them
from competing cones.

\section{Neuronal implementation of computation}
\label{sect:neuronal-implementation}
This section explores how neuron populations act as computational
operators in the algebra of convex cones. Populations are
qualitatively more powerful than individual neurons because sparse
activity can select different learned parts of a population-level
mapping for different messages.

We distinguish the adaptation phase from the subsequent computational
phase. With a diminishing learning rate, the weights are assumed to
have approximately converged to an NNLS solution. With a constant
learning rate, they may instead fluctuate within a small steady-state
neighbourhood of such a solution. In either case, weight changes are
assumed to be slow relative to message transmission, so the weights
can be regarded as effectively fixed during an individual
computational operation.

\subsection{The neuron population as a primitive operator}
\label{sect:population-as-operator}
A sequence of messages represents and transmits a cone along an axon
bundle. The axons define the coordinate axes, whereas the messages
outline the cone through correlations between axon signals or through
their point density (\figref{fig:cones}). Multiplication by any fixed
weight matrix is a linear transformation and therefore maps a convex
cone to a convex cone. For excitatory transmission, the weight matrix
is also entrywise non-negative, consistently with the biological
constraint that motivates conic rather than arbitrary signed linear
combinations.

The representation is robust against confusing different cones. Even
if individual messages from two sequences coincide, a substantial
part of the complete sequences must match, or correlate, for the cones
to be confused. Moreover, if every active component of a message is
multiplied by the same positive factor---for example, because the
overall firing-rate gain changes---the message merely moves along a
ray from the origin. Because a cone contains every non-negative scalar
multiple of each of its points, such a common gain change leaves the
represented cone unchanged.

Let $p$ denote the inhibitory input cone and $q$ the excitatory input
cone. During adaptation, the population receives a stream of
inhibitory target messages together with excitatory candidate
messages. The candidates provide possible explanations of components
of the target. A candidate that is correlated with a component not yet
explained systematically influences the synaptic weights. A candidate
that is uncorrelated with the remaining prediction error has no
systematic long-term influence. Learning therefore makes the
excitatory input reconstruct the predictable part of the inhibitory
message, while the part that cannot be reconstructed remains in the
prediction residual.

The output of the population is the reflection of this residual. To
see the sign, let $\myvec{y}$ be the inhibitory target message and let
$\widehat{\myvec{y}}$ be its reconstruction from the excitatory
message. The residual in the target coordinates is
\begin{equation}
	\myvec{e}
	=
	\myvec{y}-\widehat{\myvec{y}},
\end{equation}
whereas the population output has the opposite sign because the target
input is inhibitory:
\begin{equation}
	\myvec{z}
	=
	\widehat{\myvec{y}}-\myvec{y}
	=
	-\myvec{e}.
\end{equation}

The message representation is assumed to be sparse. At any given
time, only a small number of learned rays is active, and their
non-negative combinations occupy a low-dimensional part of the
population space. The active rays identify the component that can be
reconstructed, while components unrelated to them remain in the
residual. Different sparse ray correspondences can therefore coexist
in the same large population and can be selected by different
messages.

In the ideal algebraic description, the reconstructed messages
generate the conic projection $p\llcorner q$, and the residual
messages generate the conic rejection $p\neg q$. The generic neuron
population in \figref{fig:adaptive-filter}, having the defining
equation \eqref{eq:defining-equation}, then implements
\begin{equation}
	r
	=
	-\left(p\neg q\right),
	\label{eq:primitive-population-operation}
\end{equation}
where $r$ denotes the output cone. This equation specifies the intended
algebraic semantics. The next subsection describes how sparse ray
correspondences are learned, and the following subsection states the
conditions under which the resulting population map realizes or
approximates this semantics.

\subsection{Learning sparse ray correspondences}
\label{sect:learning-sparse-ray-correspondences}
We first recall the learning problem at the level of an individual
neuron. Let the scalar excitatory input functions
$x_1,\ldots,x_m$ generate the neuron-local cone
\begin{equation}
	c
	=
	\Conic\{x_1,\ldots,x_m\}.
\end{equation}
For an inhibitory target function $y$, adaptation solves
\begin{equation}
	\myvec{w}^{\,*}
	\in
	\argmin_{\myvec{w}\geq\myvec{0}}
	\left\|
	y-\sum_i w_i x_i
	\right\|^2.
\end{equation}
The coefficient vector $\myvec{w}^{\,*}$ need not be unique, but the
reconstructed function is unique and is given by
\begin{equation}
	\widehat{y}
	=
	\sum_i w_i^*x_i
	=
	P_cy.
\end{equation}
By Moreau decomposition, the neuron-local residual satisfies
\begin{equation}
	y-\widehat{y}
	\in
	\neg c,
	\qquad
	\left\langle
	\widehat{y},y-\widehat{y}
	\right\rangle
	=
	0.
\end{equation}
Thus, an individual neuron reconstructs the conic projection of its
target onto the cone generated by its excitatory input functions, and
the remaining prediction error is the corresponding pointwise conic
rejection.

This result also explains the role of correlation in learning. Write
the residual at output neuron $j$ as
\begin{equation}
	e_j
	=
	y_j-\widehat{y}_j.
\end{equation}
Because $z_j=-e_j$, the learning rule
\eqref{eq:conical-lms-rule}, before projection onto non-negative
weights, changes the synapse from candidate input $i$ to output $j$ in
proportion to
\begin{equation}
	x_i e_j.
\end{equation}
The expected weight change is therefore governed by the correlation
between the candidate and the current residual. A candidate correlated
with an unexplained part of the target has a systematic effect on the
weight. As that part is learned, its correlation with the residual
approaches the NNLS optimality condition. A candidate satisfying
\begin{equation}
	\mathbb{E}\{x_i e_j\}
	=
	0
\end{equation}
has no systematic update. With a diminishing learning rate its
influence can vanish asymptotically; with a constant learning rate,
small steady-state fluctuations remain, as discussed in
\sectref{sect:nnls-interpretation}.

We now lift this mechanism to sparse population messages. Let
$\myvec{a}_k$ be the sparse excitatory code for ray $k$, expressed in
the native coordinates of the excitatory population, and let
$\myvec{g}_k$ be the corresponding sparse ray expressed in the
coordinates of the inhibitory population. An idealized training
presentation has the form
\begin{equation}
	\myvec{x}^{(t)}
	=
	\alpha_t\myvec{a}_k+\myvec{d}^{(t)},
	\qquad
	\myvec{y}^{(t)}
	=
	\alpha_t\myvec{g}_k,
	\qquad
	\alpha_t\geq0,
	\label{eq:sparse-ray-training-pair}
\end{equation}
where $\myvec{d}^{(t)}$ contains candidate activity that is
uncorrelated with the relevant residual. Changes in $\alpha_t$ move
the paired messages along their respective rays without changing the
represented correspondence. Repeated presentations make the
population learn
\begin{equation}
	W^T\myvec{a}_k
	\approx
	\myvec{g}_k.
	\label{eq:learned-ray-correspondence}
\end{equation}

The two populations initially use different coordinate systems. At a
coarser level, their aligned messages may be related by
\begin{equation}
	\myvec{x}
	=
	A\myvec{y}.
	\label{eq:population-coordinate-map}
\end{equation}
For the paired ray codes, this idealized relation gives
$\myvec{a}_k=A\myvec{g}_k$.
For signed unconstrained least squares under the statistical
assumptions used in the subspace calculation above, every minimizing
weight matrix has the same action as $A^+$ on inputs of the form
$\myvec{x}=A\myvec{y}$. It therefore reconstructs
\begin{equation}
	W^T\myvec{x}
	\approx
	A^+A\myvec{y},
\end{equation}
the orthogonal projection onto the part of the inhibitory message
space recoverable through $A$.
In the conic case, the paired ray
relation in \eqref{eq:learned-ray-correspondence} gives the
corresponding non-negative alignment on the represented sparse
message family. For compactness, the notation $p\llcorner q$ and
$p\neg q$ leaves this learned coordinate alignment implicit. Thus, in
expressions such as $P_q\myvec{u}$ below, $q$ is understood to have
been expressed in the inhibitory population's coordinates.

Training presentations for different rays may be interleaved. If the
sets
\begin{equation}
	\operatorname{supp}(\myvec{a}_k)
	\times
	\operatorname{supp}(\myvec{g}_k)
\end{equation}
are disjoint for different $k$, their updates affect disjoint synaptic
blocks and the learning processes do not interfere. Exact
disjointness is not required: weak overlap or compatible constraints
on shared weights permit approximate simultaneous learning. Sparse
encoding makes such limited interference plausible but does not, by
itself, guarantee it.

Let $I\subseteq\{1,\ldots,K\}$ be the index set of ray
correspondences that are active in a given message, where $K$ is the
number of learned ray correspondences. After several ray
correspondences have been learned, linearity gives
\begin{equation}
	W^T
	\left(
	\sum_{k\in I}\alpha_k\myvec{a}_k
	\right)
	\approx
	\sum_{k\in I}\alpha_k\myvec{g}_k,
	\qquad
	\alpha_k\geq0,
	\label{eq:learned-ray-mixture}
\end{equation}
provided that the same learned blocks remain active and their shared
weights impose compatible mappings. We assume that the rays indexed
by $I$ generate a face
\begin{equation}
	F_I
	=
	\Conic
	\left\{
	\myvec{g}_k
	\,\middle|\,
	k\in I
	\right\}
\end{equation}
of the aligned cone $q$.
Sparse activity means that
\begin{equation}
	|I|
	\leq
	s
	\ll
	K,
	\qquad
	d_I
	\triangleq
	\dim\operatorname{span}(F_I)
	\leq
	|I|
	\ll
	\dim H,
	\label{eq:low-dimensional-active-face}
\end{equation}
where $s$ is the assumed sparsity level, that is, an upper bound on
the number of rays active at any one time. Thus, the selected local
mapping normally acts on a low-dimensional face even though the
ambient population space is high-dimensional.

The learning construction relies on the following assumptions:
\begin{enumerate}
	\item
	Messages use a sparse ray representation. Only a small set $I$ of
	rays is active at one time, and these rays generate a face $F_I$
	with low-dimensional span.
	
	\item
	Each relevant ray is presented sufficiently often as a consistent
	pair of excitatory and inhibitory messages. The amplitudes and
	temporal variations provide sufficient excitation for learning the
	mapping on the represented message family.
	
	\item
	Candidates belonging to a target ray are correlated with the
	unexplained part of that target, whereas unrelated candidates have
	zero or sufficiently small correlation with the residual. Their
	weight updates consequently have zero or small long-term mean.
	
	\item
	The sparse input and output supports identify the active ray set
	and hence select the appropriate low-dimensional face.
	
	\item
	Ray-specific synaptic blocks are disjoint, weakly overlapping, or
	impose mutually compatible requirements on their shared weights.
	This allows their learning processes to be interleaved with limited
	interference.
	
	\item
	On the represented message family, a target message associated
	with the active face can be decomposed as
	\begin{equation}
		\myvec{u}
		=
		\myvec{f}_I+\myvec{r},
		\qquad
		\myvec{f}_I
		=
		\sum_{k\in I}\alpha_k\myvec{g}_k,
		\qquad
		\alpha_k\geq0,
		\label{eq:matched-residual-decomposition}
	\end{equation}
	where
	\begin{equation}
		\myvec{r}
		\in
		\neg q,
		\qquad
		\left\langle
		\myvec{f}_I,\myvec{r}
		\right\rangle
		=
		0.
		\label{eq:projection-compatible-residual}
	\end{equation}
	Thus, the unmatched residual is non-positively correlated with
	every ray in $q$ and is orthogonal to the reconstructed component.
\end{enumerate}

Together with $\myvec{f}_I\in q$, equations
\eqref{eq:matched-residual-decomposition} and
\eqref{eq:projection-compatible-residual} are precisely the conditions
of Moreau decomposition. They therefore imply
\begin{equation}
	\myvec{f}_I
	=
	P_q\myvec{u},
	\qquad
	\myvec{r}
	=
	\myvec{u}-P_q\myvec{u}.
	\label{eq:learned-moreau-decomposition}
\end{equation}
Consequently, if the learned ray correspondences reconstruct
$\myvec{f}_I$, their residual is precisely the conic rejection on this
idealized sparse message family. Approximate correlation,
non-orthogonality, synaptic overlap, finite learning and departures
from the population-level compatibility condition instead produce an
approximation, which we quantify next.

\subsection{Formal population-level realization}
\label{sect:formal-population-realization}
The neuron-local NNLS result does not by itself establish a
population-level cone operation. Multiplication by a fixed matrix
remains linear, whereas projection onto a general convex cone is
non-linear. The extension to a population-level operation relies on
the local structure of projection onto a finitely generated cone,
activity-dependent selection of the effective population map, and the
sparse learning assumptions stated above.

Assume that the aligned population cone $q$ is finitely generated.
Its linear span is finite-dimensional, and $q$ is polyhedral when
regarded as a cone within this span. Moreover,
\begin{equation}
	P_q\myvec{u}
	=
	P_q
	\left(
	P_{\operatorname{span}(q)}\myvec{u}
	\right),
\end{equation}
because the component of $\myvec{u}$ orthogonal to
$\operatorname{span}(q)$ does not affect its nearest point in $q$.
Projection onto $q$ is continuous. Because $q$ is a cone, the
projection is also positively homogeneous:
\begin{equation}
	P_q\left(\alpha\myvec{u}\right)
	=
	\alpha P_q\myvec{u},
	\qquad
	\alpha\geq0.
\end{equation}
Thus, multiplying a message by a non-negative factor multiplies its
projection by the same factor. Because $q$ is polyhedral within its
finite-dimensional span, projection onto $q$ is also piecewise linear
on $H$. To describe its linear
pieces, let $F$ be a face of $q$ and define its projection region by
\begin{equation}
	\mathcal{R}_F
	\triangleq
	\left\{
	\myvec{u}\in H
	\,\middle|\,
	P_q\myvec{u}
	\in
	\operatorname{relint}(F)
	\right\}.
\end{equation}
Here, $\operatorname{relint}(F)$ denotes the {\em relative interior}
of $F$. A face may be lower-dimensional than the ambient space $H$
and may therefore have empty ordinary interior in $H$. Its relative
interior is instead computed within its own linear span. More
precisely,
\begin{equation}
	\operatorname{relint}(F)
	=
	\left\{
	\myvec{x}\in F
	\,\middle|\,
	\begin{array}{l}
		\text{there exists $\varepsilon>0$ such that }
		\myvec{x}+\myvec{z}\in F\\
		\text{for every $\myvec{z}\in\operatorname{span}(F)$ with
			$\lVert\myvec{z}\rVert<\varepsilon$}
	\end{array}
	\right\}.
\end{equation}
Thus, directions outside $\operatorname{span}(F)$ are disregarded
when deciding whether a point is interior to $F$. For example, the relative interior of a ray consists of its non-zero
points, while the relative interior of a two-dimensional wedge consists
of the points strictly between its boundary rays.

Throughout the $\mathcal{R}_F$ region,
\begin{equation}
	P_q\myvec{u}
	=
	P_{\operatorname{span}(F)}\myvec{u}.
	\label{eq:local-conic-projector}
\end{equation}
Consequently, conic projection reduces locally to orthogonal
projection onto the span of the active face.

Let $G_F$ be the matrix whose columns are the frame vectors that
generate the face $F$. The orthogonal projector onto its linear span is
\begin{equation}
	P_{\operatorname{span}(F)}
	=
	G_F G_F^{+}.
\end{equation}
When $\{\myvec{0}\}$ is a face of $q$, its corresponding projector is
the zero operator. Different faces therefore give
different linear pieces of the conic projector.

Activity sparsity can contribute to the neuronal selection of these
pieces. Each message has only a small active support, while neuronal
thresholding or rectification suppresses weak or inactive output
components. Here, an {\em active support} means the set of coordinates retained by
the activation dynamics, rather than merely the mathematical support of
an already thresholded vector. The active input support and the set of output neurons
driven above threshold thereby select an input-dependent submatrix of
$W$, and hence one of the local mappings.

Let $S(\myvec{u})$ denote the active support of the excitatory message
$A\myvec{u}$, and let $R(\myvec{u})$ denote the active support of the
reconstructed inhibitory message. Let $D_{S(\myvec{u})}$ and
$D_{R(\myvec{u})}$ be the corresponding diagonal support masks. The
effective end-to-end population map can then be written
\begin{equation}
	T(\myvec{u})
	=
	D_{R(\myvec{u})}
	W^T
	D_{S(\myvec{u})}
	A\myvec{u}.
	\label{eq:sparse-effective-population-map}
\end{equation}
Within a region in which the two supports remain fixed, $T$ is the
linear map determined by the selected submatrix of $W^T$. As the
message moves between regions, its active supports can change, thereby
selecting different local linear maps. The complete population map is
therefore piecewise linear.

Under the ideal sparse ray-learning assumptions, a message
$\myvec{u}\in\mathcal{R}_F$ activates the rays generating $F$. Its
selected excitatory message then satisfies
\begin{equation}
	D_{S(\myvec{u})}A\myvec{u}
	=
	\sum_{k\in I}\alpha_k\myvec{a}_k.
\end{equation}
Equation \eqref{eq:learned-ray-mixture} consequently reconstructs the
component
$\myvec{f}_I=P_{\operatorname{span}(F)}\myvec{u}$, which is retained
by the output support. In that ideal case,
\begin{equation}
	T(\myvec{u})
	=
	P_{\operatorname{span}(F)}\myvec{u}
	=
	P_q\myvec{u}
\end{equation}
on the represented message family. Thus, the activity-defined regions
and face-defined projection regions agree there by construction.

Let $\mathcal{M}$ denote the sparse message model represented by the
population. Departures from the ideal assumptions are described by
requiring that, for every relevant face $F$,
\begin{equation}
	\left\|
	T(\myvec{u})
	-
	P_{\operatorname{span}(F)}\myvec{u}
	\right\|
	\leq
	\epsilon\|\myvec{u}\|,
	\qquad
	\myvec{u}
	\in
	p\cap\mathcal{M}\cap\mathcal{R}_F,
	\label{eq:local-conic-projection-error}
\end{equation}
where $\epsilon\geq0$ is the relative approximation error. Sources of
this error include incomplete adaptation, finite training data,
non-zero correlations between nominally unrelated candidates,
overlap between ray-specific synaptic blocks, imperfect support
selection and deviations from the projection-compatible residual
condition.

Equation \eqref{eq:local-conic-projection-error} is deliberately
restricted to $\mathcal{M}$. It does not assert that the selected
end-to-end operator
\begin{equation}
	D_{R(\myvec{u})}
	W^T
	D_{S(\myvec{u})}
	A
\end{equation}
equals the full matrix of
$P_{\operatorname{span}(F)}$ on the entire ambient space. The
synaptic factor
$D_{R(\myvec{u})}W^TD_{S(\myvec{u})}$ is entrywise non-negative, but
no entrywise sign condition has been imposed on the coordinate map
$A$. For coordinate-ray codes, in which the generators are
proportional to distinct coordinate vectors, the relevant projector
reduces to a non-negative coordinate mask and exact realization is
possible. For more general codes, the claim concerns agreement on the
represented sparse messages, with $\epsilon$ quantifying the remaining
discrepancy.

Since \eqref{eq:local-conic-projector} gives
$P_q\myvec{u}=P_{\operatorname{span}(F)}\myvec{u}$ throughout
$\mathcal{R}_F$, equation
\eqref{eq:local-conic-projection-error} implies
\begin{equation}
	\left\|
	T(\myvec{u})
	-
	P_q\myvec{u}
	\right\|
	\leq
	\epsilon\|\myvec{u}\|
\end{equation}
on the represented message model. The prediction residual consequently
satisfies
\begin{equation}
	\left\|
	\left(
	\myvec{u}-T(\myvec{u})
	\right)
	-
	\left(
	\myvec{u}-P_q\myvec{u}
	\right)
	\right\|
	=
	\left\|
	T(\myvec{u})-P_q\myvec{u}
	\right\|
	\leq
	\epsilon\|\myvec{u}\|.
\end{equation}
Thus, the reconstruction approximates the pointwise conic projection,
and the population residual approximates the corresponding pointwise
conic rejection, on the sparse messages represented by the
population.

In the ideal case in which $\epsilon=0$ and the condition holds
throughout $p$, the closed conic hulls of the reconstructed and
residual messages satisfy
\begin{equation}
	\Conic
	\left\{
	T(\myvec{u})
	\,\middle|\,
	\myvec{u}\in p
	\right\}
	=
	p\llcorner q,
	\qquad
	\Conic
	\left\{
	\myvec{u}-T(\myvec{u})
	\,\middle|\,
	\myvec{u}\in p
	\right\}
	=
	p\neg q.
\end{equation}
For $\epsilon>0$, the preceding inequalities instead quantify the
pointwise accuracy of the realization on the represented messages.

Across the population, $W^T\myvec{x}$ is the raw linear
reconstruction of the inhibitory message, whereas sparse neuronal
activity and activation-dependent selection produce the effective map
$T$. When the learned correspondence generalizes over the represented
message model, $T(\myvec{u})$ realizes or approximates the conic
projection, and $\myvec{u}-T(\myvec{u})$ realizes or approximates the
conic rejection. Because the target input is inhibitory, the
population outputs the reflection of the residual cone. The resulting
ideal algebraic operation is therefore the primitive already stated
in \eqref{eq:primitive-population-operation}.

The neuron population serves as a {\em primitive operator} because the
basic operations defined in
\sectref{sect:definitions-of-basic-operations-on-cones} can be
constructed from this operation, as detailed below. The polar is not
one of those basic operations and is not directly implementable by
populations because it would violate the sparsity constraint
(\sectref{sect:sparsity}). Its intended effect can usually be achieved
instead by {\em conic rejection}
(\sectref{sect:intersection-by-rejection-and-span}).

\subsection{Basic operations in terms of primitive operators}
\begin{figure}[!hbt] 
	\includegraphics[width=\textwidth]{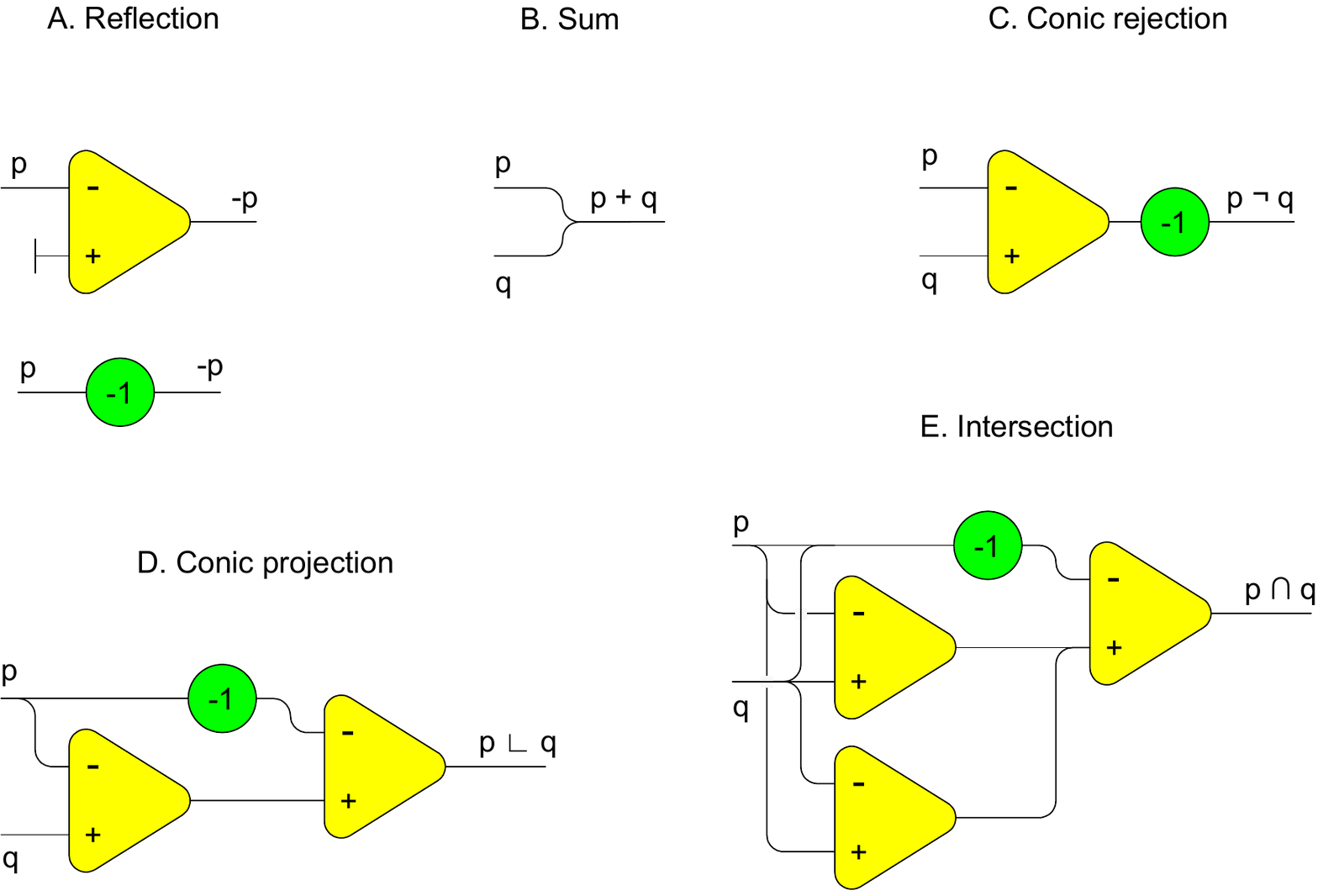}
	\caption{{\bf Basic operations as implemented by neuron populations.}
		Populations can implement all basic operations.
		For convenience, we introduce a shorthand symbol for {\em reflection} (A, bottom).
		\label{fig:cone-operations}}
\end{figure}

Each basic operation requires at most four neuron populations; sum requires none. The basis for the circuits is the formula collection provided in \sectref{sect:definitions-of-basic-operations-on-cones}, its depiction in \figref{fig:cone-operations}, and elaboration below. The figure presents generic populations as rounded triangles, distinguishing inhibitory and excitatory inputs with '-' and '+' symbols, respectively. The flow of signals progresses from left to right.

{\bf Reflection:}
A single generic neuron population can implement the {\em reflection} operation without excitatory inputs (\figref{fig:cone-operations} A, top), but for brevity, we dedicate a symbol to this function: the encircled `-1' (\figref{fig:cone-operations} A, bottom). It has a biological counterpart in local inhibitory neurons ubiquitous in the CNS. An obvious but helpful rule for minimizing the number of reflections in a circuit is
\begin{equation}
	\label{eq:mirror-propagation}
	-(y \, \neg \, x) = (-y) \, \neg \, (-x) .
\end{equation}

{\bf Sum:}
{\em Sum} does not require any populations but is achieved by simply bundling together the axons
(\figref{fig:cone-operations} B). 

{\bf Conic rejection:}
{\em Conic rejection} is the ``native'' operation of the neuron population, but needs an
appended {\em reflection} to specify the sign change of the inhibitory input
(\figref{fig:cone-operations} C).

{\bf Conic projection:}
A combination of three populations, one of which is a {\em reflection},
can implement {\em conic projection} using relation \eqref{eq:projection-by-rejection} and \eqref{eq:mirror-propagation} (\figref{fig:cone-operations}~D).

{\bf Intersection:}
{\em Intersection} has the most complex implementation. It can be realized by three {\em conic rejections} and one {\em reflection}, using relations \eqref{eq:intersection-by-rejection} and \eqref{eq:mirror-propagation} (\figref{fig:cone-operations} E).

\subsection{Conditionals}
\begin{figure}[!hbt] 
	\centering
	\includegraphics[width=0.5 \textwidth]{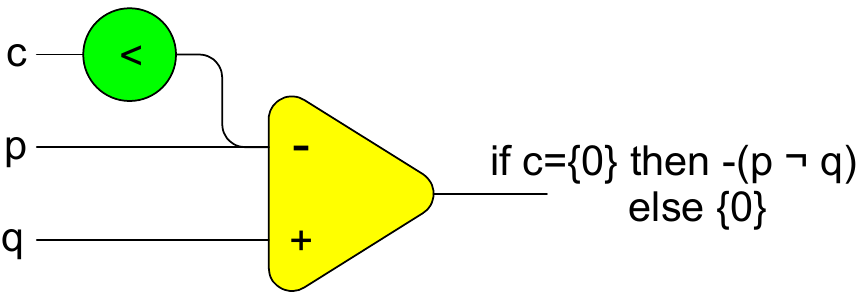}
	\caption{{\bf Conditionals.}
		By an additional, strong inhibitory input, populations can implement conditionals. 	
		\label{fig:conditionals-and-memory}}
\end{figure}
Neuron populations can execute conditionals through strong inhibitory input
(\figref{fig:conditionals-and-memory}), forcing the postsynaptic neurons into
the activation function's {\em cutoff} range. In this illustration, the circle
labeled with a “$<$” symbol indicates a control population capable of producing
the blocking signal. Basket cells \citep{McBain.2012cin} provide strong perisomatic
GABAergic inhibition and are plausible biological analogs of neurons
producing this control signal. Their action can include {\em shunting 
inhibition}: activation of postsynaptic $\mathrm{GABA_A}$ receptor
channels increases membrane conductance and thereby attenuates
concurrent excitatory input \citep{Mitchell.Silver.2003sim}. The
present population-level abstraction does not distinguish between
the shunting and hyperpolarizing components of inhibition, but uses
only their ability to suppress postsynaptic output.

The inhibitory gate acts directly on the current control message; it does not
by itself determine whether an abstract cone is empty. To implement a
cone-level emptiness test, we therefore assume that a control population
persistently represents the result of the test, or that the messages defining
the condition cone are presented in a synchronized manner. The control
population remains inactive exactly when $c=\{\myvec{0}\}$ and otherwise
produces sufficiently strong inhibition to block the output. Under this
additional representational assumption, the implemented conditional is
\begin{equation}
	\mathtt{if~}c\,=\,\{\myvec{0}\}
	\mathtt{~then~}[-(p\,\neg\,q)]
	\mathtt{~else~}\{\myvec{0}\}.
\end{equation}

\subsection{Memory operations}
\label{sect:memory-operation}
We have already described the primary {\em memory write} operation as synaptic weight updates of the neuron population in \sectref{sect:population-as-operator}.

Any feed-forward action can be treated as a {\em memory read} because any excitatory input reads the weight matrix and multiplies it with the input message. A more distinct memory read is achieved by momentarily keeping the inhibitory input $\myvec{y}$ inactive while sending excitatory messages $\myvec{x}$ to the population. This produces the outputs $W^T \myvec{x} - \myvec{y} = W^T \myvec{x}$, which equates to $p \, \llcorner \, q$.

When writing memory, the inhibitory input $p$ serves as the contents 
and the excitatory input $q$ serves as the address or lookup key. Upon readout, an exact match with $q$ is not required. The combination of sparsity and a soft-threshold activation function allows associative lookup
\citep{Kohonen.et.al.1981sap,Palm.2013nam}.

\subsection{Adaptive filter applications}
\begin{figure}[!hbt] 
	\includegraphics[width=\textwidth]{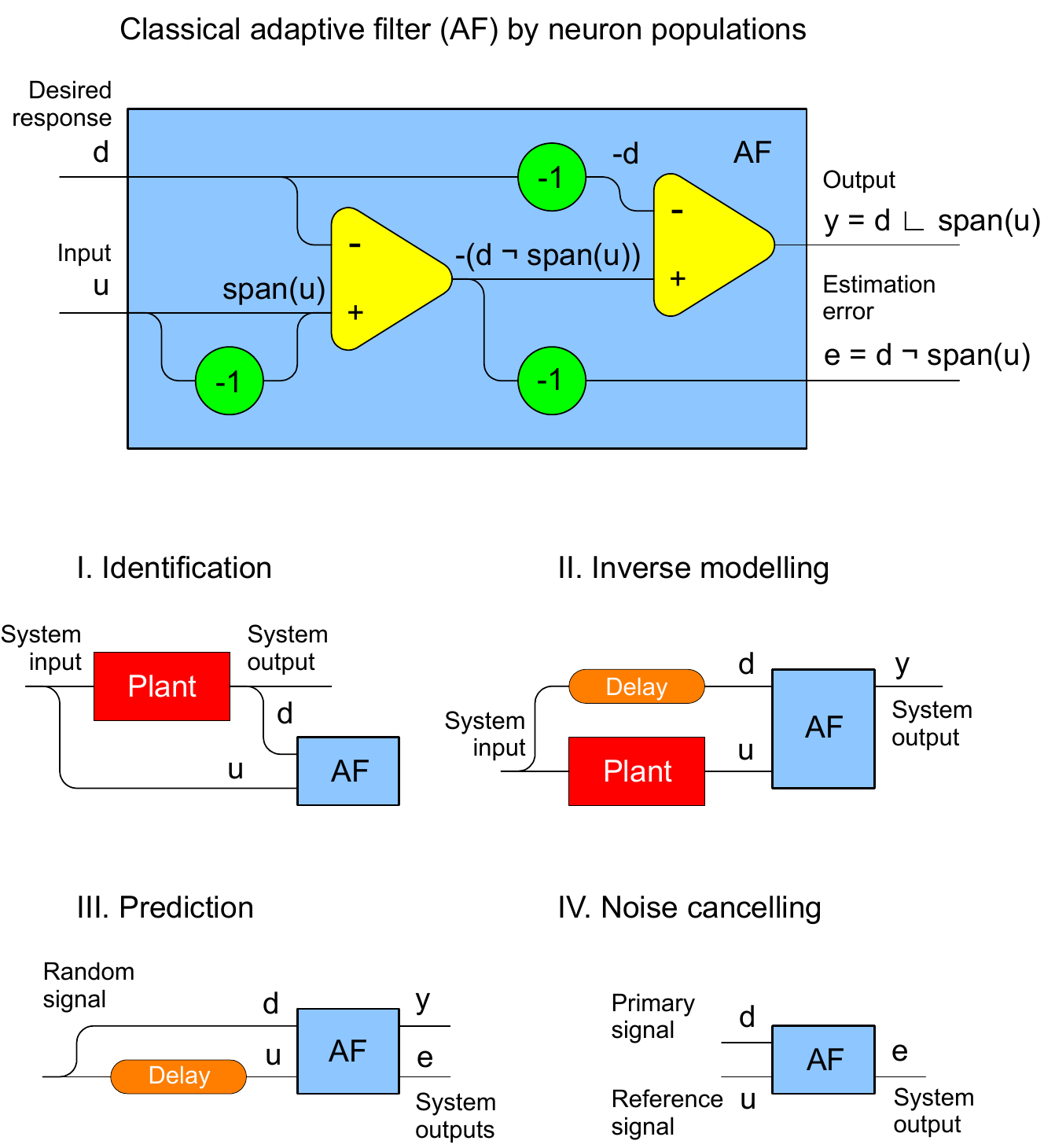}
	\caption{{\bf Classical adaptive filter applications by neuron populations.}
		Neuron populations can build a classical adaptive filter with projection 
		output $y$ and error output~$e$.
		In this configuration, neuron populations can directly realize
		the adaptive filter application classes I-IV \citep{Haykin.2002aft}. 
		``Plant'' refers to an unknown transfer function.
		\label{fig:adaptive-filter-applications}}
\end{figure}
The classical adaptive filter is a workhorse of statistical signal processing \citep{Haykin.2002aft,Sayed.2003foa,Widrow.Stearns.1985asp}. It has a main input $u$, a desired response input $d$, a main output $y$, and an estimation error output $e$. In our notation, the main output is the projection $d \, \llcorner \, \operatorname{span}(u)$, and the error output is the rejection $d \, \neg \, \operatorname{span}(u)$. An adaptive filter comprises two generic neuron populations and three local inhibitory populations (\figref{fig:adaptive-filter-applications}, top).

The versatility of the adaptive filter originates in its capability to perform many different functions depending on its connections. Applications can be divided into four classes (\figref{fig:adaptive-filter-applications}):

{\bf I. Identification:} In this class, the adaptive filter establishes a linear model that aligns best with an unfamiliar system, termed the ``plant''. It aims to identify and mirror the behavior of this unknown system.

{\bf II. Inverse Modeling:} Here, the adaptive filter's role is essentially flipped. Instead of modeling the plant directly, the filter offers an inverse representation of the system. It aims to reverse the behavior of the plant, providing a model that can counteract or undo the effects of the original system.

{\bf III. Prediction:} Within this class, the adaptive filter's primary objective is to forecast a random signal or even anticipate the prediction error itself. Analyzing past data can project future values or the discrepancy in such predictions.

{\bf IV. Noise Canceling:} For this class, the adaptive filter processes a primary signal contaminated with an interference signal. Alongside, it receives a reference signal which carries data about the noise disturbances. The adaptive filter's output is the cleaned primary signal where the interference has been nullified or substantially reduced.

\subsection{Recurrent networks}
\label{sect:recurrent-networks}
\begin{figure}[!hbt] 
	\includegraphics[width=\textwidth]{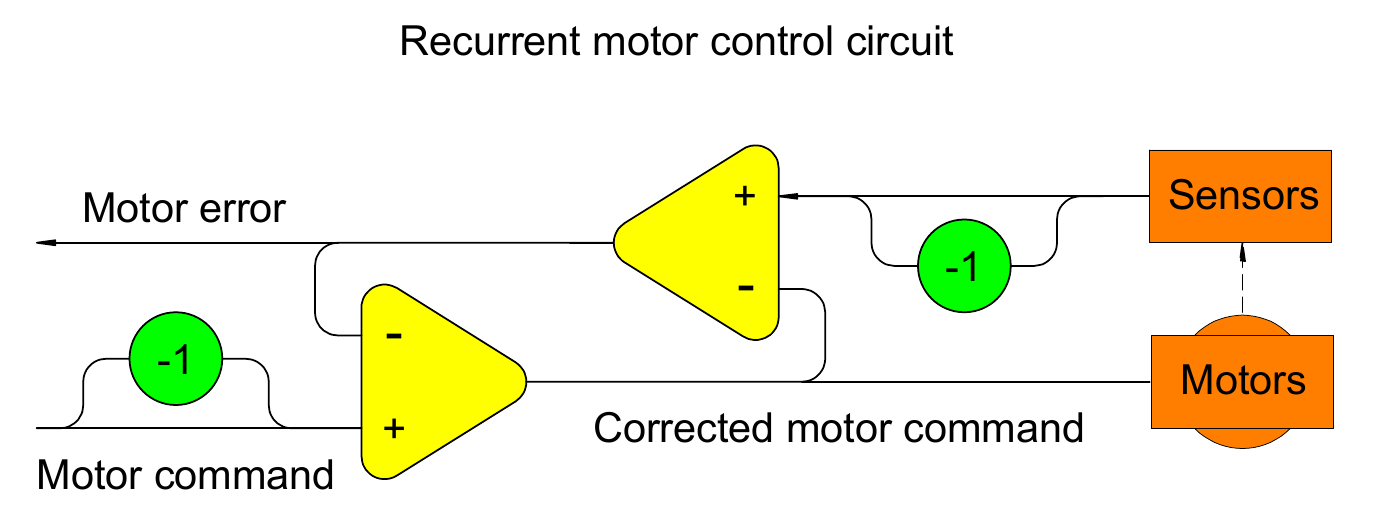}
	\caption{{\bf Recurrent sensorimotor network.}
		Two population operators form a central loop. A second
		feedback path passes through a muscle group (``Motors'') and proprioceptive sensors (``Sensors'').
		Through adaptation, the network learns the relation between motor
		commands and their sensory consequences.
		The parallel direct and reflection pathways replace each cone $a$ by
		its linear hull $a+(-a)=\operatorname{span}(a)$, thereby recovering
		the subspace representation used in the earlier control model
		\citep{Nilsson.2016noa}.
		\label{fig:recurrent-network}}
\end{figure}

Population operators need not be restricted to feed-forward networks.
Their outputs can be returned to the inputs of earlier populations,
thereby forming recurrent networks.
Figure~\ref{fig:recurrent-network} shows two population operators
connected recurrently.
This circuit can address the \emph{sensorimotor
association problem}: learning which motor commands control which
muscles and which proprioceptive signals report the resulting movements.
The Supplementary Information presents a minimal numerical realization
of the circuit in \figref{fig:recurrent-network}. The simulation uses
the incremental non-negative learning rule of
\figref{fig:adaptive-filter} during continuous recurrent operation and
examines recovery from a temporary sensor disturbance.

An earlier version of the present population operator was used to
analyse this circuit in detail \citep{Nilsson.2016noa}. The stability
and dynamics of such a circuit are properties of the complete network,
rather than of its individual population operators. An adequate
analysis requires a substantial treatment for which there is no space
in the present paper and therefore lies outside its scope. We refer the
reader to the earlier work for this analysis.

Experimental evidence indicates pathways in both directions between the red
nucleus and spinal circuitry in cats. \citet{Rathelot.Padel.1997asi} demonstrated
an ascending pathway from spinal circuitry to identified rubrospinal
neurons, whereas \citet{Hongo.et.al.1972rte} demonstrated monosynaptic
excitation of spinal interneurons by the rubrospinal tract. These
interneurons also received peripheral afferent input.
\citet{Vinay.Padel.1990sto} accordingly described this organization as a
\emph{spino-rubro-spinal loop}. These findings do not show that the same
spinal neurons participate in both directions, but this is not required
by the present model. It describes recurrence at the population level:
one projection may carry signals from a population in the red nucleus to
a spinal population, while another projection returns signals through
different neurons. Diffuse and partly random connectivity makes such an
organization natural. Thus, the absence of direct neuronal loops does not
exclude a functional recurrent loop between the populations.

\section{Results}
\label{sect:results}
Starting from a non-speculative, strictly mechanistic model of a
neuron with plasticity, we developed a mathematical characterization
of information processing by neuron populations in the CNS that is
expressible as an algebra of convex cones in a real Hilbert space.

The invariant is not carried by any single vector of instantaneous
spike rates, but by a sequence of such vectors whose conic hull forms a
convex cone. Under the biologically motivated conditions considered
here---fixed non-negative mappings, together with the stated sparsity
and geometry-preservation conditions---this cone representation is
retained as messages pass through synapses and neuron populations.
The representation is likewise retained under the corresponding fixed
linear filtering of time-variable messages. Thus, although the
individual messages and their coordinates may change, the information
they express continues to be represented by a convex-cone invariant.

Neuron populations can be conceived as operators in the algebra, taking cones as operands and delivering new cones as results. The basic operations on the cones are sum, intersection, reflection, projection, and rejection. Networks of neuron populations can form expressions implementing algebra operations that retain the sparsity of representations. They can also implement conditionals as well as memory read and write operations.

The fact that five neuron populations suffice to fully implement the classical applications of multi-dimensional adaptive filters used in statistical signal processing demonstrates the computational capacity of neuron populations. 

\section{Discussion}
\label{sect:discussion}

\subsection{Interpretation of the invariant and the operations}
\label{sect:discussion-interpretations}
Our mathematical framework for neuron information processing has been
firmly rooted in a mechanistic model, sidestepping speculation. Nevertheless, to shed light on the model's depth, one might liken a message to a ``thought'' and the cone invariant to a ``concept'' for a more intuitive grasp.

In abstract terms, we can interpret 
{\em sum} as generalization or abstraction;
{\em complement} and {\em \polar} as logical or Boolean negation, or independence;
{\em projection} as specialization or extraction of some relevant property;
and {\em intersection} as logical conjunction. 
{\em Rejection} can be interpreted as asymmetric difference.
Moreover, it has an impressive history as a statistical concept tracing back to \citet{Wold.1938asi}, who proved that every covariance-stationary time series can be written as the sum of two time series, one deterministic and associated with {\em projection}, and one stochastic and associated with {\em rejection}.
The idea inspired both Wiener and Kolmogorov \citep{Haykin.2002aft},
and \citet{Kailath.1968aia} introduced the name {\em innovation} for
the stochastic white-noise process involved. Today, it is foundational
in signal processing and control.
In neuroscience, the idea of innovation has appeared under many names, including
{\em novelty} \citep{Kohonen.1977am}, 
{\em unexpectedness} \citep{Barlow.1991vty},
{\em prediction error} \citep{Schultz.Dickinson.2000nco},
{\em decorrelation} \citep{Dean.et.al.2002dcb},
{\em surprise} \citep{Friston.et.al.2006afe},
and {\em saliency} \citep{vanPolanen.2014fih}.

To illustrate the basic operations concretely, let us consider
observations of numerous red objects, creating a cone invariant that
encapsulates the concept of {\em red}. Perceiving a spectrum of colours
similarly gives rise to the broader concept of {\em colour}.
Intuitively, {\em projection} can distil a particular attribute from a
concept: projecting {\em red umbrella} onto {\em colour} extracts
{\em red}. {\em Sum} acts as generalization: summing {\em umbrella},
{\em raincoat}, and {\em rubber boots} may produce the broader concept
of {\em rain gear}. Meanwhile, {\em rejection} of {\em rain gear} from
{\em red umbrella} leaves {\em red}, illustrating a form of logical
difference.

More generally, a semantic network represented by RDF triples of the
form ({\em subject}, {\em predicate}, {\em object}) can be embedded in
a cone representation constructed so that the predicate acts as an
attribute. Projecting the representation of an entity onto the cone
associated with that attribute then extracts the representation of the
corresponding value, or a conic combination of values when several
apply. In the example above, this corresponds to projecting the entity
{\em red umbrella} onto the attribute {\em colour} and obtaining the
value {\em red}. We are currently investigating methods for systematically
constructing such cone representations from RDF networks.

\subsection{Subspaces and algebras of concepts}
Representation of knowledge by subspaces has a long history, one of the pioneers being \citet{Watanabe.1969kag}. Substantial contributors to subspace methods are \citet{Kohonen.1977am} and  \citet{Oja.1983smo}. \citet{Tsuda.2001tsm} considered subspaces of Hilbert space. Early to propose neurons as compact operators in Hilbert space was \citet{MacLennan.1993fci}, who also realized the significance of overcomplete wavelet frames.

Generally speaking, published works on concept representations are typically not based on mechanistic neuron models but propose high-level theoretical models as hypotheses matching observations in cognitive science.

Algebras of concepts seem to have been considered mostly in the
fields of information retrieval and psychology, but generally only as 
algebras of subspaces \citep{VanRijsbergen.2004tgo,Aerts.Gabora.2005ato1,Aerts.Gabora.2005ato2}.

Hyperdimensional computing or vector symbolic architectures reviewed in \citep{Kleyko.et.al.2022aso,Kleyko.et.al.2023aso}, leverages high-dimensional representations for knowledge processing. Designed for computational efficiency, most draw top-down inspirations from cognitive science. An outlier is \citet{Stewart.et.al.2010sri}, integrating high-dimensional vectors with an empirical spiking neuron model. Our algebra of convex cones is a (matrix) symbolic architecture, but in contrast, is a strictly mechanistic model built bottom-up from the neurobiological properties of neurons.

The finitely generated convex cones considered here form a lattice under set
inclusion. Their meet is $a\wedge b=a\cap b$, and their join is the smallest
convex cone containing both, $a\vee b=\Conic(a\cup b)=a+b$. Both remain finitely
generated within the finite-dimensional span of the two frames. These operations
define hierarchies of progressively more specific or more general invariants.
Sparsity therefore does not destroy the lattice structure. It limits how large a
join a fixed neuron population can represent without thresholding, dimensionality
reduction, or recruitment of additional neurons.

Donoho (1995) showed that soft thresholding of wavelet coefficients
provides an effective method for removing noise while retaining the
significant components of a signal. Soft thresholding shrinks all
coefficients towards zero and sets those below a specified magnitude
to zero, thereby producing a sparse approximation. This principle is
particularly relevant here because messages can be regarded as
vector-valued wavelets, while the rectifying range of the neuronal
activation function provides a biologically plausible form of soft
thresholding, as discussed in \sectref{sect:activation-function}.
Small components arising from noise or from a sum operation can
therefore be suppressed, limiting the effective dimensionality of the
represented cone and helping to preserve sparse population messages.

\subsection{Adaptive filters in neuroscience}
Adaptive filters, initially proposed by \citet{Widrow.Hoff.1960asc}, have become a recurrent theme in neuroscience. Early applications by \citet{Kohonen.1977am} targeted associative memory and related functions \citep{Kohonen.et.al.1981sap}. These filters are often employed to model parts of the cerebellum
\citep{Fujita.1982afm,
	Wolpert.et.al.1998imi,
	Kawato.1999imf,
		Dean.et.al.2002dcb}.

To our knowledge, this is the first derivation in which the receptor and
ion-channel mechanisms of a biological neuron yield a projected stochastic-gradient
algorithm for NNLS \citep{Nilsson.2026meo,Lawson.Hanson.1995sls}. The claim is
narrower than saying that non-negative optimization has not appeared in
neuroscience. The novelty is that the NNLS update follows mechanistically from
this neuron model rather than being imposed on an abstract learning rule.

Notably, some studies have experimentally identified neurons exhibiting behavior akin to Fourier analysers
\citep{Pollen.et.al.1971hdt,DeValois.DeValois.1991sv}.

\subsection{Matrix embeddings}
Foundation Models or Large Language Models \citep{zhao.et.al.2023aso} of current AI-systems use high-dimensional vector embeddings for concept encoding. However, the rudimentary structure of vector spaces limits their representation capacity, for example, when trying to generalize a set of vectors.

A better alternative is a subspace representation, which offers a {\em sum} operation. A {\em matrix embedding} can accomplish this by using a positive semi-definite matrix $AA^T$ derived from the covariance matrix of observations.

We believe an even better alternative is achieved through a finite convex-cone
representation, employing the full generator matrix $A$ whose columns
are frame vectors. The conical coefficients are constrained to be
non-negative, but the entries of the frame vectors themselves need not
be non-negative. This method enables a sophisticated partial ordering of concepts, surpassing what is feasible with subspaces, as detailed in \sectref{sect:partial-order-of-subspaces} and \sectref{sect:cone-relations}.

Recent interpretability work at Anthropic \citep{Gurnee.et.al.2026vfr}
provides an independent computational
parallel. The authors define J-space using an overcomplete frame of J-lens
vectors. For a given sparsity bound $k$, J-space consists of all non-negative
combinations of at most $k$ such vectors. Geometrically, it is a generally
non-convex union of polyhedral cones, each generated by a selection of at most
$k$ J-lens vectors. The dimension of each cone is at most $k$, with equality
when its selected generators are linearly independent. For a given activation,
a J-space component may be defined as any point in this union that minimizes
the distance to the activation. Such a point exists, since the union is finite
and closed, but it need not be unique. In practice, the authors use sparse
decomposition to approximate one such component.

This construction closely parallels our use of sparse frames and non-negative
combinations. The comparison is geometric rather than biological: J-space is a
structure in transformer activation space, not evidence for the neuronal
mechanism proposed here. Its emergence nevertheless suggests that cone-based
matrix embeddings may be useful in very different information-processing
systems.

\subsection{Cognitive science aspects}
\citet{Gardenfors.2000cst} proposed {\em conceptual spaces}, a knowledge representation layer bridging the gap between neurobiology and cognition. Despite resting on a neurobiological basis, the algebra of convex cones fits well into this framework. Emphasizing this connection, a subsequent paper on conceptual spaces by \citet{Balkenius.Gardenfors.2016sit} posited that concepts ought to have convex representations.

Research has shown that individuals find it challenging to process sentences containing negations and disjunctions compared to affirmative statements and conjunctions \citep{Nordmeyer.Frank.2023pff,Sloutsky.Goldvarg.2004mro}. Our model explains this cognitive quirk because the algebra does not support low-level native representation of negation 
and disjunction, necessitating higher-level representations for these connectives.

The Gestalt theory presented by \citet{Kohler.1971tsp} faced criticism due to the absence of a biological rationale. However, the invariant introduced in this paper is founded on robust biological grounds and resonates with Köhler's notion of a {\em unit}.

The term {\em synergy}, prevalent in psychophysics, describes the intrinsic coordination of a large number of degrees of freedom acting in concert \citep{Bernstein.1967tca,Latash.2008s,Santello.et.al.2013nbo}. Synergies correspond well to the eigenvectors of the message sequence covariance matrix, the basis vectors of the subspace forming the linear hull of a convex cone.

\subsection{A programming language for the brain?}
The operations carried out by neuron populations resemble low-level assembly instructions in a computer, with data represented as convex cones. The set of operations outlined in \sectref{sect:neuronal-implementation} is notably potent, enabling sophisticated operations to be accomplished with only a few neuron populations. It raises the intriguing possibility of a higher-level language that can be compiled or interpreted through these operations.

This model introduces the possibility of a neuronal ``disassembler''---a tool capable of taking descriptions of population networks and reverse-engineering them to deduce their high-level functions.

The mechanistic model represents the widespread class of CNS neurons that receive
glutamatergic excitation through AMPA and NMDA receptors and GABAergic inhibition
through GABA receptors. Many neuronal types differ in receptor composition,
intrinsic conductances, dendritic structure, and plasticity. This diversity does
not weaken the present existence claim. If this widespread class implements the
derived operations, then the mammalian CNS has at least the capacity to express
and process information through an algebra of cones. Other neuronal classes may
implement additional operations or alternative representations.

\section{Conclusions}
\label{sect:conclusions}
This paper derives a population-level algebraic description of information processing \emph{within an explicit model class}. Under the assumptions of (i) population messages represented as rate or low-pass filtered rate vectors, and (ii) a mechanistically motivated neuron/synapse model in which effective synaptic mixing is nonnegative, the invariant objects associated with families of messages are naturally described as \emph{convex cones} (in an appropriate real Hilbert space of signals). In this sense, neuron populations in this model class implement an algebra of invariants that is conveniently expressed as cone operations.

The approach is grounded in a reduced mechanistic neuron model with plasticity developed in \citet{Nilsson.2026meo}. The present paper does not claim that this model exhausts the diversity of neuronal and synaptic mechanisms in the CNS ({\it e.g.}, spike-timing-dependent plasticity or divisive/shunting inhibition). Rather, the purpose is to show that, when the stated constraints hold, the cone-structured invariants and their algebra follow as mathematical consequences of the population mapping. In \sectref{sect:recurrent-networks}, we presented a comprehensive example grounded in
solid experimental results, demonstrating how observed neural organization
can be interpreted within the cone-algebraic framework developed here. We
look forward to further empirical data from recorded neural populations and
to investigating how such findings can be incorporated into, and used to
refine, the present theory.

The resulting mathematical structure emerges from combining (a) invariance of message families under population propagation, (b) sparsity and high-dimensional geometry, (c) plasticity as a mechanism that selects effective mappings within the admissible class, and (d) activation-function properties that support stable population-level transformations.

A further implication is that identifying an invariant requires not a single message but an ensemble (sequence) of messages. Within the present framework, matrix embeddings (subspaces and cones) provide operations---such as sum, intersection, projection, and rejection---that implement generalization, novelty isolation, and shared-structure extraction more naturally than point (vector) embeddings. In particular, cones provide a convenient representation for partial orders and hierarchical relations.

Finally, while this article does not claim that the CNS universally represents cognitive \emph{concepts} as convex cones, it suggests that populations operating in the stated regime possess the computational capacity to communicate, process, retrieve, and store cone-like invariants. Extending the framework to richer temporal codes and heterogeneous cell types remains an open and promising direction.

\section*{Acknowledgments}
This work was supported by the Air Force Office of Scientific Research under award number FA8655-25-1-7007.
The author is grateful to Dr. Sverker Janson at RISE for additional support.
OpenAI's ChatGPT, using GPT-5.6 Sol \citep{OpenAI.2026gfi}, assisted with
language editing, mathematical proof checking, and proof revision. The author
reviewed and takes responsibility for all resulting text and mathematics.

\section*{Declaration of competing interests}
The author declares that he has no known competing financial interests or
personal relationships that could have appeared to influence the work reported
in this paper.

\section*{Author contributions}
This work was conceived, conducted, and written solely by the author.

\section*{Data availability statement}
No datasets were generated or analysed during the current study. The
Supplementary Information describing the recurrent sensorimotor
simulation and the Python source code used to generate its numerical
results are provided as supplementary material accompanying this
article.

\section*{Funding declaration}
This work was supported by the Air Force Office of Scientific Research under award number FA8655-25-1-7007.


\begin{thebibliography}{90}
	\providecommand{\natexlab}[1]{#1}
	\providecommand{\url}[1]{\texttt{#1}}
	\expandafter\ifx\csname urlstyle\endcsname\relax
	\providecommand{\doi}[1]{doi: #1}\else
	\providecommand{\doi}{doi: \begingroup \urlstyle{rm}\Url}\fi
	
	\bibitem[Aerts and Gabora(2005{\natexlab{a}})]{Aerts.Gabora.2005ato1}
	D.~Aerts and L.~Gabora.
	\newblock A theory of concepts and their combinations {I}: The structure of the
	sets of contexts and properties.
	\newblock \emph{Kybernetes}, 34\penalty0 (1/2):\penalty0 167--191, Jan.
	2005{\natexlab{a}}.
	\newblock \doi{10.1108/03684920510575799}.
	
	\bibitem[Aerts and Gabora(2005{\natexlab{b}})]{Aerts.Gabora.2005ato2}
	D.~Aerts and L.~Gabora.
	\newblock A theory of concepts and their combinations {II}: A {H}ilbert space
	representation.
	\newblock \emph{Kybernetes}, 34\penalty0 (2/2):\penalty0 192--221, Jan.
	2005{\natexlab{b}}.
	\newblock \doi{10.1108/03684920510575807}.
	
	\bibitem[Averbeck et~al.(2006)Averbeck, Latham, and
	Pouget]{Averbeck.et.al.2006ncp}
	B.~B. Averbeck, P.~E. Latham, and A.~Pouget.
	\newblock Neural correlations, population coding and computation.
	\newblock \emph{Nature Rev Neurosci}, 7:\penalty0 358--366, May 2006.
	\newblock \doi{10.1038/nrn1888}.
	
	\bibitem[Balkenius and Gärdenfors(2016)]{Balkenius.Gardenfors.2016sit}
	C.~Balkenius and P.~Gärdenfors.
	\newblock Spaces in the brain: From neurons to meanings.
	\newblock \emph{Frontiers in Psychology}, 7:\penalty0 1820, nov 2016.
	\newblock \doi{10.3389/fpsyg.2016.01820}.
	
	\bibitem[Barlow(1991)]{Barlow.1991vty}
	H.~B. Barlow.
	\newblock Vision tells you more than “what is where”.
	\newblock In A.~Gorea, editor, \emph{Representations of vision: trends and
		tacit assumptions in vision research}, pages 319--329. Cambridge University
	Press, Cambridge, 1991.
	\newblock ISBN 0-521-41228-5.
	
	\bibitem[Bernstein(1967)]{Bernstein.1967tca}
	N.~A. Bernstein.
	\newblock \emph{The co-ordination and regulation of movements}.
	\newblock Pergamon Press, Oxford, {First English} edition, 1967.
	
	\bibitem[Bertsekas(1976)]{Bertsekas.1976otg}
	D.~P. Bertsekas.
	\newblock On the {Goldstein--Levitin--Polyak} gradient projection method.
	\newblock \emph{IEEE Transactions on Automatic Control}, 21\penalty0
	(2):\penalty0 174--184, Apr. 1976.
	\newblock \doi{10.1109/TAC.1976.1101194}.
	
	\bibitem[Cand{\`e}s and Tao(2005)]{Candes.Tao.2005dlp}
	E.~J. Cand{\`e}s and T.~Tao.
	\newblock Decoding by linear programming.
	\newblock \emph{IEEE Transactions on Information Theory}, 51\penalty0
	(12):\penalty0 4203--4215, 2005.
	\newblock \doi{10.1109/TIT.2005.858979}.
	
	\bibitem[Conway(1990)]{Conway.1990acf}
	J.~B. Conway.
	\newblock \emph{A Course in Functional Analysis}, volume~96 of \emph{Graduate
		Texts in Mathematics}.
	\newblock Springer-Verlag, New York, 2 edition, 1990.
	\newblock \doi{10.1007/978-1-4757-4383-8}.
	
	\bibitem[Cooke et~al.(1985)Cooke, Keane, and Moran]{Cooke.et.al.198aep}
	R.~Cooke, M.~Keane, and W.~Moran.
	\newblock An elementary proof of {Gleason's} theorem.
	\newblock \emph{Mathematical Proceedings of the Cambridge Philosophical
		Society}, 98\penalty0 (1):\penalty0 117--128, 1985.
	\newblock \doi{10.1017/S0305004100063313}.
	
	\bibitem[Dasgupta and Gupta(2003)]{Dasgupta.Gupta.2002aep}
	S.~Dasgupta and A.~Gupta.
	\newblock {An elementary proof of a theorem of Johnson and Lindenstrauss}.
	\newblock \emph{Random Structures and Algorithms}, 22\penalty0 (1):\penalty0
	60--65, 2003.
	\newblock \doi{10.1002/rsa.10073}.
	
	\bibitem[Dean et~al.(2002)Dean, Porrill, and Stone]{Dean.et.al.2002dcb}
	P.~Dean, J.~Porrill, and J.~V. Stone.
	\newblock Decorrelation control by the cerebellum achieves oculomotor plant
	compensation in simulated vestibulo-ocular reflex.
	\newblock \emph{Proceedings of the Royal Society of London. Series B:
		Biological Sciences}, 269\penalty0 (1503):\penalty0 1895--1904, sep 2002.
	\newblock \doi{10.1098/rspb.2002.2103}.
	
	\bibitem[DeValois and DeValois(1991)]{DeValois.DeValois.1991sv}
	R.~L. DeValois and K.~K. DeValois.
	\newblock \emph{Spatial Vision}.
	\newblock Oxford University Press, jun 1991.
	\newblock ISBN 9780195066579.
	\newblock \doi{10.1093/acprof:oso/9780195066579.001.0001}.
	
	\bibitem[Donoho(1995)]{Donoho.1995dnb}
	D.~L. Donoho.
	\newblock {De-Noising by Soft-Thresholding}.
	\newblock \emph{{IEEE} Transactions on Information Theory}, 41\penalty0
	(3):\penalty0 613--627, May 1995.
	\newblock \doi{10.1109/18.382009}.
	
	\bibitem[Doroslova\v{c}ki and Fan(1996)]{Doroslovacki.Fan.1996wbl}
	M.~Doroslova\v{c}ki and H.~Fan.
	\newblock Wavelet-based linear system modeling and adaptive filtering.
	\newblock \emph{IEEE Trans Signal Proc}, 44\penalty0 (5):\penalty0 1156--1167,
	May 1996.
	\newblock \doi{10.1109/78.502328}.
	
	\bibitem[Ecker et~al.(2010)Ecker, Berens, Keliris, Bethge, Logothetis, and
	Tolias]{Ecker.et.al.2010dnf}
	A.~S. Ecker, P.~Berens, G.~A. Keliris, M.~Bethge, N.~K. Logothetis, and A.~S.
	Tolias.
	\newblock Decorrelated neuronal firing in cortical microcircuits.
	\newblock \emph{Science}, 327\penalty0 (5965):\penalty0 584--587, jan 2010.
	\newblock \doi{10.1126/science.1179867}.
	
	\bibitem[Eggermont(1990)]{Eggermont.1990tcb}
	J.~J. Eggermont.
	\newblock \emph{The Correlative Brain: Theory and Experiment in Neural
		Interaction}.
	\newblock Springer Berlin Heidelberg, 1990.
	\newblock \doi{10.1007/978-3-642-51033-5}.
	
	\bibitem[Friston et~al.(2006)Friston, Kilner, and
	Harrison]{Friston.et.al.2006afe}
	K.~Friston, J.~Kilner, and L.~Harrison.
	\newblock A free energy principle for the brain.
	\newblock \emph{J Physiol Paris}, 100:\penalty0 70--87, 2006.
	\newblock \doi{10.1016/j.jphysparis.2006.10.001}.
	
	\bibitem[Fujita(1982)]{Fujita.1982afm}
	M.~Fujita.
	\newblock Adaptive filter model of the cerebellum.
	\newblock \emph{Biol Cybern}, 45\penalty0 (3):\penalty0 195--206, 1982.
	\newblock ISSN 0340-1200.
	\newblock \doi{10.1007/BF00336192}.
	
	\bibitem[Földi{\'{a}}k(1990)]{Foldiak.1990fsr}
	P.~Földi{\'{a}}k.
	\newblock Forming sparse representations by local anti-{H}ebbian learning.
	\newblock \emph{Biological Cybernetics}, 64\penalty0 (2):\penalty0 165--170,
	dec 1990.
	\newblock \doi{10.1007/bf02331346}.
	
	\bibitem[Ganguli and Sompolinsky(2012)]{Ganguli.Sompolinsky.2012css}
	S.~Ganguli and H.~Sompolinsky.
	\newblock Compressed sensing, sparsity, and dimensionality in neuronal
	information processing and data analysis.
	\newblock \emph{Annu Rev Neurosci}, 35:\penalty0 485--508, March 2012.
	\newblock \doi{10.1146/annurev-neuro-062111-150410}.
	
	\bibitem[Ganmor et~al.(2011)Ganmor, Segev, and
	Schneidman]{Ganmor.et.al.2011slo}
	E.~Ganmor, R.~Segev, and E.~Schneidman.
	\newblock Sparse low-order interaction network underlies a highly correlated
	and learnable neural population code.
	\newblock \emph{PNAS}, 108\penalty0 (23):\penalty0 9679--9684, June 2011.
	\newblock \doi{10.1073/pnas.1019641108}.
	
	\bibitem[G{\"a}rdenfors(2000)]{Gardenfors.2000cst}
	P.~G{\"a}rdenfors.
	\newblock \emph{Conceptual spaces: the geometry of thought}.
	\newblock MIT Press, Cambridge, Mass., 2000.
	\newblock ISBN 0262071991.
	
	\bibitem[Geiersbach and Pflug(2019)]{Geiersbach.Pflug.2019psg}
	C.~Geiersbach and G.~C. Pflug.
	\newblock Projected stochastic gradients for convex constrained problems in
	{Hilbert} spaces.
	\newblock \emph{SIAM Journal on Optimization}, 29\penalty0 (3):\penalty0
	2079--2099, 2019.
	\newblock \doi{10.1137/18M1200208}.
	
	\bibitem[Greer(1984)]{Greer.1984ato}
	R.~Greer.
	\newblock A tutorial on polyhedral convex cones.
	\newblock In R.~Greer, editor, \emph{Trees and Hills: Methodology for
		Maximizing Functions of Systems of Linear Relations}, volume~96 of
	\emph{North-Holland Mathematics Studies}, chapter~2, pages 15--81.
	North-Holland, 1984.
	\newblock \doi{10.1016/S0304-0208(08)72857-3}.
	
	\bibitem[Gurnee et~al.(2026)Gurnee, Sofroniew, Pearce, Piotrowski, Kauvar,
	Chen, Soligo, Bogdan, Ong, Wang, Thompson, Abrahams, Kantamneni, Ameisen,
	Batson, and Lindsey]{Gurnee.et.al.2026vfr}
	W.~Gurnee, N.~Sofroniew, A.~Pearce, M.~Piotrowski, I.~Kauvar, R.~Chen,
	A.~Soligo, P.~Bogdan, E.~Ong, R.~Wang, B.~Thompson, D.~Abrahams,
	S.~Kantamneni, E.~Ameisen, J.~Batson, and J.~Lindsey.
	\newblock Verbalizable representations form a global workspace in language
	models.
	\newblock \emph{Transformer Circuits Thread}, 2026.
	\newblock URL \url{https://transformer-circuits.pub/2026/workspace/index.html}.
	
	\bibitem[Haykin(2002)]{Haykin.2002aft}
	S.~S. Haykin.
	\newblock \emph{Adaptive filter theory}.
	\newblock Prentice Hall, Upper Saddle River, N.J., 4th edition, 2002.
	\newblock ISBN 0130901261.
	
	\bibitem[Hongo et~al.(1972)Hongo, Jankowska, and Lundberg]{Hongo.et.al.1972rte}
	T.~Hongo, E.~Jankowska, and A.~Lundberg.
	\newblock The rubrospinal tract. {IV}. effects on interneurones.
	\newblock \emph{Experimental Brain Research}, 15:\penalty0 54--78, 1972.
	\newblock \doi{10.1007/BF00234958}.
	
	\bibitem[Horn and Johnson(2012)]{Horn.Johnson.2012ma}
	R.~A. Horn and C.~R. Johnson.
	\newblock \emph{Matrix Analysis}.
	\newblock Cambridge University Press, 2 edition, 2012.
	\newblock \doi{10.1017/CBO9781139020411}.
	
	\bibitem[Ingram and Marsh(1991)]{Ingram.Marsh.1991poc}
	J.~M. Ingram and M.~Marsh.
	\newblock Projections onto convex cones in {H}ilbert space.
	\newblock \emph{Journal of Approximation Theory}, 64\penalty0 (3):\penalty0
	343--350, mar 1991.
	\newblock \doi{10.1016/0021-9045(91)90067-k}.
	
	\bibitem[Johnson and Lindenstrauss(1984)]{Johnson.Lindenstrauss.1984eol}
	W.~B. Johnson and J.~Lindenstrauss.
	\newblock Extensions of {L}ipschitz mappings into a {H}ilbert space.
	\newblock In R.~Beals, A.~Beck, A.~Bellow, and A.~Hajian, editors,
	\emph{Conference on Modern Analysis and Probability}, volume~26 of
	\emph{Contemporary Mathematics}, pages 189--206. American Mathematical
	Society, 1984.
	\newblock \doi{10.1090/conm/026/737400}.
	
	\bibitem[Kailath(1968)]{Kailath.1968aia}
	T.~Kailath.
	\newblock An innovations approach to least-squares estimation--part i: Linear
	filtering in additive white noise.
	\newblock \emph{{IEEE} Transactions on Automatic Control}, 13\penalty0
	(6):\penalty0 646--655, dec 1968.
	\newblock \doi{10.1109/tac.1968.1099025}.
	
	\bibitem[Kanerva(1988)]{Kanerva.1988sdm}
	P.~Kanerva.
	\newblock \emph{Sparse Distributed Memory}.
	\newblock M.I.T. Press, Cambridge, Mass., 1988.
	\newblock ISBN 0-262-11132-2.
	
	\bibitem[Kawato(1999)]{Kawato.1999imf}
	M.~Kawato.
	\newblock Internal models for motor control and trajectory planning.
	\newblock \emph{Curr Opin Neurobiol}, 9\penalty0 (6):\penalty0 718--727, Dec
	1999.
	\newblock \doi{10.1016/S0959-4388(99)00028-8}.
	
	\bibitem[Kleyko et~al.(2022)Kleyko, Rachkovskij, Osipov, and
	Rahimi]{Kleyko.et.al.2022aso}
	D.~Kleyko, D.~A. Rachkovskij, E.~Osipov, and A.~Rahimi.
	\newblock A survey on hyperdimensional computing aka vector symbolic
	architectures, part {I}: Models and data transformations.
	\newblock \emph{{ACM} Computing Surveys}, 55\penalty0 (6):\penalty0 130, dec
	2022.
	\newblock \doi{10.1145/3538531}.
	
	\bibitem[Kleyko et~al.(2023)Kleyko, Rachkovskij, Osipov, and
	Rahimi]{Kleyko.et.al.2023aso}
	D.~Kleyko, D.~Rachkovskij, E.~Osipov, and A.~Rahimi.
	\newblock A survey on hyperdimensional computing aka vector symbolic
	architectures, part {II}: Applications, cognitive models, and challenges.
	\newblock \emph{{ACM} Computing Surveys}, 55\penalty0 (9):\penalty0 175, jan
	2023.
	\newblock \doi{10.1145/3558000}.
	
	\bibitem[K{\"o}hler(1971)]{Kohler.1971tsp}
	W.~K{\"o}hler.
	\newblock \emph{The selected papers of {W}olfgang {K}{\"o}hler}.
	\newblock Liveright, New York, 1971.
	\newblock ISBN 0-87140-505-9.
	\newblock Edited by Mary Henle.
	
	\bibitem[Kohn et~al.(2016)Kohn, Coen-Cagli, Kanitscheider, and
	Pouget]{Kohn.et.al.2016can}
	A.~Kohn, R.~Coen-Cagli, I.~Kanitscheider, and A.~Pouget.
	\newblock Correlations and neuronal population information.
	\newblock \emph{Annual Review of Neuroscience}, 39\penalty0 (1):\penalty0
	237--256, jul 2016.
	\newblock \doi{10.1146/annurev-neuro-070815-013851}.
	
	\bibitem[Kohonen(1977)]{Kohonen.1977am}
	T.~Kohonen.
	\newblock \emph{Associative Memory: A System-Theoretical Approach}.
	\newblock Springer, 1977.
	\newblock \doi{10.1007/978-3-642-96384-1}.
	\newblock Editors K. S. Fu and W. D. Keidel and W. J. M. Levelt and H. Wolter.
	
	\bibitem[Kohonen et~al.(1981)Kohonen, Oja, and
	Lehti{\"o}]{Kohonen.et.al.1981sap}
	T.~Kohonen, E.~Oja, and P.~Lehti{\"o}.
	\newblock Storage and processing of information in distributed associative
	memory systems.
	\newblock In G.~E. Hinton and J.~A. Anderson, editors, \emph{Parallel Models of
		Associative Memory}, pages 105--143. Lawrence Erlbaum Associates, Hillsdale,
	NJ, 1981.
	
	\bibitem[Kumar et~al.(2010)Kumar, Rotter, and Aertsen]{Kumar.et.al.2010sap}
	A.~Kumar, S.~Rotter, and A.~Aertsen.
	\newblock Spiking activity propagation in neuronal networks: Reconciling
	different perspectives on neural coding.
	\newblock \emph{Nature Rev Neurosci}, 11:\penalty0 615--627, Sept 2010.
	\newblock \doi{10.1038/nrn2886}.
	
	\bibitem[Latash(2008)]{Latash.2008s}
	M.~L. Latash.
	\newblock \emph{Synergy}.
	\newblock Oxford University Press, mar 2008.
	\newblock \doi{10.1093/acprof:oso/9780195333169.001.0001}.
	
	\bibitem[Lawson and Hanson(1995)]{Lawson.Hanson.1995sls}
	C.~L. Lawson and R.~J. Hanson.
	\newblock \emph{Solving Least Squares Problems}, volume~15 of \emph{Classics in
		Applied Mathematics}.
	\newblock Society for Industrial and Applied Mathematics, Philadelphia, PA,
	1995.
	\newblock ISBN 978-0-89871-356-5.
	\newblock \doi{10.1137/1.9781611971217}.
	
	\bibitem[Luenberger(1969)]{Luenberger.1969obv}
	D.~G. Luenberger.
	\newblock \emph{Optimization by vector space methods}.
	\newblock Wiley, New York, 1969.
	\newblock ISBN 047155359X.
	
	\bibitem[Machens et~al.(2010)Machens, Romo, and Brody]{Machens.et.al.2010fbn}
	C.~K. Machens, R.~Romo, and C.~D. Brody.
	\newblock Functional, but not anatomical, separation of
	{\textquotedblleft}what{\textquotedblright} and
	{\textquotedblleft}when{\textquotedblright} in prefrontal cortex.
	\newblock \emph{The Journal of Neuroscience}, 30\penalty0 (1):\penalty0
	350--360, jan 2010.
	\newblock \doi{10.1523/jneurosci.3276-09.2010}.
	
	\bibitem[MacLennan(1993)]{MacLennan.1993fci}
	B.~J. MacLennan.
	\newblock Field computation in the brain.
	\newblock In K.~H. Pribram and K.~H. Pribram, editors, \emph{Rethinking Neural
		Networks: Quantum Fields and Biological Data}, pages 199--232. Psychology
	Press, Hillsdale, NJ, 1993.
	\newblock \doi{10.4324/9781315806570}.
	
	\bibitem[MacLennan(1995)]{MacLennan.1995cca}
	B.~J. MacLennan.
	\newblock Continuous computation and the emergence of the discrete.
	\newblock In K.~H. Pribram, editor, \emph{Origins: Brain and Self
		Organization}, pages 121--151. Lawrence Erlbaum Associates, Hillsdale, NJ,
	1995.
	\newblock \doi{10.4324/9781315789347-11}.
	
	\bibitem[Mallat(2009)]{Mallat.2009awt}
	S.~Mallat.
	\newblock \emph{A wavelet tour of signal processing: the sparse way}.
	\newblock Elsevier/Academic Press, Burlington, Mass., 3rd edition, 2009.
	\newblock ISBN 9780123743701.
	
	\bibitem[Markowitz et~al.(2018)Markowitz, Gillis, Beron, Neufeld, Robertson,
	Bhagat, Peterson, Peterson, Hyun, Linderman, Sabatini, and
	Datta]{Markowitz.et.al.2018tso}
	J.~E. Markowitz, W.~F. Gillis, C.~C. Beron, S.~Q. Neufeld, K.~Robertson, N.~D.
	Bhagat, R.~E. Peterson, E.~Peterson, M.~Hyun, S.~W. Linderman, B.~L.
	Sabatini, and S.~R. Datta.
	\newblock The striatum organizes {3D} behavior via moment-to-moment action
	selection.
	\newblock \emph{Cell}, 174\penalty0 (1):\penalty0 44--58.e17, June 2018.
	\newblock ISSN 0092-8674.
	\newblock \doi{10.1016/j.cell.2018.04.019}.
	
	\bibitem[McBain(2012)]{McBain.2012cin}
	C.~J. McBain.
	\newblock Cortical inhibitory neuron basket cells: from circuit function to
	disruption.
	\newblock \emph{The Journal of Physiology}, 590\penalty0 (4):\penalty0
	667--667, feb 2012.
	\newblock \doi{10.1113/jphysiol.2012.227967}.
	
	\bibitem[Mitchell and Silver(2003)]{Mitchell.Silver.2003sim}
	S.~J. Mitchell and R.~A. Silver.
	\newblock Shunting inhibition modulates neuronal gain during synaptic
	excitation.
	\newblock \emph{Neuron}, 38\penalty0 (3):\penalty0 433--445, May 2003.
	\newblock \doi{10.1016/S0896-6273(03)00200-9}.
	
	\bibitem[Mogensen et~al.(2019)Mogensen, Norrlid, Enander, Wahlbom, and
	Jörntell]{Mogensen.et.al.2019aor}
	H.~Mogensen, J.~Norrlid, J.~M.~D. Enander, A.~Wahlbom, and H.~Jörntell.
	\newblock Absence of repetitive correlation patterns between pairs of adjacent
	neocortical neurons {\it in vivo}.
	\newblock \emph{Frontiers in Neural Circuits}, 13:\penalty0 48, jul 2019.
	\newblock \doi{10.3389/fncir.2019.00048}.
	
	\bibitem[Moreau(1962)]{Moreau.1962dod}
	J.-J. Moreau.
	\newblock D{\'e}composition orthogonale d'un espace hilbertien selon deux
	c{\^o}nes mutuellement polaires.
	\newblock \emph{C. R. Acad. Sci. Paris}, 255:\penalty0 238--240, 1962.
	
	\bibitem[Nilsson(2016)]{Nilsson.2016noa}
	M.~Nilsson.
	\newblock Neuronal ``op-amps'' implement adaptive control in biology and
	robotics.
	\newblock In M.~Bianchi and A.~Moscatelli, editors, \emph{Human and Robot
		Hands: Sensorimotor Synergies to Bridge the Gap Between Neuroscience and
		Robotics}, chapter~6, pages 69--86. Springer International Publishing, Cham,
	2016.
	\newblock ISBN 978-3-319-26706-7.
	\newblock \doi{10.1007/978-3-319-26706-7_6}.
	
	\bibitem[Nilsson(2026)]{Nilsson.2026meo}
	M.~N.~P. Nilsson.
	\newblock Mechanistic explanation of neuroplasticity using equivalent circuits.
	\newblock \emph{Frontiers in Computational Neuroscience}, 20:\penalty0 1716559,
	Feb. 2026.
	\newblock ISSN 1662-5188.
	\newblock \doi{10.3389/fncom.2026.1716559}.
	
	\bibitem[Nilsson and J\"orntell(2021)]{Nilsson.Jorntell.2021ccf}
	M.~N.~P. Nilsson and H.~J\"orntell.
	\newblock Channel current fluctuations conclusively explain neuronal encoding
	of internal potential into spike trains.
	\newblock \emph{Phys. Rev. E}, 103:\penalty0 022407, Feb 2021.
	\newblock \doi{10.1103/PhysRevE.103.022407}.
	
	\bibitem[Nordmeyer and Frank(2023)]{Nordmeyer.Frank.2023pff}
	A.~E. Nordmeyer and M.~C. Frank.
	\newblock Pragmatic felicity facilitates the production and comprehension of
	negation.
	\newblock \emph{Collabra: Psychology}, 9\penalty0 (1):\penalty0 67931, 2023.
	\newblock \doi{10.1525/collabra.67931}.
	
	\bibitem[Oja(1983)]{Oja.1983smo}
	E.~Oja.
	\newblock \emph{Subspace methods of pattern recognition}.
	\newblock Research Studies Press, Letchworth, Herts., 1983.
	\newblock ISBN 0863800106.
	
	\bibitem[Olshausen and Field(1997)]{Olshausen.Field.1997scw}
	B.~A. Olshausen and D.~J. Field.
	\newblock Sparse coding with an overcomplete basis set: A strategy employed by
	{V}1?
	\newblock \emph{Vision Research}, 37\penalty0 (23):\penalty0 3311--3325, dec
	1997.
	\newblock \doi{10.1016/s0042-6989(97)00169-7}.
	
	\bibitem[Olshausen and Field(2004)]{Olshausen.Field.2004sco}
	B.~A. Olshausen and D.~J. Field.
	\newblock Sparse coding of sensory inputs.
	\newblock \emph{Current Opinion in Neurobiology}, 14\penalty0 (4):\penalty0
	481--487, aug 2004.
	\newblock \doi{10.1016/j.conb.2004.07.007}.
	
	\bibitem[{OpenAI}(2026)]{OpenAI.2026gfi}
	{OpenAI}.
	\newblock {GPT-5.6}: Frontier intelligence that scales with your ambition.
	\newblock \url{https://openai.com/index/gpt-5-6/}, July 2026.
	\newblock Accessed 19 July 2026.
	
	\bibitem[Palm(2013)]{Palm.2013nam}
	G.~Palm.
	\newblock Neural associative memories and sparse coding.
	\newblock \emph{Neural Networks}, 37:\penalty0 165--171, jan 2013.
	\newblock \doi{10.1016/j.neunet.2012.08.013}.
	
	\bibitem[Panzeri et~al.(2022)Panzeri, Moroni, Safaai, and
	Harvey]{Panzeri.et.al.2022tsa}
	S.~Panzeri, M.~Moroni, H.~Safaai, and C.~D. Harvey.
	\newblock The structures and functions of correlations in neural population
	codes.
	\newblock \emph{Nature Reviews Neuroscience}, 23\penalty0 (9):\penalty0
	551--567, jun 2022.
	\newblock \doi{10.1038/s41583-022-00606-4}.
	
	\bibitem[Pellionisz and Llinás(1985)]{Pellionisz.Llinas.1985tnt}
	A.~J. Pellionisz and R.~Llinás.
	\newblock Tensor network theory of the metaorganization of functional
	geometries in the central nervous system.
	\newblock \emph{Neuroscience}, 16\penalty0 (2):\penalty0 245--273, 1985.
	\newblock \doi{10.1016/0306-4522(85)90001-6}.
	
	\bibitem[Perkel and Bullock(1968)]{Perkel.Bullock.1968nc}
	D.~H. Perkel and T.~H. Bullock.
	\newblock Neural coding: A report based on an {NRP} work session.
	\newblock \emph{Neurosciences Research Program Bulletin}, 6\penalty0
	(3):\penalty0 219--349, December 1968.
	
	\bibitem[Pollen et~al.(1971)Pollen, Lee, and Taylor]{Pollen.et.al.1971hdt}
	D.~A. Pollen, J.~R. Lee, and J.~H. Taylor.
	\newblock How does the striate cortex begin the reconstruction of the visual
	world?
	\newblock \emph{Science}, 173\penalty0 (3991):\penalty0 74--77, jul 1971.
	\newblock \doi{10.1126/science.173.3991.74}.
	
	\bibitem[Quian~Quiroga et~al.(2008)Quian~Quiroga, Kreiman, Koch, and
	Fried]{QuianQuiroga.et.al.2008sbn}
	R.~Quian~Quiroga, G.~Kreiman, C.~Koch, and I.~Fried.
	\newblock Sparse but not {`Grandmother-cell'} coding in the medial temporal
	lobe.
	\newblock \emph{Trends in Cognitive Sciences}, 12\penalty0 (3):\penalty0
	87--91, mar 2008.
	\newblock \doi{10.1016/j.tics.2007.12.003}.
	
	\bibitem[Rathelot and Padel(1997)]{Rathelot.Padel.1997asi}
	J.-A. Rathelot and Y.~Padel.
	\newblock Ascending spinal influences on rubrospinal cells in the cat.
	\newblock \emph{Experimental Brain Research}, 116\penalty0 (2):\penalty0
	326--340, Sept. 1997.
	\newblock ISSN 0014-4819.
	\newblock \doi{10.1007/pl00005760}.
	
	\bibitem[Rosch and Mervis(1975)]{Rosch.Mervis.1975frs}
	E.~Rosch and C.~B. Mervis.
	\newblock Family resemblances: Studies in the internal structure of categories.
	\newblock \emph{Cognitive Psychology}, 7\penalty0 (4):\penalty0 573--605, 1975.
	\newblock \doi{10.1016/0010-0285(75)90024-9}.
	
	\bibitem[Sakurai(1996)]{Sakurai.1996pcb}
	Y.~Sakurai.
	\newblock Population coding by cell assemblies{\textemdash}what it really is in
	the brain.
	\newblock \emph{Neuroscience Research}, 26\penalty0 (1):\penalty0 1--16, sep
	1996.
	\newblock \doi{10.1016/0168-0102(96)01075-9}.
	
	\bibitem[Sanger(2003)]{Sanger.2003npc}
	T.~D. Sanger.
	\newblock Neural population codes.
	\newblock \emph{Current Opinion in Neurobiology}, 13\penalty0 (2):\penalty0
	238--249, apr 2003.
	\newblock \doi{10.1016/s0959-4388(03)00034-5}.
	
	\bibitem[Santello et~al.(2013)Santello, Baud-Bovy, and
	J\"orntell]{Santello.et.al.2013nbo}
	M.~Santello, G.~Baud-Bovy, and H.~J\"orntell.
	\newblock Neural bases of hand synergies.
	\newblock \emph{Front Comput Neurosci}, 7:\penalty0 23, 2013.
	\newblock ISSN 1662-5188.
	\newblock \doi{10.3389/fncom.2013.00023}.
	
	\bibitem[Sayed(2003)]{Sayed.2003foa}
	A.~H. Sayed.
	\newblock \emph{Fundamentals of adaptive filtering}.
	\newblock Wiley, Hoboken, N.J., 2003.
	\newblock ISBN 0471461261.
	
	\bibitem[Schultz and Dickinson(2000)]{Schultz.Dickinson.2000nco}
	W.~Schultz and A.~Dickinson.
	\newblock Neuronal coding of prediction errors.
	\newblock \emph{Annual Review of Neuroscience}, 23\penalty0 (1):\penalty0
	473--500, 2000.
	\newblock \doi{10.1146/annurev.neuro.23.1.473}.
	\newblock PMID: 10845072.
	
	\bibitem[Sloutsky and Goldvarg(2004)]{Sloutsky.Goldvarg.2004mro}
	V.~M. Sloutsky and Y.~Goldvarg.
	\newblock Mental representation of logical connectives.
	\newblock \emph{The Quarterly Journal of Experimental Psychology Section A},
	57\penalty0 (4):\penalty0 636--665, may 2004.
	\newblock \doi{10.1080/02724980343000413}.
	
	\bibitem[Stewart et~al.(2010)Stewart, Choo, and
	Eliasmith]{Stewart.et.al.2010sri}
	T.~C. Stewart, X.~Choo, and C.~Eliasmith.
	\newblock Symbolic reasoning in spiking neurons: A model of the cortex/basal
	ganglia/thalamus loop.
	\newblock In \emph{Proceedings of the Annual Meeting of the Cognitive Science
		Society ({CogSci}’10)}, volume~32, pages 1100--1105, 2010.
	\newblock URL \url{https://escholarship.org/uc/item/81k1r713}.
	
	\bibitem[Tsuda(2001)]{Tsuda.2001tsm}
	K.~Tsuda.
	\newblock The subspace method in {H}ilbert space.
	\newblock \emph{Systems and Computers in Japan}, 32\penalty0 (6):\penalty0
	55--61, 2001.
	\newblock \doi{10.1002/scj.1034}.
	
	\bibitem[van Polanen(2014)]{vanPolanen.2014fih}
	V.~van Polanen.
	\newblock \emph{Findings in Haptic (Re)search}.
	\newblock PhD thesis, {VU} University Amsterdam, 2014.
	\newblock ISBN: 978-94-6259-188-2.
	
	\bibitem[van Rijsbergen(2004)]{VanRijsbergen.2004tgo}
	C.~J. van Rijsbergen.
	\newblock \emph{The Geometry of Information Retrieval}.
	\newblock Cambridge University Press, 2004.
	\newblock \doi{10.1017/CBO9780511543333}.
	\newblock URL \url{https://doi.org/10.1017/CBO9780511543333}.
	
	\bibitem[Vinay and Padel(1990)]{Vinay.Padel.1990sto}
	L.~Vinay and Y.~Padel.
	\newblock Spatio-temporal organization of the somaesthetic projections in the
	red nucleus transmitted through the spino-rubral pathway in the cat.
	\newblock \emph{Experimental Brain Research}, 79\penalty0 (2):\penalty0
	412--426, 1990.
	\newblock \doi{10.1007/BF00608253}.
	
	\bibitem[von Neumann(1956)]{VonNeumann.1956pla}
	J.~von Neumann.
	\newblock Probabilistic logics and the synthesis of reliable organisms from
	unreliable components.
	\newblock In \emph{Automata Studies. ({AM}-34)}, pages 43--98. Princeton
	University Press, dec 1956.
	\newblock \doi{10.1515/9781400882618-003}.
	
	\bibitem[von Neumann(1958)]{VonNeumann.1958tca}
	J.~von Neumann.
	\newblock \emph{The Computer and the Brain}.
	\newblock Yale University Press, dec 1958.
	\newblock ISBN 978-0300181111.
	\newblock \doi{10.12987/9780300188080}.
	
	\bibitem[Watanabe(1969)]{Watanabe.1969kag}
	S.~Watanabe.
	\newblock \emph{Knowing and Guessing: A Quantitative Study of Inference and
		Information}.
	\newblock John Wiley, New York, 1969.
	\newblock ISBN 0-471-92130-0.
	
	\bibitem[Widrow and Hoff(1960)]{Widrow.Hoff.1960asc}
	B.~Widrow and M.~E. Hoff.
	\newblock Adaptive switching circuits.
	\newblock In \emph{{IRE WESCON Convention Record}}, pages 96--104, Los Angeles,
	CA, Aug. 1960. Institute of Radio Engineers.
	\newblock \doi{10.21236/ad0241531}.
	\newblock Part 4.
	
	\bibitem[Widrow and Stearns(1985)]{Widrow.Stearns.1985asp}
	B.~Widrow and S.~D. Stearns.
	\newblock \emph{Adaptive signal processing}.
	\newblock Prentice-Hall signal processing series. Prentice-Hall, Englewood
	Cliffs, N.J, 1985.
	\newblock ISBN 0-13-004029-0.
	
	\bibitem[Wold(1938)]{Wold.1938asi}
	H.~O. Wold.
	\newblock \emph{{A Study in the Analysis of Stationary Time Series}}.
	\newblock Almquist \& Wiksell, Uppsala, 1938.
	\newblock Doctoral thesis, Stockholm College.
	
	\bibitem[Wolpert et~al.(1998)Wolpert, Miall, and Kawato]{Wolpert.et.al.1998imi}
	D.~M. Wolpert, R.~C. Miall, and M.~Kawato.
	\newblock Internal models in the cerebellum.
	\newblock \emph{Trends in Cognitive Sciences}, 2:\penalty0 338--347, 1998.
	\newblock \doi{10.1016/S1364-6613(98)01221-2}.
	
	\bibitem[Yu et~al.(2009)Yu, Cunningham, Santhanam, Ryu, Shenoy, and
	Sahani]{Yu.et.al.2009gpf}
	B.~M. Yu, J.~P. Cunningham, G.~Santhanam, S.~I. Ryu, K.~V. Shenoy, and
	M.~Sahani.
	\newblock Gaussian-process factor analysis for low-dimensional single-trial
	analysis of neural population activity.
	\newblock \emph{Journal of Neurophysiology}, 102\penalty0 (1):\penalty0
	614--635, jul 2009.
	\newblock \doi{10.1152/jn.90941.2008}.
	
	\bibitem[Zhao et~al.(2023)Zhao, Zhou, Li, Tang, Wang, Hou, Min, Zhang, Zhang,
	Dong, Du, Yang, Chen, Chen, Jiang, Ren, Li, Tang, Liu, Liu, Nie, and
	Wen]{zhao.et.al.2023aso}
	W.~X. Zhao, K.~Zhou, J.~Li, T.~Tang, X.~Wang, Y.~Hou, Y.~Min, B.~Zhang,
	J.~Zhang, Z.~Dong, Y.~Du, C.~Yang, Y.~Chen, Z.~Chen, J.~Jiang, R.~Ren, Y.~Li,
	X.~Tang, Z.~Liu, P.~Liu, J.-Y. Nie, and J.-R. Wen.
	\newblock A survey of large language models.
	\newblock \emph{arXiv}, June 2023.
	\newblock \doi{10.48550/arXiv.2303.18223}.
	
	\bibitem[Zohary et~al.(1994)Zohary, Shadlen, and Newsome]{Zohary.et.al.1994cnd}
	E.~Zohary, M.~N. Shadlen, and W.~T. Newsome.
	\newblock Correlated neuronal discharge rate and its implications for
	psychophysical performance.
	\newblock \emph{Nature}, 370\penalty0 (6485):\penalty0 140--143, jul 1994.
	\newblock \doi{10.1038/370140a0}.
	
\end{thebibliography}

\end{document}